\documentclass[12pt,draftclsnofoot,onecolumn]{IEEEtran}
\usepackage{subfigure}
\usepackage{moreverb}
\usepackage{epsfig}
\usepackage{amsmath,amssymb,amsthm,mathrsfs,amsfonts,dsfont}
\usepackage{adjustbox,lipsum}
\usepackage{algorithm,algorithmic}
\usepackage{amsfonts}
\usepackage{epsfig}
\usepackage{amssymb}
\usepackage{amsmath}
\usepackage{amsthm}
\usepackage{subfigure}
\usepackage{multirow}
\usepackage{rotating}
\usepackage{graphicx}
\usepackage{tabularx}
\usepackage{array}
\usepackage{anyfontsize}
\usepackage{color,soul}
\usepackage{graphicx,dblfloatfix}
\usepackage{epstopdf}
\usepackage{blindtext}
\usepackage{amsmath}
\usepackage{amsthm,amssymb,amsmath,bm}
\usepackage{subfigure}
\usepackage{amsfonts}
\usepackage{epsfig}
\usepackage{amssymb}
\usepackage{amsmath}
\usepackage{cite}
\usepackage{graphicx}
\usepackage{fancyhdr}
\usepackage{subfigure}
\usepackage[subfigure]{tocloft}
\usepackage[font={small}]{caption}
\usepackage{subfigure}
\usepackage{tabularx}
\usepackage{cite}
\usepackage{color}

\usepackage{colortbl}

\begin{document}

\title{Cooperative Multi-Bitrate Video Caching and Transcoding in Multicarrier NOMA-Assisted Heterogeneous Virtualized MEC Networks}
\author{Sepehr Rezvani,
        Saeedeh Parsaeefard,
        \IEEEmembership{Senior Member,~IEEE,}
        Nader Mokari,
        \IEEEmembership{Member,~IEEE,}
        Mohammad~R.~Javan,
        \IEEEmembership{Member,~IEEE,}
        and Halim Yanikomeroglu,
        \IEEEmembership{Fellow,~IEEE}
\thanks{Sepehr~Rezvani and Nader~Mokari are with the Department of Electrical and Computer Engineering, Tarbiat Modares University, Tehran, Iran.}
\thanks{Saeedeh Parsaeefard is with Communication Technologies \& Department, ITRC, Tehran, Iran and University of Toronto (saeideh.fard@utoronto.ca, s.parsaeifard@itrc.ac.ir).}
\thanks{Mohammad~R.~Javan is with the Department of Electrical and Robotics Engineering, Shahrood University of Technology, Shahrood, Iran.}
\thanks{H. Yanikomeroglu is with the Department of Systems and Computer Engineering, Carleton University, Ottawa, ON K1S 5B6, Canada (E-mail:
halim@sce.carleton.ca).}
}

\maketitle

\begin{abstract}
Cooperative video caching and transcoding in mobile edge computing (MEC) networks is a new paradigm for future wireless networks, e.g., 5G and 5G beyond, to reduce scarce and expensive backhaul resource usage by prefetching video files within radio access networks (RANs). Integration of this technique with other advent technologies, such as wireless network virtualization and multicarrier non-orthogonal multiple access (MC-NOMA), provides more flexible video delivery opportunities, which leads to enhancements both for the network's revenue and for the end-users' service experience. In this regard, we propose a two-phase RAF for a parallel cooperative joint multi-bitrate video caching and transcoding in heterogeneous virtualized MEC networks. In the cache placement phase, we propose novel proactive delivery-aware cache placement strategies (DACPSs) by jointly allocating physical and radio resources based on network stochastic information to exploit flexible delivery opportunities. Then, for the delivery phase, we propose a delivery policy based on the user requests and network channel conditions. The optimization problems corresponding to both phases aim to maximize the total revenue of network slices, i.e., virtual networks. Both problems are non-convex and suffer from high-computational complexities. For each phase, we show how the problem can be solved efficiently. We also propose a low-complexity RAF in which the complexity of the delivery algorithm is significantly reduced. A Delivery-aware cache refreshment strategy (DACRS) in the delivery phase is also proposed to tackle the dynamically changes of network stochastic information.
Extensive numerical assessments demonstrate a performance improvement of up to $30\%$ for our proposed DACPSs and DACRS over traditional approaches.
\newline
\emph{\textbf{Index Terms--}} Adaptive bitrate streaming, cooperative caching, mobile edge computing, multi-bitrate video transcoding, multicarrier NOMA, wireless network virtualization, cache refreshment algorithm.
\end{abstract}

\section{Introduction}
Wireless edge caching has been developed as a candidate solution for next generation wireless networks, e.g., 5G, to address high data rate and/or low latency multimedia services by proactively storing contents at the edge of wireless networks and in so doing offloading scarce and expensive backhaul links \cite{CooperativeCachingDelivery,6195469,8611393}.
Among various mobile services, mobile video services and applications are expected to account for a major percentage of the global mobile data traffic in coming years \cite{6708492,Taovideotranscod}. For this reason, video caching at the network edge has drawn a lot of attention recently \cite{6195469,EnhancingMobileVideo,Collaborativemultibitrate,CollaborativeScenariosChallenges,Taovideotranscod,QoSAwareResourceAllocation,6708492}.

In practical scenarios, due to the multiple bitrate variants of each unique video file, service providers often need to transcode video files into multiple bitrates \cite{Collaborativemultibitrate,CollaborativeScenariosChallenges,EnhancingMobileVideo,QoSAwareResourceAllocation,Taovideotranscod}. To this end, adaptive bit rate (ABR) streaming techniques have been developed to enhance the quality of delivered video in radio access networks (RANs) where each video file is adjusted according to users' requests based on their display size and network channel conditions \cite{EnhancingMobileVideo,Collaborativemultibitrate,Taovideotranscod}.

Recently, mobile edge computing (MEC) networks have emerged as a promising technology for next generation wireless networks, providing cloud caching and computing capabilities within the RAN \cite{CollaborativeScenariosChallenges,7762913,DynamicComputationOffloading,8387798,PAYMARD2019160}. Thanks to this paradigm, video files could be prefetched and/or transcoded in close proximity to end-users, leading to enormous latency and backhaul traffic reductions in wireless networks. One problem with this, however, is that duplicated video caching and transcoding in multiple resource-constrained MEC servers wastes both storage and processing resources. To tackle this issue, cooperative joint multi-bitrate video caching and transcoding (CVCT) technology is proposed where each MEC server is able to receive the requested video files from neighboring MEC servers via fronthaul links \cite{Collaborativemultibitrate}. In this architecture, each MEC server is deployed side-by-side with each base station (BS) using the generic computing platforms which provides the caching and computation capabilities in heterogeneous networks (HetNets) \cite{CollaborativeScenariosChallenges,Collaborativemultibitrate,EnhancingMobileVideo,Taovideotranscod}. By sharing both the storage and processing resources among multiple MEC servers, more video files can be prefetched within RANs which results increasing the cache hit ratio \cite{CollaborativeScenariosChallenges,Collaborativemultibitrate}. However, non-simultaneous transferring and transcoding video files wastes more time and physical resources in the CVCT system, which is not beneficial for delay-sensitive services.
To cope with this challenge, parallel video transmission and transcoding capability \cite{QoSAwareResourceAllocation,4066996} can be deployed. In the parallel CVCT system, video transcoding runs in parallel with video transmission, and all the multi-hop video transmissions (between backhaul, fronthaul, and wireless access links) also run in parallel.

Non-orthogonal multiple access (NOMA) has recently considered as a promising technology to improve the spectral efficiency of 5G wireless networks \cite{8368286,7982789}. Unlike conventional orthogonal multiple access (OMA) techniques, NOMA can significantly improve the system throughput and support the massive connectivity by using successful interference cancellation (SIC) at the receivers and a mixture of multiple messages at the transmitter \cite{8368286,7982789}. The spectral efficiency can be further improved by combining NOMA with multicarrier systems, called multicarrier NOMA (MC-NOMA), which utilizes multicarrier diversity \cite{7812683}. To reduce the capital expenses (CapEx) and operation expenses (OpEx) of RANs, wireless network virtualization technology is developed where the infrastructure providers (InPs) resources are abstracted and sliced into a number of virtual networks, also known as network slices. In this technology, InPs lease the physical resources to slices according to their availability and/or service level agreements between InPs and slices. On the other hand, each slice acts as a service provider for its own users with specific QoS level agreements \cite{7406764,7384533,7795465}. In this paper, the term slice refers to virtual network unless otherwise indicated.
Exploiting MC-NOMA in virtualized networks can further reduce the wireless bandwidth cost of slices by reusing each subcarrier for multiple users owned by each slice. In this way, the integration of the aforementioned technologies enables the CVCT technology at MEC servers and network infrastructure abstraction for different cost-efficient wireless servicing.

Transcoding a large number of videos at each resource-constrained MEC server simultaneously poses another challenge for delay-sensitive services \cite{EnhancingMobileVideo,Collaborativemultibitrate,Taovideotranscod}. The significant performance gain of the CVCT system can only be achieved when a joint distributed video caching and transcoding strategy is designed \cite{Collaborativemultibitrate}. Accordingly, the major question is \emph{which bitrate variant of a video file should be cached or transcoded to another lower bitrate variant?} An efficient design of joint power and subcarrier allocation is required to achieve the benefits of MC-NOMA in the virtulized wireless networks as well as the improved throughput. Additionally, the scheduler should be fast enough to readopt the video delivery policy based on the arrival requests of users and channel state information (CSI), specifically in realistic ultra dense 5G wireless networks with a larger number of videos. To this end, the video delivery policy needs to be lightweight.

Recently, designing efficient proactive cache placement strategies (CPSs) under the consideration of transmission strategies has become more attractive to utilize the physical-layer delivery opportunities in order to have an efficient delivery performance
\cite{8611393,Taovideotranscod,6708492,CachingTransmissionDesign,cachefogkhaled}. Actually, unlike to the conventional baseline popular/bitrate video CPSs, the performance of the proactive video CPSs can be improved by utilizing the delivery opportunities, i.e., designing a proactive delivery-aware CPS (DACPS).
Despite the huge potential, designing an efficient DACPS needs to address the following challenges:
\textbf{1) A DACPS should be efficient for the long-term of the next delivery phase.} This algorithm should cover all the delivery moments and also be efficient in the long-term. Therefore, appropriate methodologies have to be devised to handle the variations of CSI and requests of users in different moments of the next delivery phase. In other words, some estimation approaches should be performed in the proactive DACPS design. A delivery-aware cache refreshment strategy (DACRS) is also useful to tackle the unpredicted changes of network stochastic information in the delivery phase.
\textbf{2) A DACPS should efficiently utilize the delivery opportunities.} To consider all the transmission and transcoding technologies in the design of an efficient DACPS with a preferable delivery observation, joint optimization of the available physical resources, such as storage, transmission, and processing with user association and request scheduling decisions is required. The output of this optimization is only the video placement.

\subsection{Related Works}
Roughly speaking, all research on wireless edge caching can be classified into two categories: 1) delivery performance analysis for different CPSs; 2) designing CPSs in order to have an efficient delivery performance.
In the first category, \cite{GreenOFDMAresourcecache} and \cite{7925732} investigate the benefits of designing joint user association and radio resource allocation for different CPSs. The authors in \cite{GreenOFDMAresourcecache} proposed a joint resource allocation and user association algorithm in orthogonal frequency division multiple access (OFDMA)-based fronthaul capacity constrained cloud-RANs (C-RANs) in order to minimize the network delivery cost. In \cite{7925732}, the authors devised a joint resource allocation and user association algorithm in HetNets with device-to-device (D2D) communications to maximize the user revenues.
In the second category, a lot of work has been done to investigate the benefits of considering transmission opportunities in the CPS design
for different cache-assisted schemes, such as backhaul-limited networks \cite{8611393,6195469,6708492,BackhaulAwareCaching,7406764},
cooperative transmission networks \cite{FastDeliveryDistributedCaching,cachefogkhaled}, cooperative caching networks
\cite{Delayperformanceanalysis}, joint cooperative caching and transmission networks \cite{CachingTransmissionDesign,CooperativeCachingDelivery},
NOMA-assisted networks \cite{8368286,7982789,8422478,8422358}, and network virtualization \cite{7384533,7795465,7406764,7410051}. However, none of these works have utilized the benefits of video transcoding in their systems.

In the context of cloud-based video transcoding, some research efforts investigate the advantages of cloud computing and devise joint processing resource
allocation and scheduling policies to reduce the transcoding delays in the delivery phase \cite{7763747,QoSAwareResourceAllocation}. In addition, \cite{EnhancingMobileVideo,Taovideotranscod,Collaborativemultibitrate} investigate joint multi-bitrate video caching and transcoding by utilizing the ABR streaming technology in C-RANs. In \cite{EnhancingMobileVideo}, a transmission-aware joint multi-bitrate video caching and transcoding policy is devised to maximize the number and quality of concurrent video requests in each time slot in a single-cell scenario. In \cite{Taovideotranscod}, the benefits of joint caching and radio resource allocation policy is investigated for a multi-cell MEC network without any cooperation between MEC servers. Additionally, \cite{Collaborativemultibitrate} investigates the design of a transcoding-aware cache replacement strategy in the online delivery phase of a non-parallel CVCT system based on the arrival video requests. Accordingly, designing an efficient proactive DACPS for the CVCT systems is still an open problem. Furthermore, the parallel transmission and transcoding capability is not applied for the CVCT system in \cite{Collaborativemultibitrate} which can avoid wasting time and physical resources. Besides, prior works do not utilize the benefits of jointly allocating physical resources for designing an efficient DACPS. In addition, based on our most up-to-date knowledge, the impact of applying MC-NOMA in virtualized wireless networks in terms of bandwidth cost reduction is not yet addressed in the related works. In our research, we address these aforementioned challenges.

\subsection{Our Contributions}
In this paper, we consider a parallel CVCT system in a MC-NOMA-assisted heterogeneous virtualized MEC (HV-MEC) network. This network consists of multiple remote radio systems (RRSs) each equipped with a BS and MEC server that enables the CVCT capability at network edge. For this setup, we propose a virtualization model with a pricing scheme where the network slices are isolated based on the QoS level agreements. In contrast to \cite{Collaborativemultibitrate}, where the main goal was only to decrease network cost, we aim to maximize the slice revenues by jointly increasing slice incomes (which is obtained by providing access data rates for subscribed users) and decreasing slice costs. In this system, allocating more radio resources can increase the user data rates. Besides, increasing data rate of users needs more processing and/or fronthaul/backhaul resources to avoid wasting resources in the parallel system. However, allocating more radio and physical resources causes a network cost increment that degrades slice revenues. Accordingly, the inherent trade-off between the slice incomes and costs should be carefully handled.

To address this, we propose a resource allocation framework (RAF) where network operational time is divided into two phases: a cache placement phase (Phase 1); and a delivery phase (Phase 2). To the best of our knowledge, this paper is the first in the literature to propose efficient DACPSs in a parallel CVCT system for a MC-NOMA-assisted HV-MEC network based on available stochastic wireless channel distribution information (CDI) and video popularity distribution (VPD). This novel strategy is designed on the basis of jointly optimizing available physical resources as storage, processing, and transmission (transmit power of RRSs, subcarriers, and backhaul and fronthaul capacities) with user association and request scheduling to maximize the slice revenues and avoid over utilizing the available network resources.

Since the benefits of each component of the CVCT system are not well evaluated numerically in the literature, via simulation results, we investigate the performance gain of each of the caching, transcoding, and cooperative technologies when it is adopted to an initial system, where no caching, transcoding, and cooperation capabilities are considered for RRSs.
Some of these performance gain results are summarized in the following (when our proposed DACPS is adopted for the system):
\begin{itemize}
  \item Non-cooperative caching with no transcoding (only caching at RRSs without any transcoding and cooperation capabilities): 459.8\% (see Fig. \ref{Fig07Rev})
  \item Non-cooperative caching and transcoding (only caching and transcoding at RRSs without any cooperation capability): 982.9\% (see Fig. \ref{Fig07Rev})
  \item Cooperative caching with no transcoding (RRSs have storage and can collaborate with each other, but they have no processing capability.): 1076.8\% (see Fig. \ref{Fig06Rev})
  \item Cooperative caching and transcoding (CVCT system): 1740.2\% (see Fig. \ref{Fig05Rev})
\end{itemize}
For the above results, the system settings are based on Table \ref{Main simulation parameters} and the simulation environment is according to Fig. \ref{Fig03}.
This work can be considered a benchmark for future HV-MEC networks.
Extensive numerical results show that MC-NOMA outperforms the system revenue nearly $21.91\%$ compared to OMA.
Moreover, we show that our proposed DACPSs have up to $30\%$ performance gain in terms of the total revenue of slices compared to the conventional baseline video popularity/bitrate strategies.

This paper also provides a novel solution to reduce the computational complexity of our delivery algorithm with on-demand and real-time cloud services. We show that our proposed low-complexity RAF (LC-RAF) can be efficiently utilized for dense environments where there are higher levels of path loss. Last but not least, we propose a DACRS to tackle the dynamically unpredicted changes of VPD and CDI during the delivery phase. It is shown that the proposed DACRS can improve the slice revenues up to $20\%$ compared to our proposed proactive DACPSs where the adopted CPS remains fixed through the whole delivery phase.

\subsection{Paper Organization}
The rest of this paper is organized as follows. Section \ref{Section system model and Problem Formulation} presents the network architecture and
formulates the cache placement and delivery optimization problems. Section \ref{Section Solution} contains the solution of the problems and the proposed LC-RAF. The numerical results are presented in Section \ref{Section simulation results}. Our concluding remarks are provided in Section \ref{Section Conclusion}. The abbreviations used in the paper are summarized in Table \ref{Table ABBREVIATION}.
\begin{table*}[tp]
\centering
\caption{Abbreviations.}
\begin{center} \label{Table ABBREVIATION}
\scalebox{1}{\begin{tabular}{|c|c||c|c|}
 \hline \rowcolor[gray]{0.850}
 \textbf{Abbreviation} & \textbf{Definition} & \textbf{Abbreviation} & \textbf{Definition} \\
\hline
\rowcolor[gray]{0.905}
 ABR & Adaptive bit rate & IPM & Interior-point method \\
 \rowcolor[gray]{0.910}
 BS & Base station & LC-RAF & Low-complexity RAF \\
 \rowcolor[gray]{0.915}
 CCNT & Cooperative caching with no transcoding & LD & Low-diversity \\
 \rowcolor[gray]{0.920}
 CDI & Channel distribution information & LP-RRS & Low-power RRS \\
 \rowcolor[gray]{0.925}
 CPS & Cache placement strategy & MBS & Macro BS \\
 \rowcolor[gray]{0.930}
 CSI & Channel state information & MC-NOMA & Multicarrier NOMA \\
 \rowcolor[gray]{0.935}
 CVCT & Cooperative video caching and transcoding & MEC & Mobile edge computing \\
 \rowcolor[gray]{0.940}
 D.C. & Difference-of-two-concave-functions & MINLP & Mixed-integer nonlinear programming \\
 \rowcolor[gray]{0.945}
 DACPS & Delivery-aware CPS & MPV & Most popular video \\
 \rowcolor[gray]{0.950}
 DACRS & Delivery-aware cache refreshment strategy & NC & No caching \\
 \rowcolor[gray]{0.955}
 DCP & Disciplined convex programming & NoCoop & Non-cooperative \\
 \rowcolor[gray]{0.960}
 FBS & Femto BS & OMA & Orthogonal multiple access \\
 \rowcolor[gray]{0.965}
 HBV & High-bitrate video & PSD & Power spectral density \\
 \rowcolor[gray]{0.970}
 HD & High-diversity & RAF & Resource allocation framework \\
 \rowcolor[gray]{0.975}
 HP-RRS & High-power RRS & RAN & Radio access network \\
 \rowcolor[gray]{0.983}
 HV-MEC & Heterogeneous virtualized MEC & RRS & Remote radio system \\
 \rowcolor[gray]{0.990}
 IDCP & Integer disciplined convex programming & SCA & Successive convex approximation \\
  \rowcolor[gray]{0.995}
 INLP & Integer nonlinear programming & SIC & Successful interference cancellation \\
  \rowcolor[gray]{1}
 InP & Infrastructure provider & VPD & Video popularity distribution \\
 \hline
\end{tabular}}
\end{center}
\end{table*}

\section{System Model and Problem Formulation}\label{Section system model and Problem Formulation}
\allowdisplaybreaks
\subsection{Network Architecture and System Settings}
\begin{figure}
\centering
\includegraphics[scale=0.3]{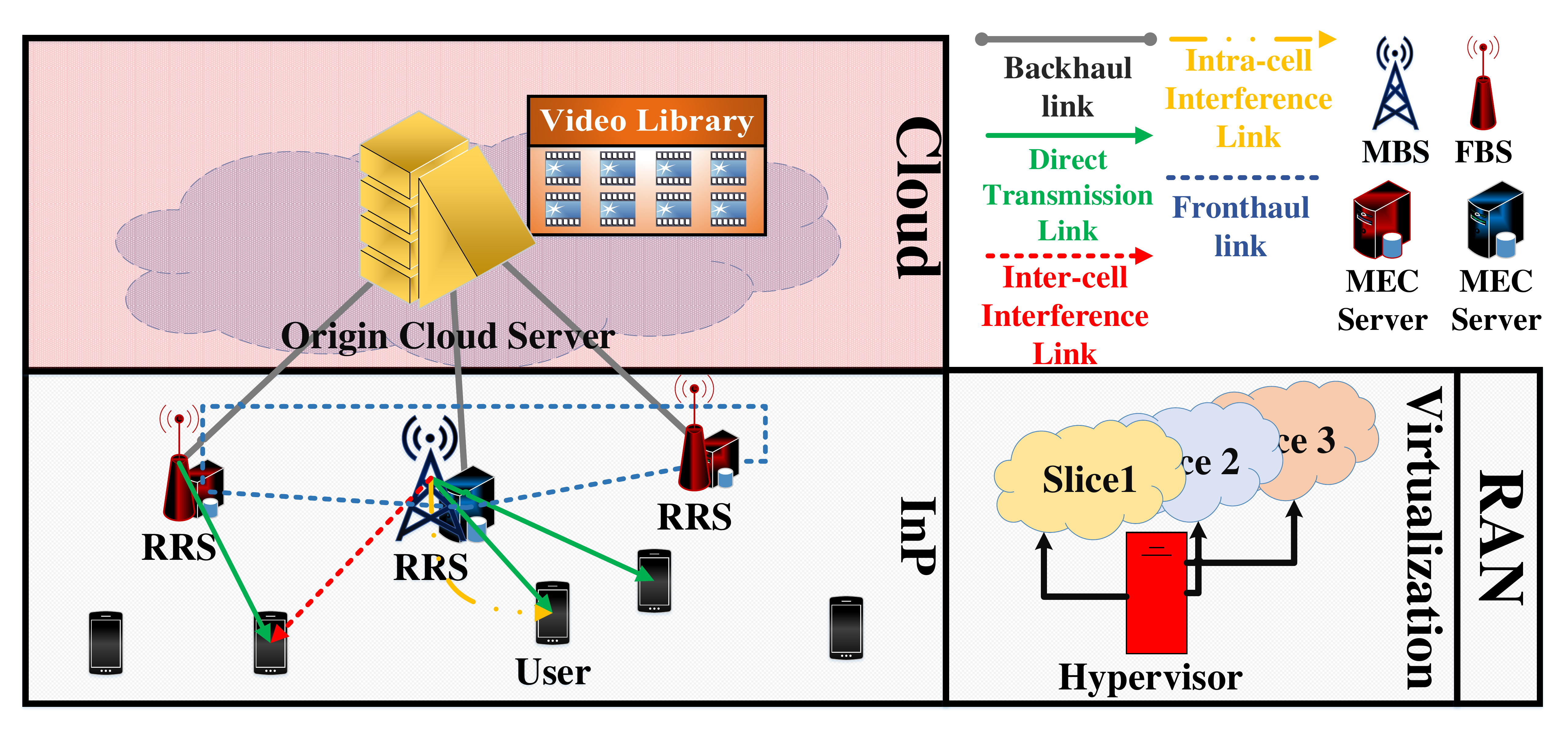}
\caption{An example of the multiuser HV-MEC architecture consisting of $3$ RRSs, $3$ slices, and one InP.}\label{Fig01SysModel}
\end{figure}

Consider a multiuser HV-MEC network consisting of multiple RRSs, each equipped with one type of access node, e.g., macro BS (MBS) or femto BS
(FBS), and a MEC server that enables video caching and transcoding capabilities at the RRS
\cite{CollaborativeScenariosChallenges,Collaborativemultibitrate}.
The set of users and RRSs are denoted by $\mathcal{U}=\{1,\dots,U\}$ and $\mathcal{B}=\{1,\dots,B\}$, respectively. All RRSs are connected by a limited wired fronthaul mesh network, which provides cooperative communication between RRSs \cite{Collaborativemultibitrate}. An origin cloud server denoted by $0$ is connected to each RRS through a limited wired backhaul link. Here, we assume that a hypervisor enables the virtualization of the network where the radio and physical resources of InP are abstracted into a set of $\mathcal{M}=\{1,\dots,M\}$ virtual networks, i.e., slices, such that slice $m$ owns a subset of users in $\mathcal{U}$, i.e., $\mathcal{U}_m$, and is responsible for providing a specific QoS for its own users \cite{7795465,7406764}. We
assume that each user is subscribed to only one slice.
Fig. \ref{Fig01SysModel} shows an illustration of this network.

Assume that there exist $V$ unique transcodable videos, each having $L$ bitrate variants\footnote{The lowest bitrate of each video type is
denoted by $1$ and the highest is denoted by $L$.} in the origin cloud server with unlimited storage capacity \cite{Collaborativemultibitrate,Taovideotranscod,cachefogkhaled}. The video library is denoted by $\mathcal{V} = \bigg\{v_l \big| v \in \{1,\dots,V\}, l \in
\{1,\dots,L\} \bigg\}$, where $v_l$ belongs to the $v^\text{th}$ video type with the $l^\text{th}$ bitrate variant with the size of $s_{v_l}$. Consider that
video $v_h$ can be transcoded to $v_l$, if $l < h$ \cite{CollaborativeScenariosChallenges,Collaborativemultibitrate,EnhancingMobileVideo}. Note that
$s_{v_l} \le s_{v_h}$, if $l \le h$ \cite{Collaborativemultibitrate}.

\begin{figure*}
\centering
\includegraphics[scale=0.30]{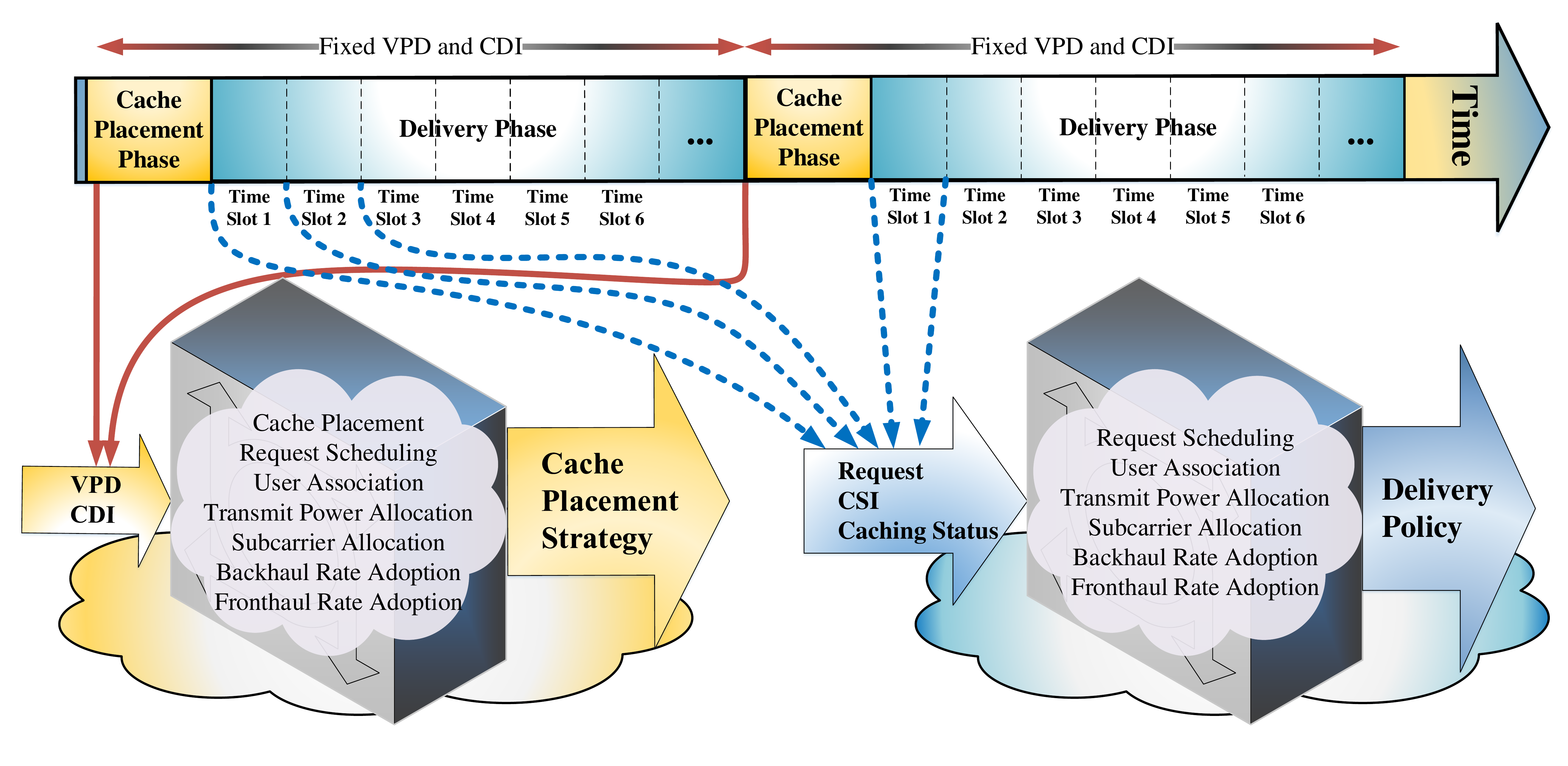}
\caption{The proposed two-phase RAF for a parallel CVCT system in HV-MECs.}\label{Fig00structure}
\end{figure*}

This network operates in two phases: Phase 1, where the scheduled video files are proactively stored in the cache of
RRSs during off-peak times \cite{Collaborativemultibitrate,EnhancingMobileVideo}; and Phase 2, where the requested videos are sent to the
end-users according to the adopted delivery policy \cite{7763747,QoSAwareResourceAllocation,4066996,Collaborativemultibitrate}.
In Phase 1, we aim to design an efficient DACPS based on the available VPD and CDI.
Phase 2, which is followed by Phase 1, is divided into multiple finite time slots where in each time slot, we propose a delivery policy based on the arrival requests of users, CSI, and the caching status. The proposed RAF is illustrated in Fig. \ref{Fig00structure}.
The main goal of this paper is proposing a CPS that considers all the existing delivery opportunities, including association of users, request scheduling, network channel conditions, user data rates, and availability and allocation of network transmission and transcoding resources in its algorithm design. To design a DACPS in Phase 1 utilizing the delivery opportunities in the system, we first need to describe the system delivery model and parameter in Phase 2. Then, we explain Phase 1 and formulate the cache placement optimization problem according to the stochastic model of Phase 2.

\subsection{Phase 2}\label{subsection System Delivery Model}
\begin{figure*}
\centering
\includegraphics[scale=0.20]{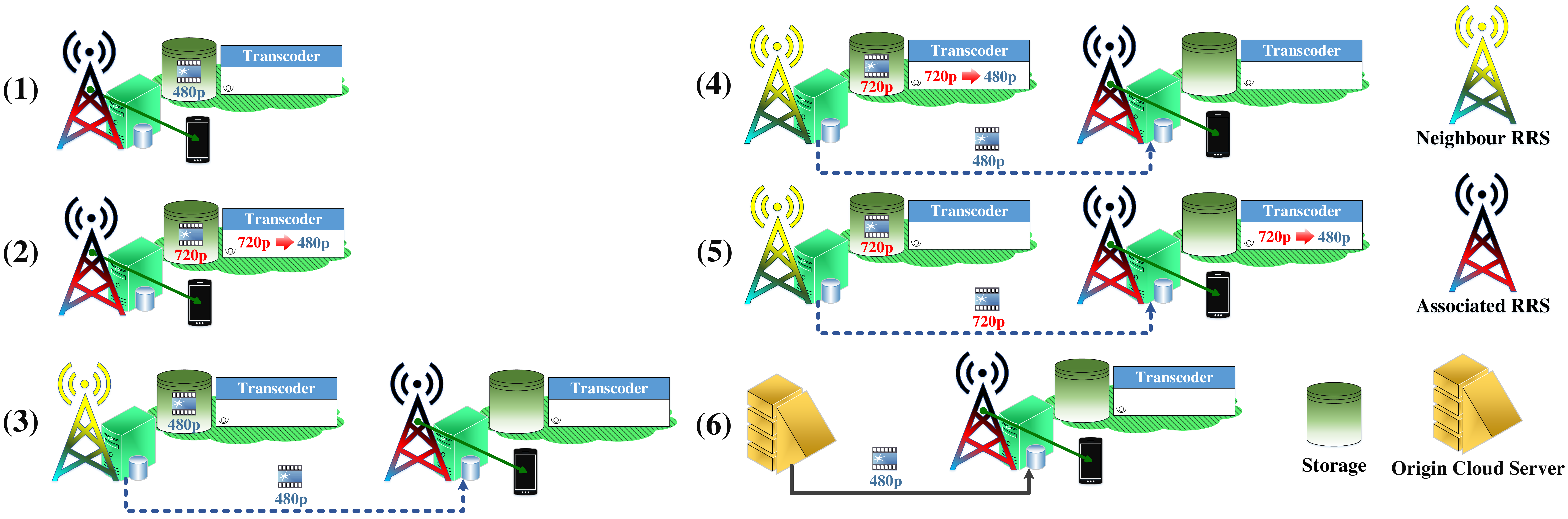}
\caption{The possible events of serving a request of a user in the CVCT system. We assume that there are two bitrate
variants of a video file at 480p and 720p where the user requests the 480p bitrate variant.}\label{Fig02ServingModel}
\end{figure*}

In this phase, we assume a time-slotted system where at the beginning of each time slot, each user requests one video file. Similar to related works  \cite{8611393,7406764,EnhancingMobileVideo,Taovideotranscod,7410051,7384533,6708492,JointOptimizationofCloud,PowerAllocationforCache,DistributedCachingforDataDissemination}, the CSI and requests of users remain fixed through a time slot and are completely independent from other time slots\footnote{Similar to previous works, the main motivation for considering this model is to simplify the video transmission model. Actually, dynamic video requesting and changes of CSI during each data transmission time cause a more complex delivery model, which is not yet investigated in the joint radio resource allocation and content placement context \cite{8611393}. Assuming a dynamic video requesting with changes of CSI during a time slot can be considered as a future work.}. All requests of users at each time slot should also be served within the time slot \cite{8611393,PowerAllocationforCache,7406764}. Hence, the adopted delivery policy for each time slot is completely independent from other time slots. Therefore, we focus on only one time slot in Phase 2. The request of user $u$ for video $v_l$ is indicated by a binary variable $\delta^{v_l}_{u} \in \{0,1\}$ such that
if user $u$ requests video $v_l$, $\delta^{v_l}_{u}=1$ and otherwise, $\delta^{v_l}_{u}=0$. Thus, we have $\sum_{v_l \in \mathcal{V}} \delta^{v_l}_{u}=1,
\forall u \in \mathcal{U}$.
The binary parameter $\theta_{b,u} \in \{0,1\}$ determines the user association indicator where if user $u$ is associated with RRS $b$, $\theta_{b,u}=1$ and otherwise, $\theta_{b,u}=0$. In this system, we assume that each user can be connected to at most one RRS, which is represented by \cite{8611393,7406764,7410051,EnhancingMobileVideo,Collaborativemultibitrate}\footnote{This parallel CVCT system can also be extended to a coordinated multi-point-enabled one in which each user is able to access to more than one transmitter. Despite the significant potential, the coordinated multi-point system increases the complexity of the delivery model. Therefore, we consider this scheme as a future work.}
\begin{align}\label{user association one BS}
\sum_{b \in \mathcal{B}} \theta_{b,u} \leq 1, \,\,\,\ \forall u \in \mathcal{U}.
\end{align}
The requests of users associated with RRS $b$ for video $v_l$ can be served by one of the following binary events denoted by \cite{Collaborativemultibitrate}:
\begin{enumerate}
  \item $x^{v_l}_{b}=1$ represents that video $v_l$ can be sent directly from cache of RRS $b$.
  \item $y^{v_h,v_l}_{b}=1$ indicates that video $v_l$ is directly served by RRS $b$ after being transcoded from a higher bitrate variant $h$ at RRS $b$.
  \item $z^{v_l}_{b',b}=1$ denotes that video $v_l$ is provided from cache of RRS $b' \neq b$ via fronthaul link.
  \item $t^{v_h,v_l}_{b',b}=1$ represents that video $v_l$ is served by transcoding from $v_h,h>l$ at RRS $b'\neq b$ and then, sending to RRS $b$ via
      fronthaul link.
  \item $w^{v_h,v_l}_{b',b}=1$ indicates that video $v_l$ is obtained by sending video $v_h,h>l$ from RRS $b'$ to RRS $b$ via fronthaul link and then,
      transcoding $v_h$ to $v_l$ at RRS $b$.
  \item $o^{v_l}_{b}=1$ represents that the requests of users of RRS $b$ for video $v_l$ are served from the origin cloud server via backhaul link $b$.
\end{enumerate}
Fig. \ref{Fig02ServingModel} shows all possible events that happen to serve requests for each video file at each RRS.
To avoid duplicated video provisioning at each RRS, we assume that all the requests for each video file $v_l$ from users associated with RRS $b$ can be served by only one type of events \cite{Collaborativemultibitrate}, i.e.,
\begin{small}
\begin{align}\label{all cases constraint}
x^{v_l}_{b} +
\sum\limits_{\scriptstyle v_h \in \mathcal{V} \atop \scriptstyle h > l} y^{v_h,v_l}_{b} +
\sum\limits_{\scriptstyle b' \in \mathcal{B} \atop \scriptstyle b' \neq b} z^{v_l}_{b',b} +
\sum\limits_{\scriptstyle b' \in \mathcal{B} \atop \scriptstyle b' \neq b} \sum\limits_{\scriptstyle v_h \in \mathcal{V} \atop \scriptstyle h > l} \left(
t^{v_h,v_l}_{b',b} + w^{v_h,v_l}_{b',b} \right) + o^{v_l}_{b} = \min\{\sum_{u=1}^{U} \delta^{v_l}_{u} \theta_{b,u},1\}, \forall b \in \mathcal{B}, v_l \in
\mathcal{V}.
\end{align}
\end{small} \vspace{-0.25cm} \\
In practical terms, the video transcoding can only be performed for higher bitrate variants of a transcodable video file to lower bitrate variants.
Accordingly, we have $y^{v_h,v_l}_{b},t^{v_h,v_l}_{b',b},w^{v_h,v_l}_{b',b} =0, \forall h \leq l$
\cite{Collaborativemultibitrate,QoSAwareResourceAllocation}.
Note that each event can be chosen if the required video exists in the target storage
\cite{Collaborativemultibitrate,EnhancingMobileVideo,QoSAwareResourceAllocation}.
Let $\rho^{v_l}_{b}$ be the binary cache placement indicator where $\rho^{v_l}_{b}=1$ if video $v_l$ is cached by RRS $b$, and $\rho^{v_l}_{b}=0$,
otherwise.
Therefore, we have
\begin{align}\label{constraint cases storage}
x^{v_l}_{b} \leq \rho^{v_l}_{b},~~~y^{v_h,v_l}_{b} \leq \rho^{v_h}_{b},~~~z^{v_l}_{b',b} \leq \rho^{v_l}_{b'},~~~t^{v_h,v_l}_{b',b} \leq
\rho^{v_h}_{b'},~~~w^{v_h,v_l}_{b',b} \leq \rho^{v_h}_{b'}.
\end{align}

In this parallel CVCT system, the MC-NOMA technology is deployed at each RRS such that the total frequency bandwidth $W$ is divided into a set of $\mathcal{N}=\{ 1,2,\dots,N \}$ orthogonal subcarriers where the frequency band of each subcarrier is $W_\text{s}$.
In this scheme, we assume that users aim to download the video files; online video servicing based on the playback rate of videos in the HV-MEC networks is considered for future work. To this end, MC-NOMA allows each orthogonal subcarrier $n$ to be shared among multiple users
in each RRS via applying a superposition coding at the transmitter side\footnote{During the data transmission time of this parallel system, the transmitter is always able to superpose the received bits of multiple video files.} and a SIC at the receiver side
\cite{OptimalPowerSubcarrierSchober,Doubleiterativewaterfilling,7812683}.
The binary subcarrier assignment indicator is also indicated by $\tau^{n}_{b,u}$, where if subcarrier $n$ is assigned to the channel from RRS $b$ to user
$u \in \mathcal{U}$, $\tau^{n}_{b,u}=1$ and otherwise, $\tau^{n}_{b,u}=0$.
Note that each user can take subcarriers from RRS $b$ if the user is associated with that RRS. Therefore, we should have \cite{8611393,7406764,Taovideotranscod}
\begin{align}\label{user association subcarrier}
\tau^{n}_{b,u} \leq \theta_{b,u}, \forall b \in \mathcal{B}, n \in \mathcal{N}, u \in \mathcal{U}.
\end{align}
We denote by $p^{n}_{b,u}$ the transmit power of RRS $b$ to user $u \in \mathcal{U}$ on subcarrier $n$, and, $h^{n}_{b,u}$ the instantaneous channel power
gain between RRS $b$ and user $u \in \mathcal{U}$ on subcarrier $n$. After performing SIC, the instantaneous signal-to-interference-plus-noise ratio (SINR) at user $u \in \mathcal{U}$ associated with
RRS $b$ on subcarrier $n$ is \cite{OptimalPowerSubcarrierSchober,7812683}
\begin{align}
\gamma^{n}_{b,u} = \frac { p^{n}_{b,u} h^{n}_{b,u} }
    { I^{\text{Intra},n}_{b,u} + I^{\text{Inter},n}_{b,u}  +  \sigma^{n}_{b,u} },
\end{align}
where
$I^{\text{Intra},n}_{b,u} = \sum\limits_{u' \in \mathcal{U}, u' \neq u \atop \scriptstyle h^{n}_{b,u'} > h^{n}_{b,u} }
p^{n}_{b,u'} h^{n}_{b,u}$ represents the induced intra-cell interference on user $u \in \mathcal{U}$ over subcarrier $n$,
$I^{\text{Inter},n}_{b,u} = \sum\limits_{\scriptstyle b'\in\mathcal{B} \hfill \atop \scriptstyle b' \neq b} \sum\limits_{\scriptstyle u' \in \mathcal{U} \hfill \atop \scriptstyle u' \neq u} p^{n}_{b',u'} h^{n}_{b',u}$ is the received inter-cell interference at user $u \in \mathcal{U}$ over subcarrier $n$, and
$\sigma^{n}_{b,u}=W_\text{s} N_0$ is the additive white Gaussian noise (AWGN) power, in which $N_0$ is the noise power spectral density (PSD).
Therefore, the instantaneous data rate at user $u \in \mathcal{U}$ from RRS $b$ on subcarrier $n$ is
$
r^{n}_{b,u}  = \tau^{n}_{b,u} W_\text{s} \log_2 ( 1 + \gamma^{n}_{b,u} ).
$
To apply the SIC technique for MC-NOMA, the following constraint should be satisfied \cite{7812683}:
\begin{small}
\begin{align}\label{SIC user}
\tau^{n}_{b,u} \frac { p^{n}_{b,u'} h^{n}_{b,u} }  { \sum\limits_{v \in \mathcal{U}, v \neq u' \atop \scriptstyle h^{n}_{b,v} > h^{n}_{b,u'}} p^{n}_{b,v}
h^{n}_{b,u} + I^{\text{Inter},n}_{b,u}  +  \sigma^{n}_{b,u} }
>
\tau^{n}_{b,u'} \gamma^{n}_{b,u'}, \forall  b \in \mathcal{B} ,
u,u' \in \mathcal{U}, n \in \mathcal{N}, \tau^{n}_{b,u},\tau^{n}_{b,u'} > 0,
h^{n}_{b,u} > h^{n}_{b,u'}.
\end{align}
\end{small} \vspace{-0.3cm}
\\
Accordingly, the instantaneous data rate at user $u \in \mathcal{U}$ assigned to RRS $b$ is
$
r^\text{Ac}_{b,u}  = \sum_{n=1}^{N} r^{n}_{b,u}.
$
Under the assumption that each RRS $b$ has a maximum transmit power $P^\text{max}_b$, we have
\begin{align}\label{Constraint maximum transmit power BS}
\sum\limits_{u \in \mathcal{U}} \sum_{n=1}^{N} \tau^{n}_{b,u} p^{n}_{b,u} \leq P^\text{max}_b, \forall b \in \mathcal{B}.
\end{align}
Consequently, the instantaneous latency of user $u$ to receive video $v_l$ from RRS $b$ can be calculated as $D^{\text{Ac},v_l}_{b,u} = \frac{ s_{v_l} }{
r^\text{Ac}_{b,u} }$.

Transcoding video $v_h$ to $v_l,\forall l < h$, is performed via the ABR streaming technique where the transcoding operation is mapped to a
$\eta^{v_h,v_l}$-bits computation-intensive task \cite{CollaborativeScenariosChallenges,Collaborativemultibitrate}.
Let $N^{v_h,v_l}_{\text{Cycle}}$ be the number of central processing unit (CPU) cycles required to compute $1$ bit of the computation-intensive task of transcoding video\footnote{$N^{v_h,v_l}_{\text{Cycle}}$ is referred to the workload of the task of transcoding video $v_h$ to $v_l$ in the ABR technique.} $v_h$ to $v_l$ \cite{7762913,DynamicComputationOffloading}.
Each RRS performs all scheduled computation tasks in parallel by efficiently allocating its computation resources \cite{8387798,PAYMARD2019160}.
The number of CPU cycles per second allocated to RRS $b$ for transcoding video $v_h$ to $v_l$ is also indicated by $\phi^{v_h,v_l}_b \in \{0,1,2,\dots\}$
\cite{7762913,7763747,QoSAwareResourceAllocation}. Let $\chi^\text{max}_b$ be the maximum processing capacity of RRS $b$. Therefore, the per-RRS maximum processing capacity constraint can be expressed as
\begin{align}\label{processing capacity constraint eRRH}
\sum\limits_{\scriptstyle v_h \in \mathcal{V} \atop \scriptstyle h > l} \sum\limits_{v_l \in \mathcal{V}} \left( y^{v_h,v_l}_{b} + \sum\limits_{b' \in \mathcal{B}/\{b\}}  t^{v_h,v_l}_{b,b'} + \sum\limits_{b' \in \mathcal{B}/\{b\}}  w^{v_h,v_l}_{b',b} \right) \phi^{v_h,v_l}_b \leq \chi^\text{max}_b, \forall
b \in \mathcal{B}.
\end{align}
The speed of the transcoding process is obtained by the video transrating, i.e., transcoding bit rate, which is the number of bits transcoded by the
processor per second \cite{4066996}. Therefore, the delay of transcoding video $v_h$ to $v_l$ at RRS $b$ can be obtained by $D^{\text{TC},v_h,v_l}_{b} =
\frac{ \eta^{v_h,v_l} N^{v_h,v_l}_{\text{Cycle}} } { \phi^{v_h,v_l}_b }$ \cite{DynamicComputationOffloading,8387798,PAYMARD2019160}.

For this setup, we consider $R^\text{max}_{0,b}$ and $R^\text{max}_{b',b}$ as the
maximum capacity of backhaul link $b$, and fronthaul link from RRS $b'$ to RRS $b$, respectively
\cite{JointOptimizationofCloud}. We also denote $r^{v_l}_{0,b}$ and $r^{v_l}_{b',b}$ the adopted data rate for RRS $b$ to receive video $v_l$ from the origin cloud server and from the neighboring RRS $b'$, respectively. Hence, the following maximum channel capacity constraints should be satisfied:
\begin{align}\label{channel capacity constraint BH}
\sum\limits_{v_l \in \mathcal{V}}  o^{v_l}_{b}  r^{v_l}_{0,b} \leq R^\text{max}_{0,b}, \forall b \in \mathcal{B},
\end{align}
and
\begin{align}\label{channel capacity constraint FH BS}
\sum\limits_{v_l \in \mathcal{V}} \left( z^{v_l}_{b',b} + \sum\limits_{\scriptstyle v_h \in \mathcal{V} \atop \scriptstyle h > l}  t^{v_h,v_l}_{b',b}  \right) r^{v_l}_{b',b} + \sum\limits_{v_l \in \mathcal{V}} \sum\limits_{\scriptstyle v_h \in \mathcal{V} \atop \scriptstyle h > l}  w^{v_h,v_l}_{b',b} r^{v_h}_{b',b} \leq R^\text{max}_{b',b}, \forall b,b' \in \mathcal{B}, b' \neq b.
\end{align}
Accordingly, the delays of receiving video $v_l$ from the origin cloud server and RRS $b'$ at RRS $b$ are represented as $D^{\text{BH},v_l}_{0,b} = \frac{
s_{v_l} } { r^{v_l}_{0,b} }$ and $D^{\text{FH},v_l}_{b',b} = \frac{ s_{v_l} } { r^{v_l}_{b',b} }$, respectively.

From the isolation perspective in slicing context, to guarantee the QoS at users owned by each slice $m$, i.e., $\mathcal{U}_m$, we apply a minimum data
rate constraint as\footnote{In this paper, similar to the prior works, e.g., \cite{8611393,7406764,7410051,7384533,AResourceAllocationPerspective,6708492}, it is assumed that the set of active users and their QoS requirement are fixed in Phase 2 and also available in Phase 1. Since users are subscribed to the slices based on their QoS level agreements, the subscribed user set $\boldsymbol{U}_m$ is also fixed in Phase 2 and available in Phase 1.}
\begin{align}\label{constraint minimum rate access}
\sum_{b \in \mathcal{B}} r^\text{Ac}_{b,u} \geq R^\text{min}_{m}, \forall m \in \mathcal{M}, u \in \mathcal{U}_m,
\end{align}
where $R^\text{min}_{m}$ represents the minimum required access data rate of users in $\mathcal{U}_m$ \cite{7410051}.
Satisfying \eqref{constraint minimum rate access} for a non-zero variable $R^\text{min}_{m}$ needs allocating at least one subcarrier to each user in $\mathcal{U}_m$. According to \eqref{user association subcarrier}, this condition can only be applied if each user be associated with at least one RRS. Therefore, for each non-zero variable $R^\text{min}_{m}$, all users in $\mathcal{U}_m$ should be associated with at least one RRS. Accordingly, based on constraints \eqref{user association one BS}, \eqref{user association subcarrier}, and \eqref{constraint minimum rate access}, if $R^\text{min}_{m} >0$, the inequality in \eqref{user association one BS} turns into an equality for all users in $\mathcal{U}_m$. It is noteworthy that $R^\text{min}_{m}=0$ means slice $m$ does not guarantee any QoS for its own users. In this regard, this slice provides a best-effort service in which the requests of its own users are not guaranteed to be served.

The video transcoding process runs in parallel with the video transmission, where the delay of each transcoding in the system is measured by transcoding
the first several segments of a video file \cite{QoSAwareResourceAllocation,4066996}. This is negligible compared to the corresponding wireless transmission
delay. To efficiently allocate physical resources in this parallel system, some transmission/transcoding delay constraints should be held for each multi-hop scheduling event. For instance, in Event 2, the delay of transcoding video $v_h$ to $v_l$ at RRS $b$ should not be greater than the access latency of user $u$ to receive video $v_l$ from RRS $b$ \cite{EnhancingMobileVideo,4066996}, i.e.,
\begin{align}\label{constraint event 2 access transc}
y^{v_h,v_l}_{b} D^{\text{TC},v_h,v_l}_{b}  \leq \delta^{v_l}_{u} D^{\text{Ac},v_l}_{b,u}.
\end{align}
For Event $3$, the delay of fronthaul transmission should not be greater than the access delay
\cite{JointOptimizationofCloud,7410051,PowerAllocationforCache}. Therefore, we have
\begin{align}\label{constraint event 3 access front}
z^{v_l}_{b',b} D^{\text{FH},v_l}_{b',b}  \leq \delta^{v_l}_{u} D^{\text{Ac},v_l}_{b,u}.
\end{align}
For Event $4$, the delay of transcoding and fronthaul transmission should not be greater than the fronthaul and access delays, respectively. These practical constraints can be represented as
\begin{align}\label{constraint event 4 transcoding fronthaul}
t^{v_h,v_l}_{b',b} D^{\text{TC},v_h,v_l}_{b'}  \leq  t^{v_h,v_l}_{b',b} D^{\text{FH},v_l}_{b',b},
\end{align}
\begin{align}\label{constraint event 4 fronthaul access}
t^{v_h,v_l}_{b',b} D^{\text{FH},v_l}_{b',b}  \leq  \delta^{v_l}_{u} D^{\text{Ac},v_l}_{b,u}.
\end{align}
For Event $5$, fronthaul and transcoding delays should not be greater than the transcoding and access delays,
respectively. Hence, we have
\begin{align}\label{constraint event 5 fronthaul transcoding}
w^{v_h,v_l}_{b',b} D^{\text{FH},v_h}_{b',b}  \leq  w^{v_h,v_l}_{b',b} D^{\text{TC},v_h,v_l}_{b},
\end{align}
\begin{align}\label{constraint event 5 transcoding access}
w^{v_h,v_l}_{b',b} D^{\text{TC},v_h,v_l}_{b}  \leq \delta^{v_l}_{u} D^{\text{Ac},v_l}_{b,u}.
\end{align}
Finally, for Event $6$, the backhaul delay should be equal or less than the access delay for each video transmission. Accordingly, we have
\begin{align}\label{constraint event 6 backhaul access}
o^{v_l}_{b} D^{\text{BH},v_l}_{0,b} \leq \delta^{v_l}_{u} D^{\text{Ac},v_l}_{b,u}.
\end{align}
If all conditions in \eqref{constraint event 2 access transc}-\eqref{constraint event 6 backhaul access} hold, the total latency of each user comes from
its access delay (wireless transmission delay) \cite{7410051,cachefogkhaled,4066996}. This parallel system prevents the extra fronthaul/backhaul  transmission and video transcoding delays in the network.

Since slices lease the physical resources of InP, we intend to propose a new pricing model to represent this structure. For the access transmission resources, the unit price of transmit power and spectrum of RRS $b$ are
indicated by $\alpha^\text{Pow}_b$ per Watt and $\alpha^\text{Sub}_b$ per Hz, respectively \cite{7384533,7410051}. Moreover, the unit price of backhaul and
fronthaul rates are defined as $\alpha^\text{BH}$ and $\alpha^\text{FH}$ per bps \cite{7410051}. For the storage resources, each slice pays
$\mu^\text{Cache}_b$ per bit to utilize the memory of RRS $b$ \cite{7384533}.
The price of the processing resources usage at RRS $b$ is also defined as $\mu^\text{Proc}_b$ per CPU cycle.
On the other hand, each slice $m$ gets rewards from its own users due to providing their access data rates \cite{7406764,7384533,7410051}.
We define $\psi_{m}$ as the reward of slice $m$ from each user $u \in \mathcal{U}_m$ per unit of received data rate (bit/s).
Let us consider that $\psi_m$ is an increasing function of $R^\text{min}_{m}$, i.e., for $m \neq m' \in \mathcal{M}$, for $R^\text{min}_{m} \geq R^\text{min}_{m'}$, we have  $\psi_{m} \geq \psi_{m'}$.
In this scheme, we aim to maximize the revenue of slices, which can be defined as the reward minus cost of each slice. The reward of each slice $m$ is $\sum\limits_{b \in \mathcal{B}} r^\text{Ac}_{b,u} \psi_m$.
To define the cost of each slice, we first formulate the cost of provisioning video $v_l$ to RRS $b$ caused by one of the scheduling events as
\begin{multline}\label{cost providingBS}
\$^{\text{Cost,RRS},v_l}_{b} = x^{v_l}_{b} \left( s_{v_l} \mu^\text{Cache}_b \right) +
\sum\limits_{\scriptstyle v_h \in \mathcal{V} \atop \scriptstyle h > l} y^{v_h,v_l}_{b} \left( s_{v_h} \mu^\text{Cache}_b + \phi^{v_h,v_l}_b \mu^\text{Proc}_b \right) +
\sum\limits_{\scriptstyle b' \in \mathcal{B} \atop \scriptstyle b' \neq b} z^{v_l}_{b',b} \left( s_{v_l} \mu^\text{Cache}_{b'} + r^{v_l}_{b',b}
\alpha^\text{FH} \right) +  \\
\sum\limits_{\scriptstyle b' \in \mathcal{B} \atop \scriptstyle b' \neq b} \sum\limits_{\scriptstyle v_h \in \mathcal{V} \atop \scriptstyle h > l}
t^{v_h,v_l}_{b',b} \left( s_{v_h} \mu^\text{Cache}_{b'} + \phi^{v_h,v_l}_{b'} \mu^\text{Proc}_{b'} + r^{v_l}_{b',b} \alpha^\text{FH} \right) +
\sum\limits_{\scriptstyle b' \in \mathcal{B} \atop \scriptstyle b' \neq b} \sum\limits_{\scriptstyle v_h \in \mathcal{V} \atop \scriptstyle h > l}
w^{v_h,v_l}_{b',b} \big( s_{v_h} \mu^\text{Cache}_{b'} + r^{v_h}_{b',b} \alpha^\text{FH} +     \\
\phi^{v_h,v_l}_{b} \mu^\text{Proc}_{b} \big) +
o^{v_l}_{b} \left( r^{v_l}_{0,b} \alpha^\text{BH} \right).
\end{multline}
Furthermore, the cost of the access transmission resource usage for transferring the requested video file to user $u$ is
$
\$^{\text{Cost,Ac}}_{u} = \sum\limits_{b \in \mathcal{B}} \sum\limits_{ n \in \mathcal{N} } \left( p^{n}_{b,u} \alpha^\text{Pow}_b + \tau^{n}_{b,u}
W_\text{s} \alpha^\text{Sub}_b \right).
$
Therefore, the cost of serving the request of user $u$ for video $v_l$ is
$
\$^{\text{Cost,User},v_l}_{u} = \delta^{v_l}_{u} \left( \$^{\text{Cost,Ac}}_{u} + \sum\limits_{b \in \mathcal{B}} \theta_{b,u} \$^{\text{Cost,RRS},v_l}_{b}
\right)
$
which should be paid by slice $m$, if $u \in \mathcal{U}_m$. Hence, the revenue of each slice for serving video $v_l$ to user $u \in \mathcal{U}_m$ is
$\left( \sum\limits_{b \in \mathcal{B}} r^\text{Ac}_{b,u} \psi_m - \$^{\text{Cost,User},v_l}_{u} \right)$.

Based on \eqref{all cases constraint}, each slice pays the cost of storage, processing, backhaul, and fronthaul resource usages for each video provisioning to each cell only once. Moreover, each slice $m$ pays the usage cost of each subcarrier only once in each cell, even when the subcarrier is shared among users in $\mathcal{U}_m$ in that cell via the MC-NOMA technology. Accordingly, the revenue of slice $m$ in Phase 2 can be defined as
\begin{multline}\label{revenue MVNO}
\textbf{\$}^\text{slice}_m = \sum_{ u \in \mathcal{U}_m } \sum\limits_{b \in \mathcal{B}} r^\text{Ac}_{b,u} \psi_m
- \sum_{ u \in \mathcal{U}_m } \sum\limits_{b \in \mathcal{B}} \sum_{ n \in \mathcal{N} } \left( p^{n}_{b,u} \alpha^\text{Pow}_b \right)
- \sum\limits_{b \in \mathcal{B}} \sum_{ n \in \mathcal{N} } \left( \max\limits_{ u \in \mathcal{U}_m } \{ \tau^{n}_{b,u} \} W_\text{s} \alpha^\text{Sub}_b
\right)     \\
- \sum_{b \in \mathcal{B}} \sum\limits_{v_l \in \mathcal{V}} \min \{ \sum_{u \in \mathcal{U}_m} \theta_{b,u} \delta^{v_l}_{u} , 1\}
\$^{\text{Cost,RRS},v_l}_{b}.
\end{multline}

In Phase 2, with the objective of maximizing the total delivery revenue of slices denoted by $\textbf{\$}_\text{tot} = \sum_{m \in \mathcal{M}}
\textbf{\$}^\text{slice}_m$ under the QoS requirements of users, we jointly optimize the user association, access transmit power and subcarrier allocation,
fronthaul and backhaul rate adaption, processing resource allocation, and request scheduling to have an efficient delivery performance. For ease of notations, we denote $\boldsymbol{\theta}=[\theta_{b,u}]$, $\boldsymbol{\phi}=[\phi^{v_h,v_l}_b]$, $\boldsymbol{p}=[p^{n}_{b,u}]$,
$\boldsymbol{\tau}=[\tau^{n}_{b,u}]$, $\boldsymbol{r}^\text{BH}=[r^{v_l}_{0,b}]$, $\boldsymbol{r}^\text{FH}=[r^{v_l}_{b',b}]$,
$\boldsymbol{\Upsilon}=[\boldsymbol{x},\boldsymbol{y},\boldsymbol{z},\boldsymbol{t},\boldsymbol{w},\boldsymbol{o}]$,
$\boldsymbol{x}=[x^{v_l}_{b}]$, $\boldsymbol{y}=[y^{v_h,v_l}_{b}]$, $\boldsymbol{z}=[z^{v_l}_{b',b}]$, $\boldsymbol{t}=[t^{v_h,v_l}_{b',b}]$,
$\boldsymbol{w}=[w^{v_h,v_l}_{b',b}]$, and $\boldsymbol{o}=[o^{v_l}_{b}]$.
The delivery optimization problem can be formulated as
\begin{subequations}\label{Main Problem delivery}
\begin{align}\label{obf Main Problem delivery}
&\max_{ \boldsymbol{\theta} , \boldsymbol{\phi} , \boldsymbol{p} , \boldsymbol{\tau} , \boldsymbol{r}^\text{BH}, \boldsymbol{r}^\text{FH},
\boldsymbol{\Upsilon} }\hspace{.0 cm} 	
\textbf{\$}_\text{tot}
\\
\textrm{s.t.}\hspace{.0cm}~~
& \text{\eqref{user association one BS}-\eqref{user association subcarrier}, \eqref{SIC user}, \eqref{Constraint maximum transmit power
BS}-\eqref{constraint event 6 backhaul access},}  \nonumber   \\
\label{constraint untranscodable}
&
y^{v_h,v_l}_{b}=0, t^{v_h,v_l}_{b',b}=0, w^{v_h,v_l}_{b',b}=0, \forall h \leq l, \\
\label{constraint exclusive subcarrier}
& \sum\limits_{u \in \mathcal{U}} \tau^{n}_{b,u} \leq \Psi_b, \forall b \in \mathcal{B},  n \in \mathcal{N},
\\
\label{constraint subcarrier binary}
& \theta_{b,u},\tau^{n}_{b,u},x^{v_l}_{b},y^{v_h,v_l}_{b},z^{v_l}_{b',b},t^{v_h,v_l}_{b',b},w^{v_h,v_l}_{b',b},o^{v_l}_{b} \in \{0,1\},~ \phi^{v_h,v_l}_b
\in \{0,1,2,\dots\},~p^{n}_{b,u},r^{v_l}_{0,b},r^{v_l}_{b',b} \geq 0,
\end{align}
\end{subequations}
where \eqref{constraint exclusive subcarrier} represents that in each cell $b$, each subcarrier $n$ can be assigned to at most $\Psi_b$ users \cite{7812683}.

\subsection{Phase 1}
In this phase, we aim to design a proactive DACPS by utilizing the delivery opportunities in Phase 2. Note that the users' request and CSI for Phase 2 are not available in this phase. To utilize the delivery model in the DACPS design, we need some stochastic information about videos popularity and wireless channel conditions. Similar to \cite{8611393,JointOptimizationofCloud,DistributedCachingforDataDissemination,cachefogkhaled}, we assume that the VPD changes slowly compared to the instantaneous requests of users and remains fixed for the entirety of the network's operational time. Also, this parameter can be estimated by the operators by collecting prior set of requests of users \cite{8611393,JointOptimizationofCloud,DistributedCachingforDataDissemination}. Accordingly, we assume that VPD is available at the scheduler in Phase 1 and does not change during Phase 2, i.e., is valid for Phase 2. Similarly, the CDI can be averaged over various CSIs in different time slots of prior Phase 2 \cite{8611393}.
Assume that the VPD follows the Zipf distribution with the Zipf parameter $\lambda$ and is the same among all users in the network \cite{8611393,cachefogkhaled,Delayperformanceanalysis}.
Therefore, the popularity of requesting video $v_l$ with rank\footnote{Consider that the videos in $\mathcal{V}$ are randomly sorted from
$1$ to $VL$.} $\Lambda_{v_l}$ is given by
$
\Delta_{v_l} = \frac{ 1/{(\Lambda_{v_l})}^{\lambda} }{ \sum_{v_l \in \mathcal{V}} 1/{(\Lambda_{v_l})}^{\lambda} } , \forall v_l \in \mathcal{V}.
$
To design a DACPS which covers the whole Phase 2, we propose an averaged based joint cache placement and ergodic resource allocation based on the VPD and CDI. Our proactive DACPS is valid until the VPD and/or the CDI changes \cite{8611393}.

In Phase 1, we aim to formulate the stochastic problem of maximizing the total revenue of slices based on the available VPD and CDI to have an efficient delivery performance in Phase 2. In other words, since requests of users and CSI are not available in Phase 1, \eqref{Main Problem delivery} should be reformulated based on the VPD and CDI. In this regard, the average or ergodic data rate of wireless access link between user $u$ and RRS $b$ is $\bar{r}^\text{Ac}_{b,u} =
\mathbb{E}_{\boldsymbol{h}} \big{ \{ } r^\text{Ac}_{b,u} \big{ \} }$, where $\mathbb{E}_{\boldsymbol{h}} \{ \cdot  \} $ is the expectation operator on the
channel power gains \cite{8611393}. This expectation is necessary even in slow fading scenarios, since the whole of Phase 2 has a much longer time length compared to
Phase 1. In this phase, in contrast to the instantaneous access delays formulated in Subsection
\ref{subsection System Delivery Model}, the average access delay for receiving video $v_l$ from RRS $b$ at user $u$ obtained by
$\bar{D}^{\text{Ac},v_l}_{b,u} = \frac{ s_{v_l} }{ \bar{r}^\text{Ac}_{b,u} }$ is considered. Moreover, in order to apply SIC, the average SIC constraint
should be satisfied as\footnote{In contrast to Phase 2 where the SIC of MC-NOMA is applied based on the CSI, in Phase 1, we apply SIC based on available CDI.}
\begin{multline}\label{SIC user ergodic}
\mathbb{E}_{\boldsymbol{h}} \bigg{ \{ } \tau^{n}_{b,u} \frac { p^{n}_{b,u'} h^{n}_{b,u} }  { \sum\limits_{v \in \mathcal{U}, v \neq u' \atop \scriptstyle
h^{n}_{b,v} > h^{n}_{b,u'}} p^{n}_{b,v} h^{n}_{b,u} + I^{\text{Inter},n}_{b,u}  +  \sigma^{n}_{b,u} } \bigg{ \} }
>
\mathbb{E}_{\boldsymbol{h}} \big{ \{ }  \tau^{n}_{b,u'} \gamma^{n}_{b,u'}  \big{ \} }, \forall  b \in \mathcal{B} ,
u,u' \in \mathcal{U}, n \in \mathcal{N},  \\
\tau^{n}_{b,u},\tau^{n}_{b,u'} > 0, \mathbb{E}_{\boldsymbol{h}} \big{ \{ }  h^{n}_{b,u} \big{ \} } > \mathbb{E}_{\boldsymbol{h}} \big{ \{ } h^{n}_{b,u'}
\big{ \} }.
\end{multline}
Furthermore, since the requests of users ($\delta^{v_l}_{u} \in \left\{0,1\right\}$) are unknown in Phase 1, constraint \eqref{all cases constraint} should be reformulated based on the VPD ($\Delta^{v_l} \in [0,1]$) which is non-achievable. To tackle this challenge and cover all possible situations in Phase 2, we assume that all videos should be served by each RRS based on one of the events described in Fig. \ref{Fig02ServingModel} if at least one user is
associated with that RRS. Hence, we have
\begin{multline}\label{all cases constraint average}
x^{v_l}_{b} +
\sum\limits_{\scriptstyle v_h \in \mathcal{V} \atop \scriptstyle h > l} y^{v_h,v_l}_{b} +
\sum\limits_{\scriptstyle b' \in \mathcal{B} \atop \scriptstyle b' \neq b} z^{v_l}_{b',b} +
\sum\limits_{\scriptstyle b' \in \mathcal{B} \atop \scriptstyle b' \neq b} \sum\limits_{\scriptstyle v_h \in \mathcal{V} \atop \scriptstyle h > l} \left(
t^{v_h,v_l}_{b',b} + w^{v_h,v_l}_{b',b} \right) + o^{v_l}_{b} = \min\{\sum_{u=1}^{U} \theta_{b,u},1\}, \forall b \in \mathcal{B}, v_l \in \mathcal{V}.
\end{multline}
Although this constraint consumes more physical resources, it covers all possible situations. In other words, it guarantees that various sets of arrival
requests in different time slots of Phase 2 can be served \cite{Collaborativemultibitrate}.
With this assumption and available CDI, constraints \eqref{constraint event 2 access transc}, \eqref{constraint event 3 access front}, \eqref{constraint
event 4 fronthaul access}, \eqref{constraint event 5 transcoding access}, and \eqref{constraint event 6 backhaul access} in Phase 2 are
reformulated as
\begin{multline}\label{constraint event 2 average access transc}
y^{v_h,v_l}_{b} D^{\text{TC},v_h,v_l}_{b}  \leq \bar{D}^{\text{Ac},v_l}_{b,u},~~~~~
z^{v_l}_{b',b} D^{\text{FH},v_l}_{b',b}  \leq \bar{D}^{\text{Ac},v_l}_{b,u},~~~~~
t^{v_h,v_l}_{b',b} D^{\text{FH},v_l}_{b',b}  \leq  \bar{D}^{\text{Ac},v_l}_{b,u},~~~~~    \\
w^{v_h,v_l}_{b',b} D^{\text{TC},v_h,v_l}_{b}  \leq \bar{D}^{\text{Ac},v_l}_{b,u},~~~~~
o^{v_l}_{b} D^{\text{BH},v_l}_{0,b} \leq \bar{D}^{\text{Ac},v_l}_{b,u},
\end{multline}
respectively.
Let $C^\text{max}_b$ be the maximum storage capacity of RRS $b$.
In contrast to Phase 2, in this phase, we add a cache size constraint for each RRS as follows:
\begin{align}\label{constraint cache size BS}
\sum\limits_{v_l \in \mathcal{V}} \rho^{v_l}_{b} s_{v_l} \leq C^\text{max}_b, \forall b \in \mathcal{B}.
\end{align}

In Phase 1, our pricing model presented in Subsection \ref{subsection System Delivery Model} is also averaged based on both VPD and CDI.
In this line, the average provisioning cost of a video file at RRS $b$ can be formulated as
$
\bar{\$}^{\text{Cost,RRS}}_{b}  = \sum\limits_{v_l \in \mathcal{V}} \Delta_{v_l} \$^{\text{Cost,RRS},v_l}_{b}.
$
Moreover, the average reward of each slice $m$ for providing a video file to user $u \in \mathcal{U}_m$ based on the considered average data rate $\bar{r}^\text{Ac}_{b,u}$ can be obtained by $\sum\limits_{b \in \mathcal{B}} \bar{r}^\text{Ac}_{b,u} \psi_m$. Therefore, the average revenue of slice $m$ is
\begin{multline}\label{average revenue each slice}
\bar{\textbf{\$}}^\text{slice}_m = \sum_{ u \in \mathcal{U}_m } \sum\limits_{b \in \mathcal{B}} \bar{r}^\text{Ac}_{b,u} \psi_m
- \sum_{ u \in \mathcal{U}_m } \sum\limits_{b \in \mathcal{B}} \sum_{ n \in \mathcal{N} } \left( p^{n}_{b,u} \alpha^\text{Pow}_b \right)
- \sum\limits_{b \in \mathcal{B}} \sum_{ n \in \mathcal{N} } \left( \max\limits_{ u \in \mathcal{U}_m } \{ \tau^{n}_{b,u} \} W_\text{s} \alpha^\text{Sub}_b
\right)   -  \\
\sum_{b \in \mathcal{B}} f_{b,m}\left( \boldsymbol{\theta} \right) \bar{\$}^{\text{Cost,RRS}}_{b},
\end{multline}
where $f_{b,m}\left( \boldsymbol{\theta} \right)$ is a function of $\theta_{b,u}$ which represents the number of specific videos in $\mathcal{V}$ requested by users
in $\mathcal{U}_m$ associated with RRS $b$. The main challenge of \eqref{average revenue each slice} is obtaining an exact closed-form representation for
$f_{b,m}\left( \boldsymbol{\theta} \right)$ based on the available VPD which cannot be obtained, since the available VPD is independent from the user
association process.
Obviously, the value of $f_{b,m}\left( \boldsymbol{\theta} \right)$ is upper-bounded by $\sum_{u \in \mathcal{U}_m} \theta_{b,u}$ and also is lower-bounded
by $\min \left\{ \sum_{u \in \mathcal{U}_m} \theta_{b,u} , 1 \right\}$. Generally, the diversity of requesting video files affects $f_{b,m}\left(
\boldsymbol{\theta} \right)$.
Specifically, if users have more diverse requests, i.e., $\lambda \rightarrow 0$, $f_{b,m}\left( \boldsymbol{\theta} \right)$ increases which degrades the
revenue of slices. Besides, if users have less diverse requests, i.e., $\lambda \rightarrow \infty$ or only a few video files are requested among all
users, $f_{b,m}\left( \boldsymbol{\theta} \right)$ decreases. Therefore, to obtain a closed-form representation for $f_{b,m}\left( \boldsymbol{\theta}
\right)$ in \eqref{average revenue each slice}, we propose two low-diversity (LD) and high-diversity (HD) schemes. In the LD scheme, each slice assumes that all of its own users have the same requests based on the VPD. Hence, this scheme considers the best requesting situation which provides the maximum achievable revenue of slices. Conversely, in the HD scheme, each slice assumes that all of its users have different requests, i.e., the worst requesting situation is considered.
To handle the different requesting diversity situations in the CPS design, we propose two baseline diversity CPSs, namely LD and HD. In the LD strategy, we consider the upper-bound value of the average revenue of each slice formulated as
\begin{multline}\label{revenue MVNO average LD}
\bar{\textbf{\$}}^\text{slice,UB}_m = \sum_{ u \in \mathcal{U}_m } \sum\limits_{b \in \mathcal{B}} \bar{r}^\text{Ac}_{b,u} \psi_m
- \sum_{ u \in \mathcal{U}_m } \sum\limits_{b \in \mathcal{B}} \sum_{ n \in \mathcal{N} } \left( p^{n}_{b,u} \alpha^\text{Pow}_b \right)
- \sum\limits_{b \in \mathcal{B}} \sum_{ n \in \mathcal{N} } \left( \max\limits_{ u \in \mathcal{U}_m } \{ \tau^{n}_{b,u} \} W_\text{s} \alpha^\text{Sub}_b
\right)     \\
- \sum_{b \in \mathcal{B}} \min \left\{ \sum_{u \in \mathcal{U}_m} \theta_{b,u} , 1 \right\} \bar{\$}^{\text{Cost,RRS}}_{b},
\end{multline}
which is compatible with the LD scheme. On the other hand, for the HD strategy, we consider the lower-bound average revenue of each slice
which is expressed as
\begin{multline}\label{revenue MVNO average HD}
\bar{\textbf{\$}}^\text{slice,LB}_m = \sum_{ u \in \mathcal{U}_m } \sum\limits_{b \in \mathcal{B}} \bar{r}^\text{Ac}_{b,u} \psi_m
- \sum_{ u \in \mathcal{U}_m } \sum\limits_{b \in \mathcal{B}} \sum_{ n \in \mathcal{N} } \left( p^{n}_{b,u} \alpha^\text{Pow}_b \right)
- \sum\limits_{b \in \mathcal{B}} \sum_{ n \in \mathcal{N} } \left( \max\limits_{ u \in \mathcal{U}_m } \{ \tau^{n}_{b,u} \} W_\text{s} \alpha^\text{Sub}_b
\right)     \\
- \sum_{b \in \mathcal{B}} \sum_{ u \in \mathcal{U}_m } \theta_{b,u} \bar{\$}^{\text{Cost,RRS}}_{b}.
\end{multline}
This strategy is also compatible with the HD scheme.
In this phase, we design DACPSs in the LD and HD schemes to maximize the total estimated average revenue of slices which are formulated as $\bar{\textbf{\$}}^\text{LD}_\text{tot} = \sum_{m \in \mathcal{M}} \bar{\textbf{\$}}^\text{slice,UB}_m$ and $\bar{\textbf{\$}}^\text{HD}_\text{tot} = \sum_{m \in \mathcal{M}} \bar{\textbf{\$}}^\text{slice,LB}_m$, respectively.
The cache placement optimization problem in the LD scheme is
\begin{subequations}\label{Main Problem caching}
\begin{align}\label{obf Main Problem caching}
&\max_{ \boldsymbol{\rho}, \boldsymbol{\theta} , \boldsymbol{\phi} , \boldsymbol{p} , \boldsymbol{\tau} , \boldsymbol{r}^\text{BH},
\boldsymbol{r}^\text{FH}, \boldsymbol{\Upsilon} }\hspace{.0 cm} 	
\bar{\textbf{\$}}^\text{LD}_\text{tot}
\\
\textrm{s.t.}\hspace{.0cm}~~
& \text{\eqref{user association one BS}, \eqref{constraint cases storage}, \eqref{user association subcarrier}, \eqref{Constraint maximum transmit power
BS}-\eqref{channel capacity constraint FH BS}, \eqref{constraint event 4 transcoding fronthaul}, \eqref{constraint event 5 fronthaul transcoding},
\eqref{constraint untranscodable}-\eqref{constraint subcarrier binary}, \eqref{SIC user ergodic}-\eqref{constraint cache size BS},}
\nonumber   \\
\label{constraint minimum rate access erg}
& \sum_{b \in \mathcal{B}} \bar{r}^\text{Ac}_{b,u} \geq R^\text{min}_{m}, \forall m \in \mathcal{M}, u \in \mathcal{U}_m,
\\
\label{constraint rho binary}
& \rho^{v_l}_{b} \in \{0,1\}.
\end{align}
\end{subequations}
The cache placement optimization problem in the HD scheme is
\begin{subequations}\label{Main Problem caching HD}
\begin{align}\label{obf Main Problem caching HD}
\max_{ \boldsymbol{\rho}, \boldsymbol{\theta} , \boldsymbol{\phi} , \boldsymbol{p} , \boldsymbol{\tau} , \boldsymbol{r}^\text{BH},
\boldsymbol{r}^\text{FH}, \boldsymbol{\Upsilon} }\hspace{.0 cm} 	&
\bar{\textbf{\$}}^\text{HD}_\text{tot}
\\
\textrm{s.t.}\hspace{.0cm}~~
& \text{\eqref{user association one BS}, \eqref{constraint cases storage}, \eqref{user association subcarrier}, \eqref{Constraint maximum transmit power
BS}-\eqref{channel capacity constraint FH BS}, \eqref{constraint event 4 transcoding fronthaul}, \eqref{constraint event 5 fronthaul transcoding},
\eqref{constraint untranscodable}-\eqref{constraint subcarrier binary}, \eqref{SIC user ergodic}-\eqref{constraint cache size BS},
\eqref{constraint minimum rate access erg}, \eqref{constraint rho binary}.}  \nonumber
\end{align}
\end{subequations}
Table \ref{table main notations} summarizes the main notations used in each phase.
\begin{table*}[tp]
\centering
\caption{Main notations.}
\begin{center} \label{table main notations}
\scalebox{0.63}{\begin{tabular}{|c|c|c||c|c|c|}
    \hline \rowcolor[gray]{0.80}
    \hline \rowcolor[gray]{0.80}
    \hline \rowcolor[gray]{0.80}
    \hline \rowcolor[gray]{0.80}
    \textbf{Description} & \textbf{Notation} & \textbf{Phase} & \textbf{Description} & \textbf{Notation} & \textbf{Phase} \\
    \hline \rowcolor[gray]{0.855}
    \hline \rowcolor[gray]{0.855}
    \hline \rowcolor[gray]{0.855}
    \hline  \rowcolor[gray]{0.855}
    Number of users & $U$ & 1 \& 2 & Number of RRSs & $B$ & 1 \& 2 \\
    \hline \rowcolor[gray]{0.86}
    Number of slices & $M$ & 1 \& 2 & Number of unique videos & $V$ & 1 \& 2 \\
    \hline \rowcolor[gray]{0.865}
    Number of bitrate variants & $L$ & 1 \& 2 & The $v^\text{th}$ video type with the $l^\text{th}$ bitrate variant & $v_l$ & 1 \& 2 \\
    \hline \rowcolor[gray]{0.87}
    Size of video $v_l$ & $s_{v_l}$ & 1 \& 2 & Request of user $u$ for video $v_l$ & $\delta^{v_l}_{u}$ & 2  \\
    \hline \rowcolor[gray]{0.875}
    User association indicator & $\theta_{b,u}$ & 1 \& 2 & Scheduling events & $\left(x^{v_l}_{b},y^{v_h,v_l}_{b},z^{v_l}_{b',b},t^{v_h,v_l}_{b',b},w^{v_h,v_l}_{b',b},o^{v_l}_{b}\right)$ & 1 \& 2 \\
    \hline \rowcolor[gray]{0.88}
    Cache placement indicator & $\rho^{v_l}_{b}$ & 1 \& 2 & Number of subcarriers & $N$ & 1 \& 2 \\
    \hline \rowcolor[gray]{0.885}
    Subcarrier assignment indicator & $\tau^{n}_{b,u}$ & 1 \& 2 & Transmit power & $p^{n}_{b,u}$ & 1 \& 2 \\
    \hline \rowcolor[gray]{0.89}
    Instantaneous channel power gain & $h^{n}_{b,u}$ & 2 & Instantaneous SINR & $\gamma^{n}_{b,u}$ & 2 \\
    \hline \rowcolor[gray]{0.895}
    Subcarrier bandwidth & $W_\text{s}$ & 1 \& 2 & Instantaneous data rate & $r^{n}_{b,u}$ & 2 \\
    \hline \rowcolor[gray]{0.90}
    AWGN noise power & $\sigma^{n}_{b,u}$ & 1 \& 2 & Instantaneous data rate & $r^{n}_{b,u}$ & 2 \\
    \hline \rowcolor[gray]{0.915}
    Maximum transmit power of each RRS & $P^\text{max}_b$ & 1 \& 2 & Instantaneous access latency & $D^{\text{Ac},v_l}_{b,u}$ & 2 \\
    \hline \rowcolor[gray]{0.92}
    Size of transcoding task & $\eta^{v_h,v_l}$ & 1 \& 2 & Workload of each transcoding task & $N^{v_h,v_l}_{\text{Cycle}}$ & 1 \& 2 \\
    \hline \rowcolor[gray]{0.925}
    rocessing resource of each task & $\phi^{v_h,v_l}_b$ & 1 \& 2 & Maximum processing capacity of each RRS & $\chi^\text{max}_b$ & 1 \& 2 \\
    \hline \rowcolor[gray]{0.93}
    Transcoding delay & $D^{\text{TC},v_h,v_l}_{b}$ & 1 \& 2 & Backhaul capacity for each video & $r^{v_l}_{0,b}$ & 1 \& 2 \\
    \hline \rowcolor[gray]{0.935}
    Fronthaul capacity for each video & $r^{v_l}_{b',b}$ & 1 \& 2 & Maximum capacity of each backhaul link & $R^\text{max}_{0,b}$ & 1 \& 2 \\
    \hline \rowcolor[gray]{0.94}
    Maximum capacity of each fronthaul link & $R^\text{max}_{b',b}$ & 1 \& 2 & Backhaul delay & $D^{\text{BH},v_l}_{0,b}$ & 1 \& 2 \\
    \hline \rowcolor[gray]{0.945}
    Fronthaul delay & $D^{\text{FH},v_l}_{b',b}$ & 1 \& 2 & Contracted minimum data rate at each slice & $R^\text{min}_{m}$ & 1 \& 2 \\
    \hline \rowcolor[gray]{0.95}
    Unit price of transmit power & $\alpha^\text{Pow}_b$ & 1 \& 2 & Unit price of subcarrier & $\alpha^\text{Sub}_b$ & 1 \& 2 \\
    \hline \rowcolor[gray]{0.955}
    Unit price of backhaul capacity & $\alpha^\text{BH}$ & 1 \& 2 & Unit price of fronthaul capacity & $\alpha^\text{FH}$ & 1 \& 2 \\
    \hline \rowcolor[gray]{0.96}
    Unit price of storage & $\mu^\text{Cache}_b$ & 1 \& 2 & Unit price of processing & $\mu^\text{Proc}_b$ & 1 \& 2 \\
    \hline \rowcolor[gray]{0.965}
    Unit reward of each slice & $\psi_{m}$ & 1 \& 2 & Provisioning cost of each video at each slice & $\$^{\text{Cost,RRS},v_l}_{b}$ & 2 \\
    \hline \rowcolor[gray]{0.97}
    Revenue of each slice & $\textbf{\$}^\text{slice}_m$ & 2 & Total revenue of slices & $\textbf{\$}_\text{tot}$ & 2 \\
    \hline \rowcolor[gray]{0.975}
    Total revenue of slices & $\textbf{\$}_\text{tot}$ & 2 &  Maximum number of each user for each subcarrier & $\Psi_b$ & 1 \& 2 \\
    \hline \rowcolor[gray]{0.98}
    Popularity of requesting videos & $\Delta_{v_l}$ & 1 & Average rate of each user & $\bar{r}^\text{Ac}_{b,u}$ & 1 \\
    \hline \rowcolor[gray]{0.985}
    Average latency at each user & $\bar{D}^{\text{Ac},v_l}_{b,u}$ & 1 & Maximum storage capacity of each RRS & $C^\text{max}_b$ & 1 \\
    \hline \rowcolor[gray]{0.99}
    Average provisioning cost of a video file at each RRS & $\bar{\$}^{\text{Cost,RRS}}_{b}$ & 1 & Average revenue of each slice &  $\bar{\textbf{\$}}^\text{slice}_m$ & 1 \\
    \hline \rowcolor[gray]{0.995}
    Number of videos requested by users in $\mathcal{U}_m$ associated with RRS $b$ & $f_{b,m}\left( \boldsymbol{\theta} \right)$ & 1 & Upper-bound of total average revenue of slices & $\bar{\textbf{\$}}^\text{LD}_\text{tot}$ & 1 \\
    \hline \rowcolor[gray]{0.999}
    Lower-bound of total average revenue of slices & $\bar{\textbf{\$}}^\text{HD}_\text{tot}$ & 1 & - & - & - \\
    \hline
  \end{tabular}}
\end{center}
\end{table*}

The mixed-integer nonlinear programming (MINLP) problems \eqref{Main Problem delivery}, \eqref{Main Problem caching}, and \eqref{Main Problem caching HD} are completely NP-hard, which are mathematically proved in \cite{JointOptimizationPowerChannel} in the transmit power and subcarrier allocation problem for MC-NOMA in order to maximize the downlink data rate of users. In addition, in \cite{Jointsubchannelassignment}, it is mentioned that data rate maximization optimization problems with interference are completely NP-hard for OFDMA-based wireless networks and it is non-achievable to find the global optimum joint transmit power and subcarrier allocation policy by any existing method. Since the transmit power and subcarrier allocation in OFDMA is a special case of MC-NOMA \cite{JointOptimizationPowerChannel}, the result follows.
The exhaustive search requires the examination of all $(2S^\text{Power})^{BUN}. (2S^\text{FH})^{B^2VL}. (8S^\text{BH})^{BVL}. (2S^\text{Process})^{BVL^2}. 2^{BU}.4^{B^2VL^2}$ possible situations where $S^\text{Power}$, $S^\text{FH}$, $S^\text{BH}$, and $S^\text{Process}$ are the number of values that each variable in $\boldsymbol{p}$, $\boldsymbol{r}^\text{FH}$, $\boldsymbol{r}^\text{BH}$, and $\boldsymbol{\phi}$ can take, respectively.
Accordingly, it is very challenging and impractical to find the global optimum solution for such
large-scale and NP-hard optimization problems, since the number of optimization variables and constraints grow exponentially
\cite{FastDeliveryDistributedCaching,cachefogkhaled,BackhaulAwareCaching,JointOptimizationPowerChannel,7925732}.

The main structure of optimization problems \eqref{Main Problem delivery} and \eqref{Main Problem caching} are the same for a fixed
cache placement variable $\boldsymbol{\rho}$. Therefore, we provide a local optimum resource allocation algorithm that can be adopted for both problems
\eqref{Main Problem delivery} and \eqref{Main Problem caching}.
However, \eqref{Main Problem caching HD} and \eqref{Main Problem caching} differ in objective function. Accordingly, we modify the solution algorithm proposed for \eqref{Main Problem caching} to be applied to \eqref{Main Problem caching HD}.

\section{Delivery-Aware Cooperative Multi-bitrate Video Caching and Transcoding Algorithms}\label{Section Solution}
This section provides solution algorithms for the main problems \eqref{Main Problem delivery}, \eqref{Main Problem caching} and \eqref{Main Problem caching
HD}.

\subsection{DACPSs and Delivery Algorithm}\label{Subsection Deliveryaware algorithms}
Here, we propose an efficient solution algorithm for \eqref{Main Problem caching} by utilizing the alternate optimization algorithm
\cite{6708492,Jointsubchannelassignment,Doubleiterativewaterfilling}. This algorithm consists of two main steps:
1) finding joint $\boldsymbol{p}$, $\boldsymbol{\phi}$, $\boldsymbol{r}^\text{FH}$, and $\boldsymbol{r}^\text{BH}$;
2) finding joint $\boldsymbol{\Upsilon}$, $\boldsymbol{\rho}$, $\boldsymbol{\tau}$, and $\boldsymbol{\theta}$.
We repeat the aforementioned steps until we have $ \| \bar{\textbf{\$}}^\text{LD}_\text{tot} (\kappa_1) - \bar{\textbf{\$}}^\text{LD}_\text{tot}
(\kappa_1-1) \| \le \varepsilon_1$ where $\varepsilon_1$ is a positive small value and $\kappa_1$ is the iteration index, or the number of
main iterations exceeds a pre-defined threshold $\Gamma_1$. The proposed approach is summarized in Algorithm \ref{Alg iterative main}.
\begin{algorithm}[tp]
 \caption{The main alternate method.}\label{Alg iterative main}
 \begin{algorithmic}[1]
  \STATE Initialize $\boldsymbol{\Upsilon}_0$, $\boldsymbol{\rho}_0$, $\boldsymbol{p}_0$, $\boldsymbol{\phi}_0$, $\boldsymbol{r}^\text{FH}_0$,
  $\boldsymbol{r}^\text{BH}_0$, $\boldsymbol{\tau}_0$, and $\boldsymbol{\theta}_0$. 
  \\ \textbf{repeat}
  \FOR {$\kappa_1=1$ to $\Gamma_1$}
  \STATE \textbf{Step $1$}: Find $\left( \boldsymbol{p}_{\kappa_1}, \boldsymbol{\phi}_{\kappa_1}, \boldsymbol{r}^\text{FH}_{\kappa_1},
  \boldsymbol{r}^\text{BH}_{\kappa_1} \right)$ by solving \eqref{Main Problem caching} for a fixed
  $\left( \boldsymbol{\Upsilon}_{\kappa_1-1} , \boldsymbol{\rho}_{\kappa_1-1}, \boldsymbol{\tau}_{\kappa_1-1}, \boldsymbol{\theta}_{\kappa_1-1} \right)$.
  \STATE \textbf{Step $2$}: Find $\left( \boldsymbol{\Upsilon}_{\kappa_1} , \boldsymbol{\rho}_{\kappa_1}, \boldsymbol{\tau}_{\kappa_1},
  \boldsymbol{\theta}_{\kappa_1} \right)$ by solving \eqref{Main Problem caching} for a given $\left( \boldsymbol{p}_{\kappa_1},
  \boldsymbol{\phi}_{\kappa_1}, \boldsymbol{r}^\text{FH}_{\kappa_1}, \boldsymbol{r}^\text{BH}_{\kappa_1} \right)$.
  \STATE \textbf{Until} $ | \bar{\textbf{\$}}^\text{LD}_\text{tot} (\kappa_1) - \bar{\textbf{\$}}^\text{LD}_\text{tot} (\kappa_1-1) | \le \varepsilon_1$ or
  $\kappa_1=\Gamma_1$.
  \STATE Set $\kappa_1=\kappa_1+1$.
  \ENDFOR
  \STATE $\boldsymbol{\Upsilon}_{\kappa_1}$, $\boldsymbol{\rho}_{\kappa_1}$, $\boldsymbol{p}_{\kappa_1}$, $\boldsymbol{\phi}_{\kappa_1}$,
  $\boldsymbol{r}^\text{FH}_{\kappa_1}$, $\boldsymbol{r}^\text{BH}_{\kappa_1}$, $\boldsymbol{\tau}_{\kappa_1}$, and $\boldsymbol{\theta}_{\kappa_1}$ are
  adopted for the network.
 \end{algorithmic}
 \end{algorithm}

\subsubsection{Step $1$}\label{Subsection Step1 algorithm}
In the first step, we obtain $\boldsymbol{p}$, $\boldsymbol{\phi}$, $\boldsymbol{r}^\text{FH}$ and $\boldsymbol{r}^\text{BH}$ jointly, by solving the
following subproblem for fixed $\boldsymbol{\Upsilon}$, $\boldsymbol{\rho}$, $\boldsymbol{\tau}$, and $\boldsymbol{\theta}$ as
\begin{subequations}\label{Main Problem jointpower}
\begin{align}\label{obf Main Problem jointpower}
&\max_{ \boldsymbol{\phi} , \boldsymbol{p} , \boldsymbol{r}^\text{BH}, \boldsymbol{r}^\text{FH} }\hspace{.0 cm} 	
\bar{\textbf{\$}}^\text{LD}_\text{tot}
\\
\textrm{s.t.}\hspace{.0cm}~~
& \text{\eqref{Constraint maximum transmit power BS}-\eqref{channel capacity constraint FH BS}, \eqref{constraint event 4 transcoding fronthaul},
\eqref{constraint event 5 fronthaul transcoding}, \eqref{constraint subcarrier binary}, \eqref{SIC user ergodic}, \eqref{constraint event 2 average access
transc}, \eqref{constraint minimum rate access erg}.}  \nonumber
\end{align}
\end{subequations}
Problem \eqref{Main Problem jointpower} is still MINLP, which is NP-hard, due to non-concavity of objective function \eqref{obf Main Problem jointpower} and non-convexity of constraints \eqref{constraint event 4 transcoding fronthaul}, \eqref{constraint event 5 fronthaul transcoding}, \eqref{constraint subcarrier binary}, \eqref{SIC user ergodic}, \eqref{constraint event 2 average access transc}, and \eqref{constraint minimum rate access erg}.
In order to deal with the aforementioned challenges, we first relax $\phi^{v_h,v_l}_{b}$ in \eqref{constraint subcarrier binary} to be a non-negative real value \cite{7762913,PAYMARD2019160}, which is an acceptable approach in this context since the maximum number of CPU cycles in each processor is on the order of $10^9$ \cite{DynamicComputationOffloading,8387798,PAYMARD2019160}. For the non-convex constraints, we apply transformation methods to tackle their non-convexity (please see Appendix \ref{appendix transformation power problem}). Then, to tackle the non-concavity of $\bar{r}^\text{Ac}_{b,u}$ in \eqref{obf Main Problem jointpower} and \eqref{constraint minimum rate access erg}, and its non-convexity in constraints \eqref{constraint event 2 average access transc transform}-\eqref{constraint event 6 backhaul average access transform} of the transformed problem (presented in Appendix \ref{appendix transformation power problem}), we use the successive convex approximation (SCA) approach based on the difference-of-two-concave-functions (D.C.) approximation method \cite{Jointsubchannelassignment,LimitedFeedback,JointOptimizationofCloud}.
In this regard, we first initialize the approximation parameters. Then, we solve the convex approximated problem to find $\left( \boldsymbol{\phi} , \boldsymbol{p} , \boldsymbol{r}^\text{BH}, \boldsymbol{r}^\text{FH} \right)$. These iterations are repeated until the stopping criterion is satisfied. The derivations of the proposed SCA algorithm are presented in Appendix \ref{appendix SCA power}.
Additionally, the pseudo code of the SCA algorithm with the D.C. approximation method is summarized in Algorithm \ref{Alg SCA power caching}.
\begin{algorithm}[tp]
 \caption{The proposed SCA algorithm with the D.C. approximation method.} \label{Alg SCA power caching}
 \begin{algorithmic}[1]
 \STATE Initialize $\boldsymbol{p}_{0}$.
 \\ \textbf{repeat}
  \STATE Update the convex approximated forms of \eqref{constraint minimum rate access erg}, \eqref{constraint event 2 average access transc transform}-\eqref{constraint event 6 backhaul average access transform} and substitute them into \eqref{Main Problem jointpower transform}.
  \STATE Find $\left( \boldsymbol{\phi}_{\kappa_2} , \boldsymbol{p}_{\kappa_2} , \boldsymbol{r}^\text{BH}_{\kappa_2}, \boldsymbol{r}^\text{FH}_{\kappa_2} \right)$ by solving the convex approximated problem of \eqref{Main Problem jointpower transform}.
  \STATE Set $\kappa_2=\kappa_2 + 1$
  \\ \textbf{Until} Convergence of $\boldsymbol{p}$.
  \STATE The variables $\boldsymbol{p}$, $\boldsymbol{r}^\text{BH}$, $\boldsymbol{r}^\text{FH}$, and $\boldsymbol{\phi}$ are the outputs of the algorithm.
 \end{algorithmic}
\end{algorithm}

\subsubsection{Step $2$}\label{Subsection Step1 algorithm}
Here, we jointly find the binary variables $\boldsymbol{\Upsilon}$, $\boldsymbol{\rho}$, $\boldsymbol{\theta}$, and $\boldsymbol{\tau}$ for the given $\left( \boldsymbol{p}, \boldsymbol{\phi}, \boldsymbol{r}^\text{FH}, \boldsymbol{r}^\text{BH} \right)$ by solving the following subproblem:
\begin{subequations}\label{Main Problem caching subcarrier}
\begin{align}\label{obf Main Problem caching subcarrier}
\min_{ \boldsymbol{\Upsilon}, \boldsymbol{\rho}, \boldsymbol{\theta}, \boldsymbol{\tau}  }\hspace{.0 cm}
& ~~	
\bar{\textbf{\$}}^\text{LD}_\text{tot}
\\
\textrm{s.t.}\hspace{.0cm}~~
& \text{\eqref{user association one BS}, \eqref{constraint cases storage}, \eqref{user association subcarrier}, \eqref{processing capacity constraint eRRH}-\eqref{channel capacity constraint FH BS}, \eqref{constraint event 4 transcoding fronthaul}, \eqref{constraint event 5 fronthaul transcoding}, \eqref{constraint untranscodable}-\eqref{constraint subcarrier binary}, \eqref{SIC user ergodic}-\eqref{constraint cache size BS}, \eqref{constraint minimum rate access erg}, \eqref{constraint rho binary}.} \nonumber
\end{align}
\end{subequations}
Problem \eqref{Main Problem caching subcarrier} is classified as integer nonlinear programming (INLP) due to
combinatorial constraints \eqref{constraint subcarrier binary} with respect to $\boldsymbol{\Upsilon}, \boldsymbol{\theta}$, and $ \boldsymbol{\tau}$,
combinatorial constraints in \eqref{constraint rho binary} with respect to $\boldsymbol{\rho}$,
nonlinear SIC constraint \eqref{SIC user ergodic} with respect to $\boldsymbol{\tau}$,
the term $\min\{\sum_{u=1}^{U} \theta_{b,u},1\}$ in \eqref{all cases constraint average},
both the terms $\max\limits_{ u \in \mathcal{U}_m } \{ \tau^{n}_{b,u} \}$ and $\min \left\{ \sum_{u \in \mathcal{U}_m} \theta_{b,u} , 1 \right\}$ in \eqref{obf Main Problem caching subcarrier} (please refer to \eqref{revenue MVNO average LD}),
multiplication of $\min \left\{ \sum_{u \in \mathcal{U}_m} \theta_{b,u} , 1 \right\}$ in scheduling variables in $\boldsymbol{\Upsilon}$ in \eqref{obf Main Problem caching subcarrier} (please refer to \eqref{revenue MVNO average LD} and \eqref{cost providingBS}), nonlinearity of fractional delay functions in \eqref{constraint event 4 transcoding fronthaul} and \eqref{constraint event 5 fronthaul transcoding}, and average access delay functions in \eqref{constraint event 2 average access transc} with respect to $\boldsymbol{\tau}$.
To solve \eqref{Main Problem caching subcarrier}, our aim is to transform \eqref{Main Problem caching subcarrier} into an integer disciplined convex programming (IDCP). This can be easily solved by utilizing the efficient standard optimization software CVX with the internal solver MOSEK that applies the Branch\&bound\&cut algorithm \cite{8611393,CVXmatlab,MOSEKsolver,LimitedFeedback,7812683}.

\emph{\textbf{Proposition 1}}: The INLP problem \eqref{Main Problem caching subcarrier} can be equivalently transformed into an IDCP form
which is presented in Appendix \ref{appendix Proposition 1}.

To solve \eqref{Main Problem delivery} for Phase 2, we can also apply Algorithm \ref{Alg iterative main}, since when $\boldsymbol{\rho}$ is fixed,  \eqref{Main Problem delivery} and \eqref{Main Problem caching} have a similar structure. Due to space limitations, the solution method for problem \eqref{Main Problem delivery} is not included here. To solve \eqref{Main Problem caching HD} in the HD scheme, we again apply a similar method to Algorithm \ref{Alg iterative main} (See Appendix \ref{appendix transformation HD cache placement problem}).

\subsection{Convergence of The Proposed Algorithms}\label{SubSection convergence}
Here, we discuss the convergence of our proposed Algorithm \ref{Alg iterative main} for solving \eqref{Main Problem caching}. This discussion is presented in the format of the following two propositions.

\emph{\textbf{Proposition 2}}: The objective function \eqref{obf Main Problem caching} is upper-bounded by the total average reward of slices which is obtained by $\sum_{ u \in \mathcal{U}_m } \sum\limits_{b \in \mathcal{B}} \bar{r}^\text{Ac}_{b,u} \psi_m$ that is a nonnegative finite term, due to limited access bandwidth and constraint \eqref{Constraint maximum transmit power BS}. Therefore, for a feasible problem \eqref{Main Problem caching}, the proposed alternate Algorithm \ref{Alg iterative main} will converge to a locally optimal solution.
\begin{proof}
Please see Appendix \ref{appendix proposition 2}.
\end{proof}

\emph{\textbf{Proposition 3}}: The SCA approach with the D.C. approximation method generates a sequence of improved feasible solutions. Therefore, the proposed algorithm for solving \eqref{Main Problem jointpower} will converge to a locally optimal solution when the number of SCA iterations are large enough.
\begin{proof}
Please see Appendix \ref{appendix proposition 3}.
\end{proof}
Since all proposed algorithms have similar structures, the convergence of the other algorithms can be proved same as the proposed Algorithm \ref{Alg iterative main} for solving \eqref{Main Problem caching}.

\subsection{Computational Complexity of the Proposed Algorithm}\label{Section computational complexity}
Here, we aim to obtain the computational complexity of the proposed solution algorithms for problems \eqref{Main Problem delivery}, \eqref{Main Problem caching}, and \eqref{Main Problem caching HD}.
Since the proposed alternate algorithms for solving \eqref{Main Problem delivery}, \eqref{Main Problem caching}, and \eqref{Main Problem caching HD} have basically the same structure, we only present the details of obtaining the computational complexity of solving  \eqref{Main Problem caching} using Algorithm \ref{Alg iterative main}. After that, the computational complexity of solving \eqref{Main Problem delivery} and \eqref{Main Problem caching HD} is investigated.

In the first step of Algorithm \ref{Alg iterative main}, we first use a transformation method presented in Appendix \ref{appendix transformation power problem} to solve \eqref{Main Problem jointpower}. We then solve the equivalent result problem \eqref{Main Problem jointpower transform} using the iterative SCA approach based on the D.C. approximation method. In each iteration $\kappa_2$ of SCA, the approximated disciplined convex programming (DCP) problem of \eqref{Main Problem jointpower transform} is solved by CVX, which employs the geometric programming with the interior-point method (IPM) \cite{LimitedFeedback,CVXmatlab}. Therefore,
the computational complexity of solving the approximated problem of \eqref{Main Problem jointpower transform} is on the order of
$\Omega^\text{1}_\text{LD} = \frac {  \log\left(  T^\text{1}_\text{LD} /\left(t^0 \varrho\right) \right) }   {  \log \epsilon^0 }$,
where $T^\text{1}_\text{LD} = 3B + B^2 + U + M + BU^2N + BVLU \left( 1+B +L \right) + B^2VL^2 \left( 2+2U \right)$ is the total number of constraints in the approximated problem of \eqref{Main Problem jointpower transform}, $t^0$ is the initial point for approximating the accuracy of the IPM, $0< \varrho \ll \infty$ is the stopping criterion for the IPM, and $\epsilon^0$ is for updating the accuracy of the IPM \cite{LimitedFeedback,CVXmatlab}. Note that $\Omega^\text{1}_\text{LD}$ is only for one iteration of the SCA approach. The total complexity of solving \eqref{Main Problem jointpower transform} by using the SCA method mainly depends on the number of optimization variables, constraints, and accuracy of the algorithm. In the second step of Algorithm \ref{Alg iterative main}, we find $\left( \boldsymbol{\Upsilon} , \boldsymbol{\rho} , \boldsymbol{\theta}, \boldsymbol{\tau} \right)$ by solving \eqref{Main Problem caching subcarrier}. In fact, we transform \eqref{Main Problem caching subcarrier} into an equivalent IDCP problem \eqref{Main Problem caching subcarrier transf epi2} using the epigraph technique. The resulting IDCP problem \eqref{Main Problem caching subcarrier transf epi2} is also solved by utilizing CVX with the MOSEK solver. The complexity of solving \eqref{Main Problem caching subcarrier transf epi2} can thus be obtained by
$\Omega^\text{2}_\text{LD} = \frac {  \log\left(  T^\text{2}_\text{LD} /\left(t^0 \varrho\right) \right) }   {  \log \epsilon^0 }$,
where $T^\text{2}_\text{LD} = 3B + 2U + M + B^2 + 2BU + BN + 2BUN + BU^2N + BVL \big( 2+B+L+6M+4UN+3LM+3BM+4LUN+4BUN \big) + B^2VL^2 \big( 4+6M+8UN \big) $ is the total number of constraints in \eqref{Main Problem caching subcarrier transf epi2}. Accordingly, the total computational complexity of solving \eqref{Main Problem caching} is on the order of
$\Omega^\text{tot}_\text{LD} = \Xi^\text{Main} \left( \Xi^\text{SCA} \Omega^\text{1}_\text{LD} + \Omega^\text{2}_\text{LD} \right)$ where $\Xi^\text{Main}$ and $\Xi^\text{SCA}$ are the total number of main and SCA iterations, respectively.

The computational complexity of solving the cache placement optimization problem \eqref{Main Problem caching HD} can also be obtained using the same method as \eqref{Main Problem caching}. In this line, the complexity of finding joint $\boldsymbol{p}$, $\boldsymbol{\phi}$, $\boldsymbol{r}^\text{FH}$ and $\boldsymbol{r}^\text{BH}$ at each iteration of the SCA approach is on the order of
$\Omega^\text{1}_\text{HD} = \frac {  \log\left(  T^\text{1}_\text{HD} /\left(t^0 \varrho\right) \right) }   {  \log \epsilon^0 }$,
where $T^\text{1}_\text{HD}$
is exactly equal to $T^\text{1}_\text{LD}$. On the other hand, the complexity of solving the second step of the proposed alternate algorithm for solving \eqref{Main Problem caching HD} is given by
$\Omega^\text{2}_\text{HD} = \frac {  \log\left(  T^\text{2}_\text{HD} /\left(t^0 \varrho\right) \right) }   {  \log \epsilon^0 }$,
where $T^\text{2}_\text{HD} = 3B + 2U + M + B^2 + BU + BN + 2BUN + BU^2N + BVL \big( 2+B+L+6U+4UN+3LU+3BU+4LUN+4BUN \big) + B^2VL^2 \big( 4+6U+8UN \big) $.

The complexity of solving \eqref{Main Problem delivery} can be obtained as the same way of obtaining the complexity of solving \eqref{Main Problem caching}, since both the proposed algorithms have the same structure for a fixed $\boldsymbol{\rho}$. In this way, the complexity of solving \eqref{Main Problem delivery} and finding $\boldsymbol{p}$, $\boldsymbol{\phi}$, $\boldsymbol{r}^\text{FH}$ and $\boldsymbol{r}^\text{BH}$ is exactly on the order of $\Omega^\text{1}_\text{LD}$. Moreover, the computational complexity of finding $\left( \boldsymbol{\Upsilon} , \boldsymbol{\theta}, \boldsymbol{\tau} \right)$ is obtained by
$\Omega^\text{2}_\text{Del} = \frac {  \log\left(  T^\text{2}_\text{Del} /\left(t^0 \varrho\right) \right) }   {  \log \epsilon^0 }$,
where $T^\text{2}_\text{Del} = 2B + 2U + M + B^2 + 2BU + BN + 2BUN + BU^2N + BVL \big( 2+B+L+6M+4UN+3LM+3BM+4LUN+4BUN \big) + B^2VL^2 \big( 4+6M+8UN \big) $.

\subsection{Designing a Low-Complexity Resource Allocation Framework}\label{Section low complexity framework}
It is very important that the central scheduler be fast enough to readopt the delivery policy in each time slot of Phase 2 based on the arrival of instantaneous requests from users and CSI, especially in realistic ultra dense 5G wireless networks with a large number of unique videos.
For this reason, here we propose another RAF that has a lower computational complexity in each time slot of Phase 2 than that of our proposed RAF in Fig. \ref{Fig00structure}. The main part of the complexity of the proposed framework in Fig. \ref{Fig00structure} is caused by reallocating the radio resources as well as re-associating users to RRSs at each time slot of Phase 2 to readopt the access transmission strategy based on the arrival requests and CSI. However, some environments with higher path loss heavily limit the flexibility of the user association process and reduce the impact of the wireless small-scale fading on the SINR of users. On the other hand, we aim to utilize the benefits of radio resource allocation and user association policies in the MC-NOMA system to improve user data rates and correspondingly improve user revenues.
Thus, in this novel framework, we adapt the obtained radio resource allocation, i.e., transmit power and subcarrier allocation, and user association policies in Phase 1 to all time slots of Phase 2. Hence, the system only reallocates the processing, fronthaul and backhaul resources as well as the request scheduling at the beginning of each time slot of Phase 2 based on the arrival requests from users.
Accordingly, in this framework, the LD and HD cache placement problems are exactly the same as \eqref{Main Problem caching} and \eqref{Main Problem caching HD} whereas the delivery optimization problem at each time slot of Phase 2 is formulated as
\begin{subequations}\label{Main Problem delivery lowcomplexity}
\begin{align}\label{obf Main Problem delivery lowcomplexity}
&\max_{ \boldsymbol{\phi} ,\boldsymbol{r}^\text{BH}, \boldsymbol{r}^\text{FH}, \boldsymbol{\Upsilon} }\hspace{.0 cm} 	
\textbf{\$}_\text{tot}
\\
\textrm{s.t.}\hspace{.0cm}~~
& \text{\eqref{all cases constraint}, \eqref{constraint cases storage}, \eqref{processing capacity constraint eRRH}-\eqref{channel capacity constraint FH BS}, \eqref{constraint event 2 access transc}-\eqref{constraint event 6 backhaul access}.}  \nonumber
\end{align}
\end{subequations}
The optimization problem \eqref{Main Problem delivery lowcomplexity} is a MINLP which can be efficiently solved by utilizing an alternate algorithm in which \eqref{Main Problem delivery lowcomplexity} is divided into two subproblems as: 1) finding joint $\boldsymbol{\phi}$, $\boldsymbol{r}^\text{BH}$, and $\boldsymbol{r}^\text{FH}$; 2) finding $\boldsymbol{\Upsilon}$. The problem of finding joint $\boldsymbol{\phi}$, $\boldsymbol{r}^\text{BH}$, and $\boldsymbol{r}^\text{FH}$ is a linear programming (LP) and thus, the globally optimal solution can be  found by utilizing the CVX software or the Lagrange dual method.
On the other hand, the IDCP problem of finding $\boldsymbol{\Upsilon}$ is solved by using the MOSEK solver.

The computational complexity of solving \eqref{Main Problem delivery lowcomplexity} can be obtained as the same way of obtaining the complexity of solving \eqref{Main Problem delivery}. In this way, the complexity of finding $\boldsymbol{\phi}$, $\boldsymbol{r}^\text{BH}$, and $\boldsymbol{r}^\text{FH}$ is on the order of $\Omega^\text{1}_\text{Del,LowComplex} = \frac {  \log\left(  T^\text{1}_\text{Del,LowComplex} /\left(t^0 \varrho\right) \right) }   {  \log \epsilon^0 }$, where $T^\text{1}_\text{Del,LowComplex} = 2B + B^2 + BVL \big( 1+B+L \big) + 4B^2VL^2$. Moreover, the complexity of finding $\boldsymbol{\Upsilon}$ is obtained by $\Omega^\text{2}_\text{Del,LowComplex} = \frac {  \log\left(  T^\text{2}_\text{Del,LowComplex} /\left(t^0 \varrho\right) \right) }   {  \log \epsilon^0 }$, where $T^\text{2}_\text{Del,LowComplex} = 2B + B^2 + BVL \big( 3+2B+2L \big) + 6B^2VL^2$.

\subsection{Cache Refreshment Strategy}\label{Subsection cache refreshment strategy}
As we mentioned in our proposed two-phase RAF shown in Fig. \ref{Fig00structure}, Phase 1 occurs only at off-peak times where storages will be updated based on the changes of CDI and/or VPD. It means that the system cannot update its storages upon CDI and/or VPD are changed during Phase 2 (peak-traffic times), due to the scarcity of backhaul capacity. To tackle this issue, we propose a cache refreshment framework where each time slot of Phase 2 is split into three phases: video request, video transmission, and cache refreshment, respectively \cite{7384533,7406764}. In the cache refreshment phase, the storage of RRSs will be updated based on the adopted DACRS according to the instantaneous requests of users and CSI in prior phases, and also availability of video files at the storages. The cache refreshment phase is occurred only at the end of the video transmission phase, since the new video files can be replaced only when they are fully available at storages \cite{7384533,7406764}. Since each cache refreshment decision is made only based on the last requests of users and CSI, the DACRS is very useful when CDI and/or VPD are not available in Phase 1 of the proposed RAF in Fig. \ref{Fig00structure}.

In the proposed DACRS, similar to our proposed DACPSs, we jointly find variables $\boldsymbol{\rho}$, $\boldsymbol{\theta}$, $\boldsymbol{\phi}$, $\boldsymbol{p}$, $\boldsymbol{\tau}$, $\boldsymbol{r}^\text{BH}$, $\boldsymbol{r}^\text{FH}$, and $\boldsymbol{\Upsilon}$ while the output is only $\boldsymbol{\rho}$ that determines the cache update decision. Actually, the cache refreshment optimization problem is very similar to the delivery optimization problem \eqref{Main Problem delivery} in which $\boldsymbol{\rho}$ is not pre-defined. However, each cache refreshment operation can be occurred when the target video be available at the storage. Actually, the following constraint is added to \eqref{Main Problem delivery} as
\begin{multline}\label{constraint video availability}
\rho^{v_l}_{b} \leq \rho^{\text{Old},v_l}_{b} + \sum\limits_{\scriptstyle v_h \in \mathcal{V} \atop \scriptstyle h > l} y^{\text{Old},v_h,v_l}_{b} + \sum\limits_{\scriptstyle b' \in \mathcal{B} \atop \scriptstyle b' \neq b} z^{\text{Old},v_l}_{b',b} +
\sum\limits_{\scriptstyle b' \in \mathcal{B} \atop \scriptstyle b' \neq b} \sum\limits_{\scriptstyle v_h \in \mathcal{V} \atop \scriptstyle h > l} \bigg(
t^{\text{Old},v_h,v_l}_{b',b} + w^{\text{Old},v_h,v_l}_{b',b} \bigg) + \sum\limits_{\scriptstyle b' \in \mathcal{B} \atop \scriptstyle b' \neq b} \sum\limits_{\scriptstyle v_k \in \mathcal{V} \atop \scriptstyle k < l} ( w^{\text{Old},v_l,v_k}_{b',b} )
\\
+ o^{\text{Old},v_l}_{b}, \forall b \in \mathcal{B},
v_l \in \mathcal{V}.
\end{multline}
Video $v_l$ can be cached at RRS $b$ at the cache refreshment phase if it is available on the buffer of RRS $b$. In \eqref{constraint video availability}, the binary parameters $\rho^{\text{Old},v_l}_{b}$, $y^{\text{Old},v_h,v_l}_{b}$, $z^{\text{Old},v_l}_{b',b}$, $t^{\text{Old},v_h,v_l}_{b',b}$, $w^{\text{Old},v_h,v_l}_{b',b}$, and $o^{\text{Old},v_l}_{b}$ are defined during the video transmission phase and fixed at the cache refreshment phase. The cache refreshment optimization problem is formulated by
\begin{subequations}\label{Main Problem refreshment}
\begin{align}\label{obf Main Problem refreshment}
&\max_{ \boldsymbol{\rho}, \boldsymbol{\theta} , \boldsymbol{\phi} , \boldsymbol{p} , \boldsymbol{\tau} , \boldsymbol{r}^\text{BH}, \boldsymbol{r}^\text{FH},
\boldsymbol{\Upsilon} }\hspace{.0 cm} 	
\textbf{\$}_\text{tot}
\\
\textrm{s.t.}\hspace{.0cm}~~
& \text{\eqref{user association one BS}-\eqref{user association subcarrier}, \eqref{SIC user}, \eqref{Constraint maximum transmit power
BS}-\eqref{constraint event 6 backhaul access}, \eqref{constraint untranscodable}-\eqref{constraint subcarrier binary}, \eqref{constraint cache size BS}, \eqref{constraint rho binary}, \eqref{constraint video availability}.}  \nonumber
\end{align}
\end{subequations}

Due to the similarity of \eqref{Main Problem refreshment} to problems \eqref{Main Problem caching} and \eqref{Main Problem delivery}, the proposed Algorithm \ref{Alg iterative main} can be utilized to solve \eqref{Main Problem refreshment}. To avoid duplicated discussions, the presentation of our proposed solution algorithm for solving \eqref{Main Problem refreshment} is not included here.

\section{Simulation Results}\label{Section simulation results}
\begin{figure}
\centering
\includegraphics[scale=0.48]{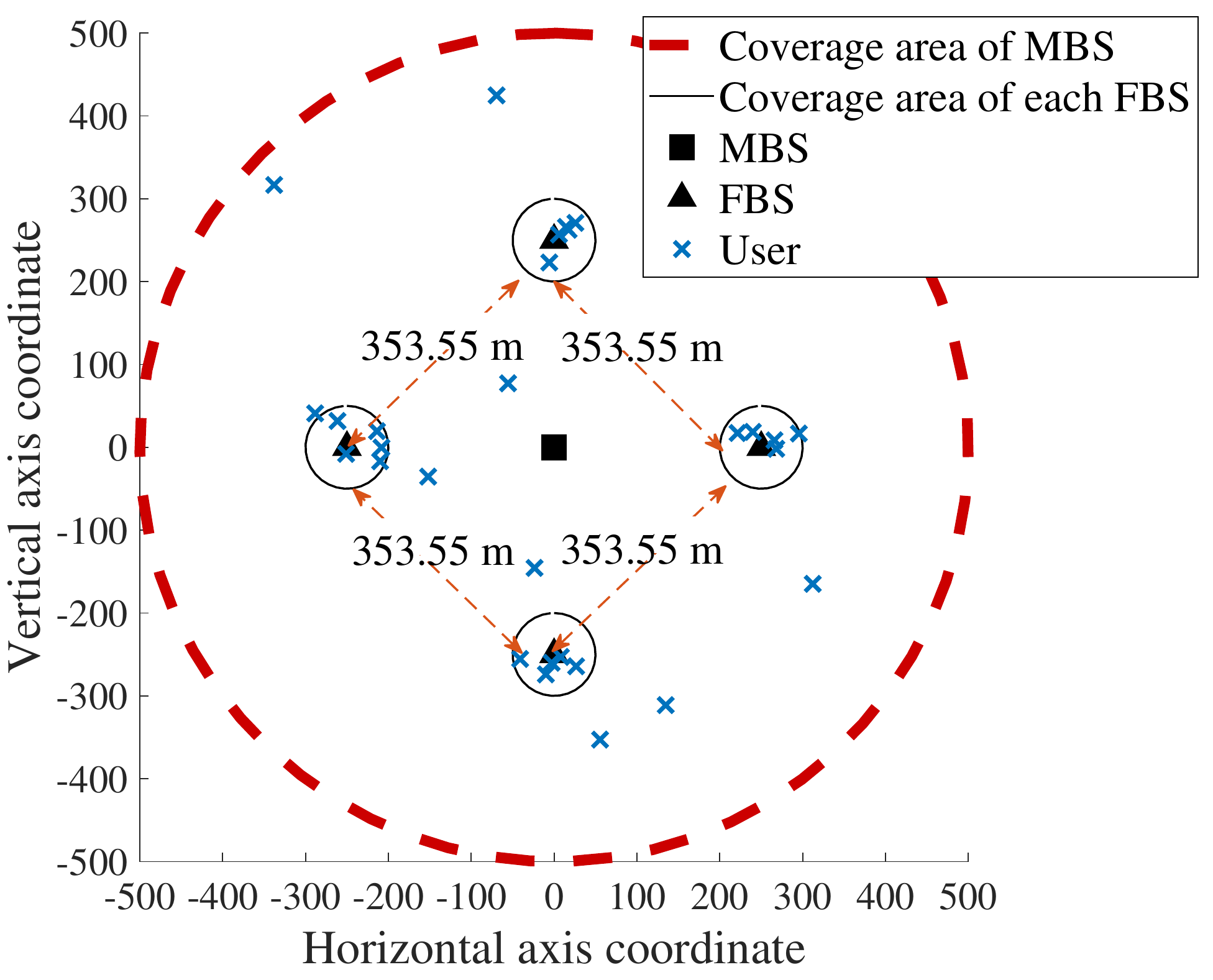}
\caption{Network setup and user placement in the simulation results.}
\label{Fig03}
\end{figure}

In this section, we present simulation results that demonstrate the performance of our proposed CPSs via MATLAB Monte Carlo simulations through $500$ network realizations \cite{7384533}. The network topology and user placement is shown in Fig. \ref{Fig03} where a single high-power RRS (HP-RRS), i.e., MBS, is located in the center of a circular area with radius $500$ m and $4$ low-power RRSs (LP-RRS), i.e., FBS, are located in a distance of $250$ m from the HP-RRS \cite{Jointsubchannelassignment}.
Ten users are uniformly deployed in the coverage area of HP-RRS while $5$ users are located in the coverage area of each LP-RRS with radii of $50$ m \cite{Jointsubchannelassignment}.
As seen in Fig. \ref{Fig03}, the shortest distance from a FBS to another FBS is $250\sqrt{2}\approx353.55$ m.

The wireless channel is centered at a frequency of $2.5$ GHz with bandwidth $W=5$ MHz, $N=64$ and $W_\text{s}=78.125$ KHz \cite{OptimalPowerSubcarrierSchober}. The combined path loss and shadowing is modeled as $128.1 + 37.6 \log_{10} d_{b,u} + z_{b,u}$ in dB in which
$z_{b,u}$ is a log-normal shadowing random variable with the standard deviation of 8 dB, and $d_{b,u} > 0$ denotes the distance between RRS $b$ and user $u$ in km \cite{GreenTouch,Doubleiterativewaterfilling,Taovideotranscod}.
The small-scale fading of the wireless channel is modeled as independent and identically distributed Rayleigh fading with variance $1$ \cite{Doubleiterativewaterfilling}.
The CDI is averaged over $1000$ CSI samples for a fixed channel power loss, since the location of users and RRSs are fixed in our numerical examples.
The PSD of AWGN noise is set to $N_0=-174$ dBm/Hz with a noise figure of $9$ dB at each user \cite{GreenOFDMAresourcecache}. In the MC-NOMA technology, we set $\Psi_b=2,\forall b \in \mathcal{B}$ \cite{7812683,Doubleiterativewaterfilling,OptimalPowerSubcarrierSchober}.
For the HP-RRS, we set $P^\text{max}_1=47$ dBm whereas for each LP-RRS $b$, $P^\text{max}_b=27$ dBm \cite{Jointsubchannelassignment}.

We assume that there exists $V=10$ unique videos, each having $L=4$ bitrate variants. In our simulations, we set the relative bitrates of the four variants to be $0.45$, $0.55$, $0.67$, and $0.82$ of the original video bitrate $2$ Mbps (HD quality) \cite{Collaborativemultibitrate,EnhancingMobileVideo}. Besides, all video variants have equal length of $10$ minutes \cite{Collaborativemultibitrate}. The skew parameter of the Zipf distribution is set to $\lambda=0.8$ \cite{Collaborativemultibitrate,EnhancingMobileVideo}.
Similar to \cite{Collaborativemultibitrate,EnhancingMobileVideo} we assume that the processing workload $\eta^{v_h,v_l}$ is proportional to $s_{v_l}$.
In this regard, we set $\eta^{v_h,v_l}= s_{v_l}$ \cite{Collaborativemultibitrate}.

For the proposed virtulization model, we assume that there exists $M=2$ slices in the infrastructure with $R^\text{min}_{1}=1$ Mbps and $R^\text{min}_{2}=2$ Mbps. Each slice also has $U/2=15$ users such that each user subscribes to any slices with a probability of $1/M=50\%$ \cite{7384533,7406764}.
In our proposed pricing scheme, we take
$\mu^\text{Cache}_1=2$ units/Gbit, $\mu^\text{Cache}_b=1.6$ units/Gbit, $\forall b \in \mathcal{B}/\{1\}$ \cite{7384533},
$\mu^\text{Proc}_1=0.8$ units/GHz, $\mu^\text{Proc}_b=0.7$ units/GHz, $\forall b \in \mathcal{B}/\{1\}$,
$\alpha^\text{Pow}_1=6$ units/Watt, $\alpha^\text{Pow}_b=4$ units/Watt, $\forall b \in \mathcal{B}/\{1\}$,
$\alpha^\text{Sub}_1=60$ units/MHz, $\alpha^\text{Sub}_b=40$ units/MHz, $\forall b \in \mathcal{B}/\{1\}$ \cite{7384533,7410051},
$\alpha^\text{FH}=2$ units/Mbps, $\alpha^\text{BH}=5$ units/Mbps \cite{7410051}.
Moreover, the unit reward of slices are $\psi_{1}=8.75$ units/Mbps and $\psi_{2}=9$ units/Mbps \cite{7406764,7384533,7410051}.

For processing capacities, we set $\chi^\text{max}_1=50$ GHz (maximum number of CPU cycles per seconds in HP-RRS is $50 \times 10^9$) and $\chi^\text{max}_b=25$ GHz, $\forall b \in \mathcal{B}/\{1\}$.
Moreover, the required number of CPU cycles per byte at each RRS $b$ is set to $N^{v_h,v_l}_{\text{Cycle}}=5900$ \cite{DynamicComputationOffloading,Energyefficiencyofmobileclients}.
For the storage capacities, we set $C^\text{max}_1=0.2\sum_{v_l \in \mathcal{V}} s_{v_l}$ and $C^\text{max}_b=0.1\sum_{v_l \in \mathcal{V}} s_{v_l}, \forall b \in \mathcal{B}/\{1\}$.
For the fronthaul and backhaul capacities, we set $R^\text{max}_{b',b}=40$ Mbps and $R^\text{max}_{0,b}=80$ Mbps, respectively.
The simulation settings are summarized in Table \ref{Main simulation parameters}.
\begin{table*}
\centering
\caption{Simulation settings.}
\begin{center} \label{Main simulation parameters}
\scalebox{0.63}{\begin{tabular}{|c|c|c|c|c|c|}
    \hline \rowcolor[gray]{0.70}
    \hline \rowcolor[gray]{0.70}
    \hline \rowcolor[gray]{0.70}
    \hline \rowcolor[gray]{0.70}
    \textbf{Description} & \textbf{Notation} & \textbf{Value} & \textbf{Description} & \textbf{Notation} & \textbf{Value} \\
    \hline \rowcolor[gray]{0.82}
    \hline \rowcolor[gray]{0.82}
    \hline \rowcolor[gray]{0.82}
    \hline \rowcolor[gray]{0.82}
    \multicolumn{3}{|c|}{\textbf{System Parameters}} & \multicolumn{3}{|c|}{\textbf{Virtualization and Pricing Parameters}} \\
    \hline \rowcolor[gray]{0.940}
    Carrier frequency & $\times$ & $2.5$ GHz & Number of slices & $M$ & $2$ \\
    \hline \rowcolor[gray]{0.945}
    Wireless channel bandwidth & $W$ & $5$ MHz & Number of users of each slice & $\times$ & 15 \\
    \hline \rowcolor[gray]{0.950}
    Number of subcarriers & $N$ & $64$ & User selection distribution & $\times$ & Uniform \\
    \hline \rowcolor[gray]{0.955}
    Subcarrier bandwidth & $W_\text{s}$ & $78.125$ KHz & Unit price of transmit power & $\alpha^\text{Pow}_b$ & HP-RRS: $6$ units/Watt, LP-RRS: $4$ units/Watt \\
    \hline \rowcolor[gray]{0.960}
    Path loss model & $\times$ & $128.1 + 37.6 \log_{10} d_{b,u} + z_{b,u}$ (dB), $d_{b,u}$ (Km) & Unit price of subcarrier bandwidth & $\alpha^\text{Sub}_b$ & HP-RRS: $60$ units/MHz, LP-RRS: $40$ units/MHz \\
    \hline \rowcolor[gray]{0.965}
    Shadowing standard deviation & $z_{b,u}$ & $8$ dB & Unit price of storage resources & $\mu^\text{Cache}_b$ & HP-RRS: $2$ units/Gbit, LP-RRS: $1.6$ units/Gbit \\
    \hline \rowcolor[gray]{0.970}
    Small-scale fading model & $\times$ & Rayleigh fading with variance $1$ & Unit price of processing resources & $\mu^\text{Proc}_b$ & HP-RRS: $0.8$ units/GHz, LP-RRS: $0.7$ units/GHz \\
    \hline \rowcolor[gray]{0.975}
    PSD of AWGN & $N_0$ & $-174$ dBm/Hz & Unit price of fronthaul capacity & $\alpha^\text{FH}$ & $2$ units/Mbps \\
    \hline \rowcolor[gray]{0.980}
    Noise figure at users & $\times$ & $9$ dB & Unit price of backhaul capacity & $\alpha^\text{BH}$ & $5$ units/Mbps \\
    \hline \rowcolor[gray]{0.82}
    \multicolumn{3}{|c|}{\textbf{Video Parameters}} & \cellcolor[gray]{0.990} Unit reward of slices & \cellcolor[gray]{0.990} $\psi_{m}$ & \cellcolor[gray]{0.990} $\psi_{1}=8.75$ units/Mbps and $\psi_{2}=9$ units/Mbps \\
    \hline \rowcolor[gray]{0.82}
    \cellcolor[gray]{0.940} Number of unique videos & \cellcolor[gray]{0.940} $V$ & \cellcolor[gray]{0.940} $10$ & \multicolumn{3}{|c|}{\textbf{Threshold Parameters}} \\
    \hline \rowcolor[gray]{0.945}
    Number of bitrate variants & $L$ & $4$ & \cellcolor[gray]{0.940} Transmit power of RRSs & \cellcolor[gray]{0.940} $P^\text{max}_b$ & \cellcolor[gray]{0.950} HP-RRS: $47$ dBm, LP-RRS: $27$ dBm \\
    \hline \rowcolor[gray]{0.955}
    Original video bitrate & $\times$ & $2$ Mbps (HD quality) & \cellcolor[gray]{0.945} Processing capacity of RRSs & \cellcolor[gray]{0.945} $\chi^\text{max}_b$ & \cellcolor[gray]{0.945} HP-RRS: $50$ GHz, LP-RRS: $25$ GHz \\
    \hline \rowcolor[gray]{0.965}
    Relative bitrates & $\times$ & $\left[0.45,0.55,0.67,0.82\right]$ & \cellcolor[gray]{0.950}  Storage capacity of RRSs & \cellcolor[gray]{0.950} $C^\text{max}_b$ & \cellcolor[gray]{0.950} HP-RRS: $20\%$, LP-RRS: $10\%$ \\
    \hline \rowcolor[gray]{0.975}
    Video length & $\times$ & $10$ minutes & \cellcolor[gray]{0.960} Maximum fronthaul capacity & \cellcolor[gray]{0.960} $R^\text{max}_{b',b}$ & \cellcolor[gray]{0.960} $40$ Mbps \\
    \hline \rowcolor[gray]{0.985}
    Zipf parameter & $\lambda$ & $0.8$ & \cellcolor[gray]{0.973} Maximum backhaul capacity & \cellcolor[gray]{0.973} $R^\text{max}_{0,b}$ & \cellcolor[gray]{0.973} $80$ Mbps \\
     \hline \rowcolor[gray]{0.990}
    Task workload & $N^{v_h,v_l}_{\text{Cycle}}$ & $5900$ & \cellcolor[gray]{0.985} Maximum number of NOMA users & \cellcolor[gray]{0.985} $\Psi_b$ & \cellcolor[gray]{0.985} $2$ \\
    \hline \rowcolor[gray]{0.999}
    Size of transcoding task & $\eta^{v_h,v_l}$ & $s_{v_l}$ & \cellcolor[gray]{0.999} Minimum data rate of each slice & \cellcolor[gray]{0.999} $R^\text{min}_{m}$ & \cellcolor[gray]{0.999} $R^\text{min}_{1}=1$ Mbps, $R^\text{min}_{2}=2$ Mbps \\
    \hline
\end{tabular}}
\end{center}
\end{table*}

To investigate the benefits of each technology in the system, we compare the CVCT system according to the following schemes:
1) No Caching (NC), where the storage and processing capacity of RRSs are equal to zero \cite{JointOptimizationofCloud,GreenOFDMAresourcecache};
2) Non-Cooperative (NoCoop), where there is no cooperation between RRSs;
3) Cooperative Caching with No Transcoding (CCNT), where only the cooperative caching technology is considered for RRSs and the processing capacity of all RRSs are equal to zero;
4) The OMA technology (OMA), where the OFDMA technology is used for the downlink transmission of wireless access channels.

We also compare the performance of our proposed DACPSs in each scheme with two conventional baseline popular/bitrate CPSs: 1) Most Popular Video (MPV), where each RRS caches the most popular videos until its storage is full \cite{JointOptimizationofCloud,GreenOFDMAresourcecache}; 2) High-Bitrate Video (HBV), where each RRS caches the high-bitrate variants of video files randomly until its storage is full.

As noted previously, obtaining the globally optimal solution of each cache placement and delivery optimization problem via the exhaustive search method would take an unrealistically long time for $U=30$, $N=64$, $B=5$, and $VL=40$ \cite{JointOptimizationofCloud}. Thus, we limit our simulations to only evaluate the performance of our proposed solution algorithms.
We also investigate the delivery performance gain achieved by each of our proposed RAF in terms of total delivery revenue of slices and computational complexity of the delivery algorithm.

\subsection{Convergence of the Delivery Algorithm}
\begin{figure}
\centering
\includegraphics[scale=0.48]{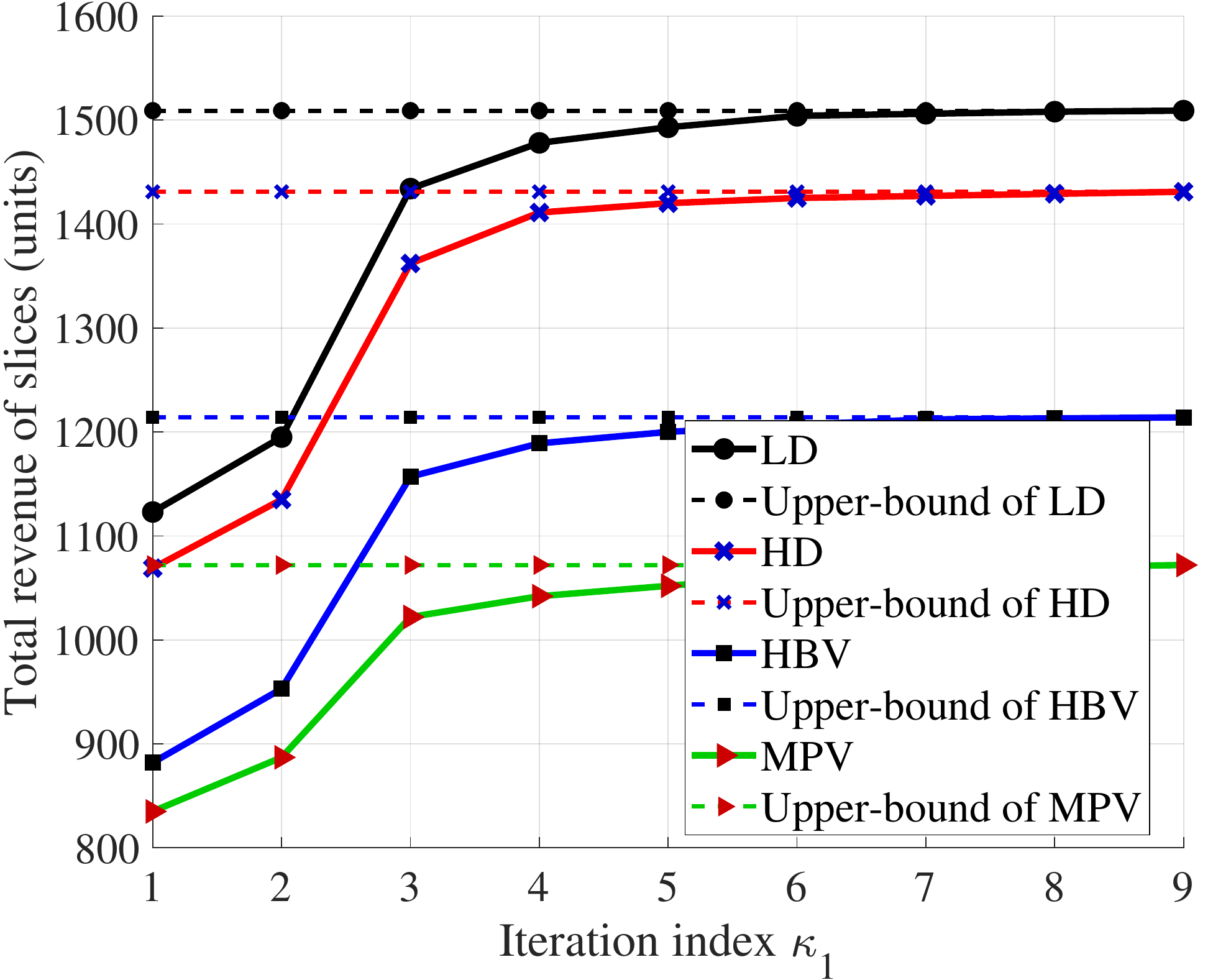}
\caption{The convergence in terms of total revenue of slices over the number of main iterations.}
\label{Fig04}
\end{figure}

Fig. \ref{Fig04} demonstrates the convergence of the proposed delivery algorithm for different CPSs.
As shown, for all heuristic and the LD and HD strategies, the proposed delivery algorithm converge to stable values in maximum $6$ iterations. In this figure, the dash lines refer to the upper-bound solutions when the algorithm converges. From Fig. \ref{Fig04}, it can bee seen that after $3$ iterations, the proposed approach will achieve up to 95\% of its upper-bound value for different CPSs which ensures us the proposed algorithm can be applied in practical scenarios in multiuser HV-MECs.

\subsection{Impact of the Storage, Processing, and Fronthaul Capacities}
\subsubsection{Impact of the storage capacity of RRSs}
\begin{figure}
\centering
\subfigure[Total provisioning cost of slices vs. storage capacity of LP-RRSs.]{
   \includegraphics[scale=0.41]{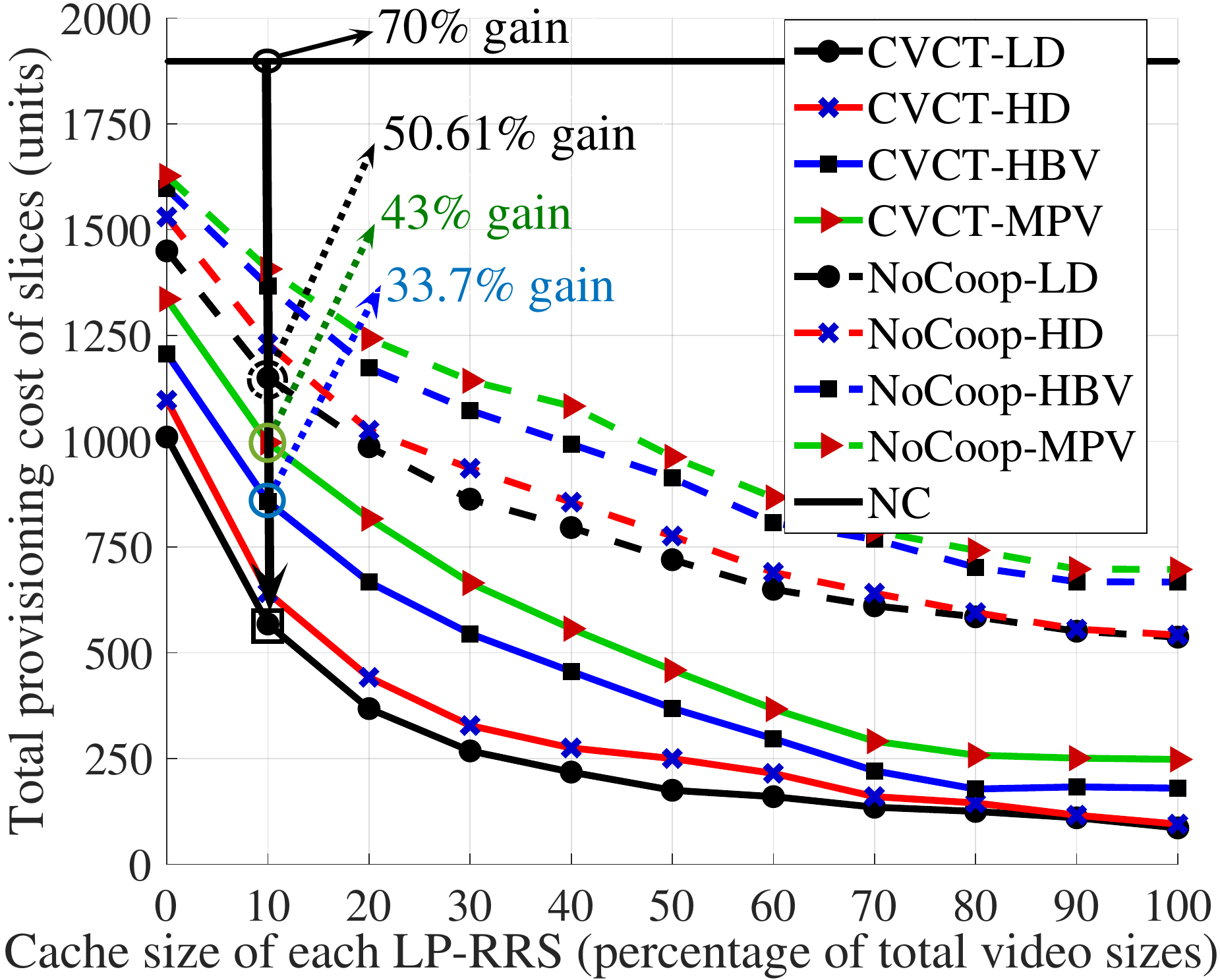}
   \label{Fig05BSCost}
}
\subfigure[Total revenue of slices vs. storage capacity of LP-RRSs.]{
   \includegraphics[scale=0.41]{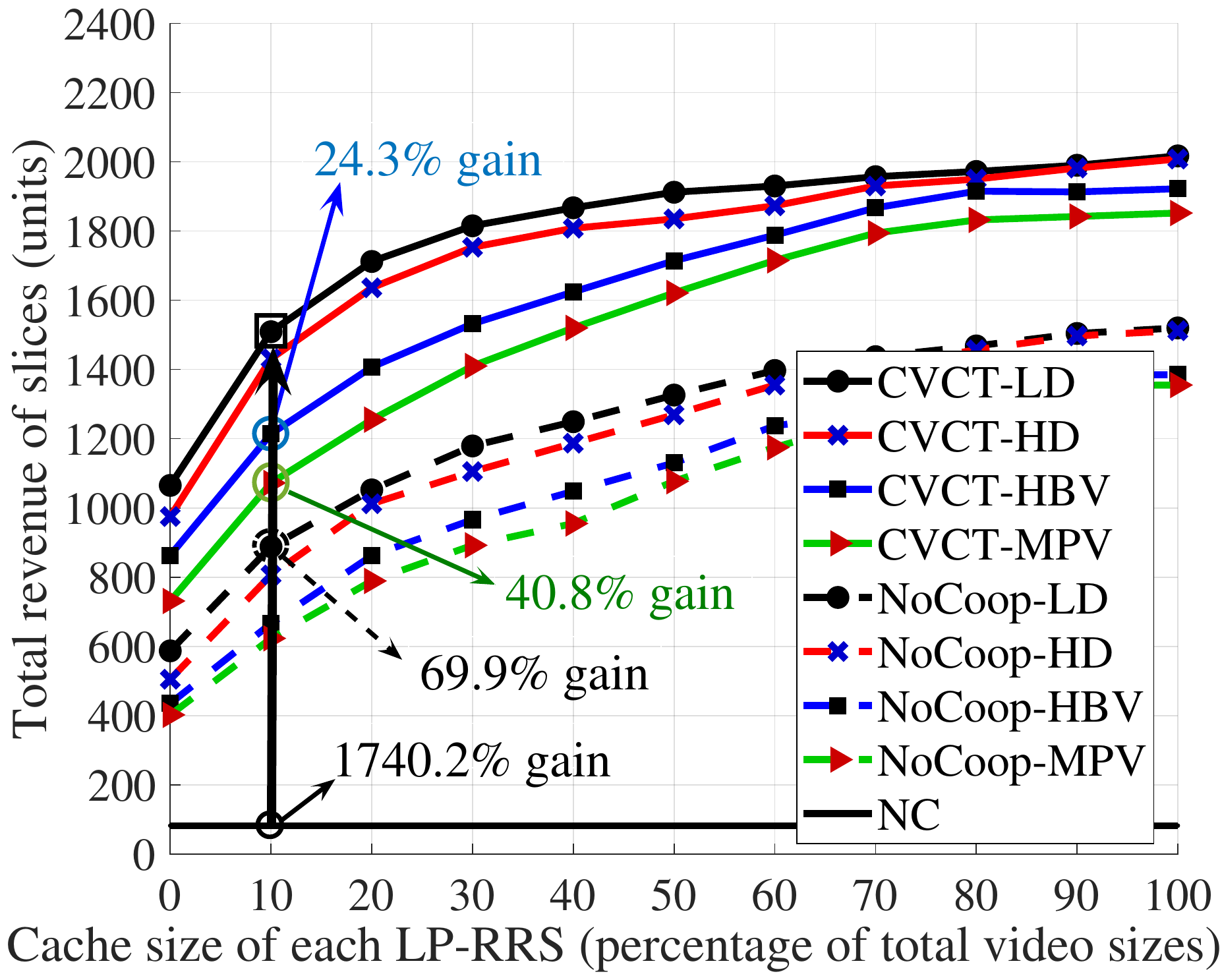}
   \label{Fig05Rev}
}
\caption{Impact of storage capacity of LP-RRSs on the performance of CPSs in different schemes.}
\label{Fig05}
\end{figure}

In Fig. \ref{Fig05}, we investigate the impact of storage capacity limitation at LP-RRSs on the performance of CPSs in cooperative and non-cooperative schemes.
As expected, more storage capacities lead to more stored videos, which increases the cache hit ratio and subsequently provides more transcoding and cooperative communication opportunities. Therefore, backhaul resource usage is significantly reduced, which decreases the total provisioning cost of slices (see Fig. \ref{Fig05BSCost}) and correspondingly improves the total revenue shown in Fig. \ref{Fig05Rev}.

From Fig. \ref{Fig05Rev}, it is observed that the NC scheme provides the lower-bound of total revenue of slices. In addition, the CVCT system with the LD strategy reduces the total provisioning cost of slices by nearly $70\%$ compared to the NC scheme, which correspondingly improves the total revenue of slices around 17-fold. On the other hand, the cooperation between RRSs improves the total revenue of slices around $69.9\%$ compared to the NoCoop scheme.

In these schemes when the storage capacities are low, the HBV and MPV strategies have lower performances than those of our proposed LD and HD strategies, since they do not consider the flexible delivery opportunities.
For instance, when the cache size percentage is $10\%$, the LD strategy outperforms the performance of the CVCT system by nearly $24.3\%$ and $40.8\%$ compared to HBV and MPV, respectively.

\subsubsection{Impact of the processing capacity of RRSs}
\begin{figure}
\centering
\subfigure[Total provisioning cost of slices vs. processing capacity of LP-RRSs.]{
   \includegraphics[scale=0.41]{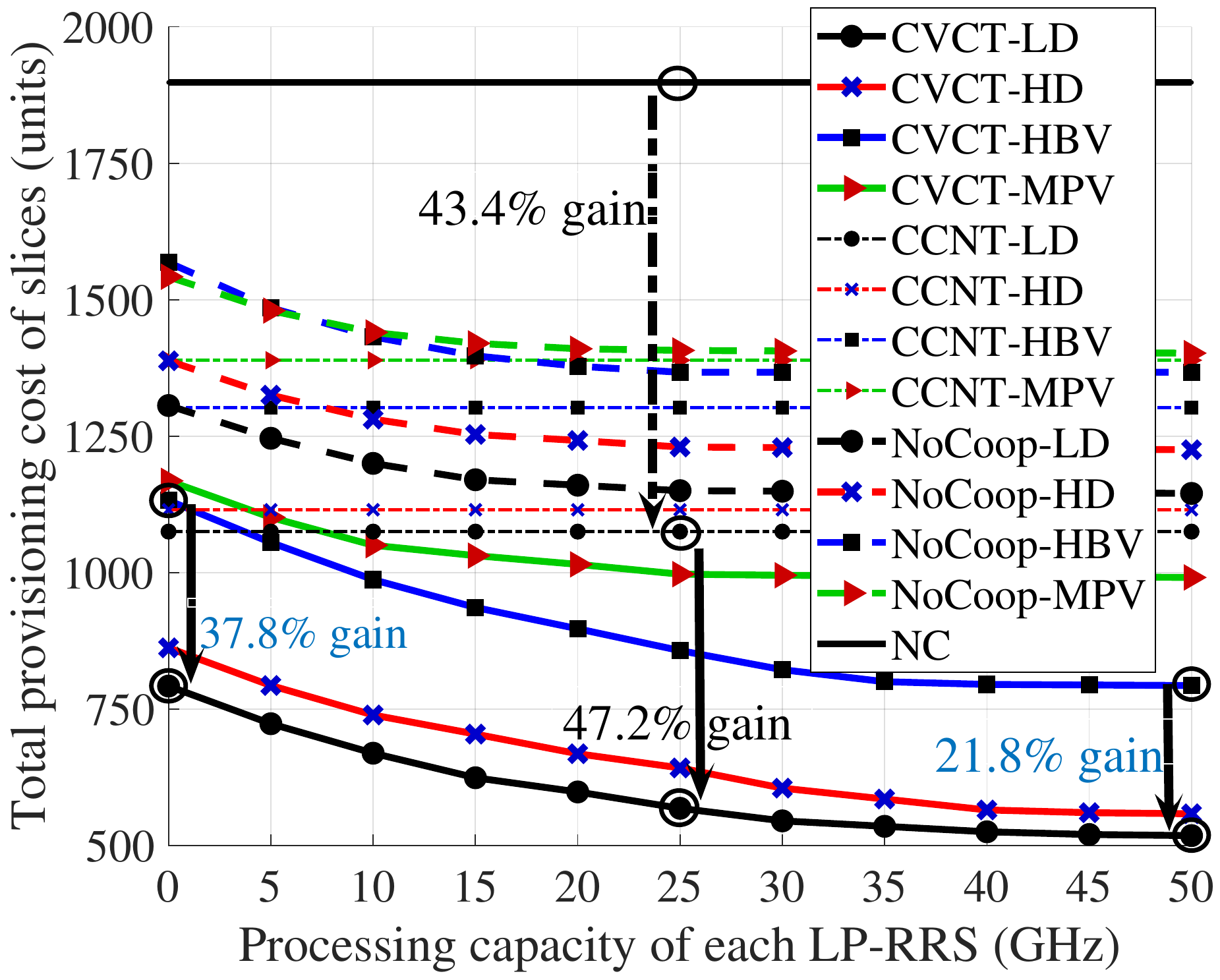}
   \label{Fig06BSCost}
}
\subfigure[Total revenue of slices vs. processing capacity of LP-RRSs.]{
   \includegraphics[scale=0.41]{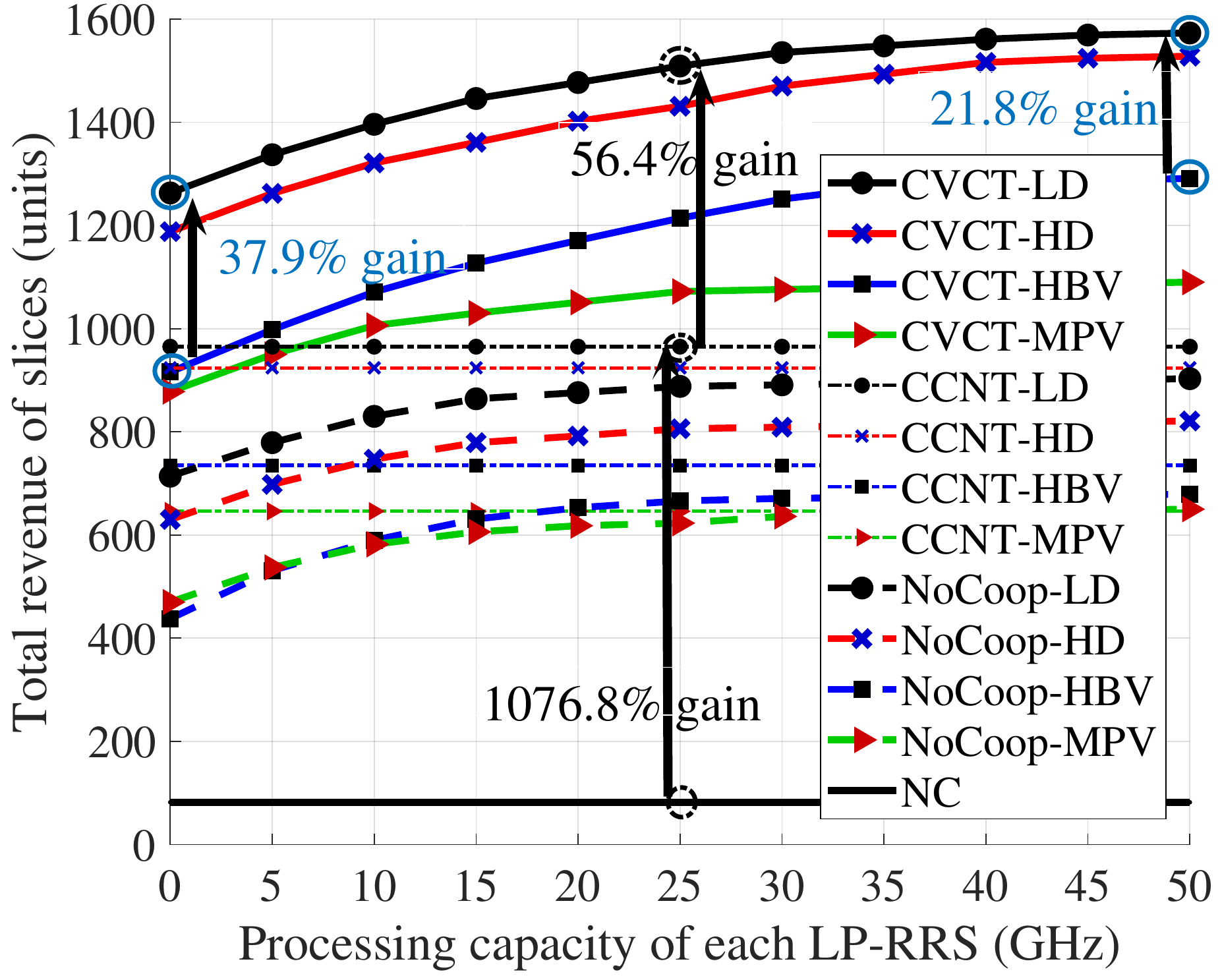}
   \label{Fig06Rev}
}
\caption
{Impact of processing capacity of LP-RRSs on the performance of CPSs in different schemes.}
\label{Fig06}
\end{figure}

Fig. \ref{Fig06} shows the impact of processing capacity limitation at LP-RRSs on the performance of CPSs in different schemes.
Obviously, larger processing capacities provide more video transcoding opportunities, which alleviate the backhaul resource usage. It is noted that two transcoding types exist in the system as self-transcoding and cooperative transcoding. The self-transcoding in the parallel transmission and transcoding system mainly depends on the wireless channel, storage, and processing capacities. In other words, for larger processing capacities, increasing them alone cannot significantly improve the self-transcoding opportunities, since higher bitrate variants should be stored and the wireless channel capacities are limited. Besides, the cooperative transcoding mainly depends on the wireless channel, storage, processing and fronthaul link capacities. In other words, the cooperative transcoding operations cannot be successfully performed if the fronthaul capacities between RRSs are insufficient. Accordingly, based on the available storage and fronthaul capacities, it can be seen that the performance of all CPSs have slow changes when the processing capacity of LP-RRSs exceed $35$ GHz and $15$ GHz for the CVCT and NoCoop schemes, respectively. These results are shown in Fig. \ref{Fig06}.

In Fig. \ref{Fig06Rev}, it is shown that the LD strategy in the CCNT scheme performs 10 times better than NC.
Moreover, the cooperative transcoding technology improves the systems performance closed to $56.4\%$ alone in the cooperative schemes when the processing capacity of LP-RRSs is $25$ GHz.

As shown in Fig. \ref{Fig06}, the performance of HBV is more affected by the amount of the processing capacity, since HBV randomly selects the highest bitrate variants in order to increase the self and (empirically) cooperative transcoding opportunities.
Interestingly, when the relative processing capacities increase, the performance gain between the HBV and LD strategies decreases from $37.9\%$ to $21.8\%$, as shown in Fig. \ref{Fig06Rev}. Therefore, the HBV strategy can be a good candidate when processing capacities are large enough.
Besides, the performance of MPV changes slowly according to variations in processing capacity, since it does not consider the transcoding of video files, specifically in the event that low bitrate variants turn out to be more popular. MPV also stores the same popular videos, which significantly degrades the cooperative transcoding opportunity in the system.

\subsubsection{Impact of the fronthaul capacity of RRSs}
\begin{figure*}
\centering
\subfigure[Total provisioning cost of slices vs. fronthaul capacity between RRSs.]{
   \includegraphics[scale=0.41]{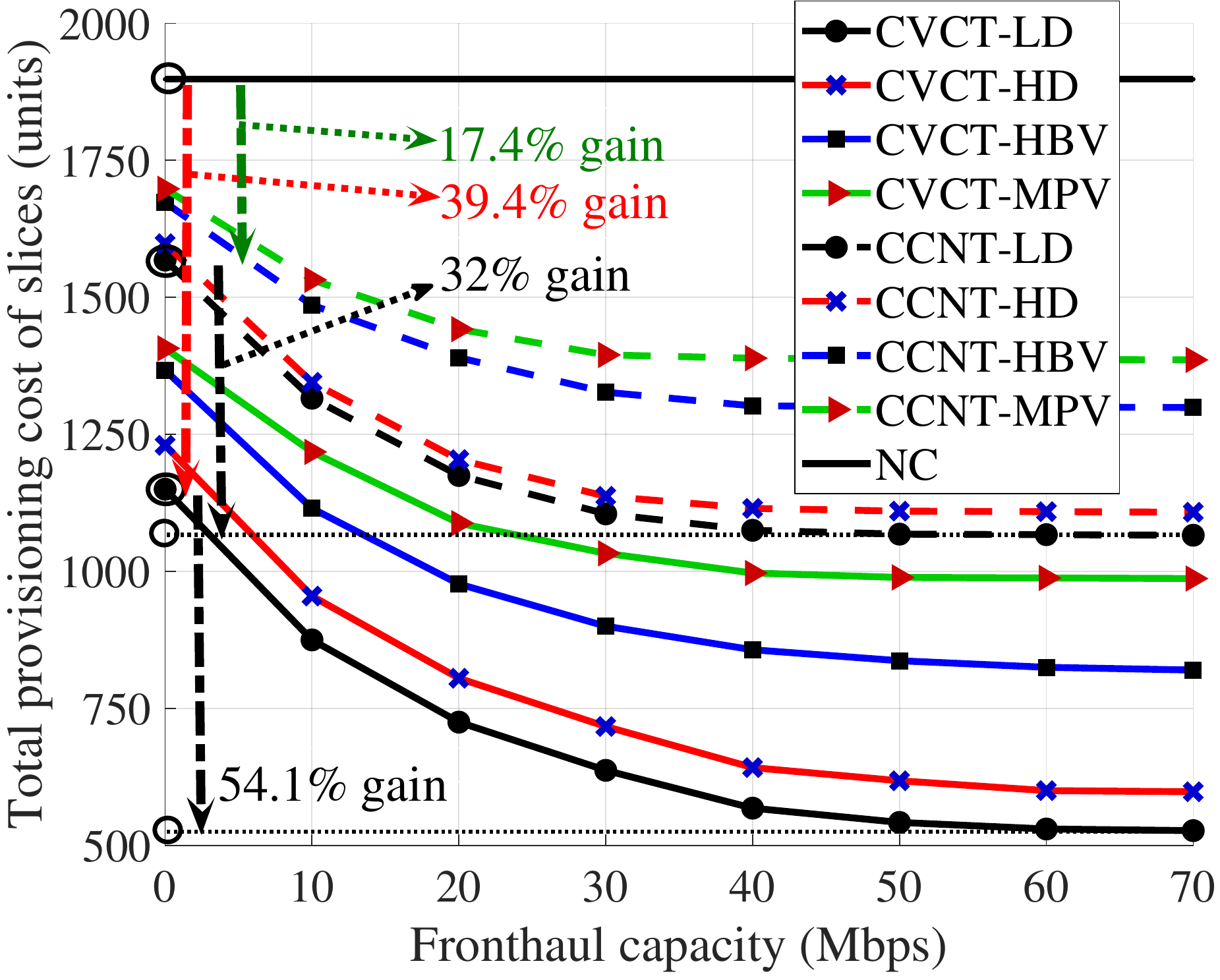}
   \label{Fig07BSCost}
}
\subfigure[Total revenue of slices vs. fronthaul capacity between RRSs.]{
   \includegraphics[scale=0.41]{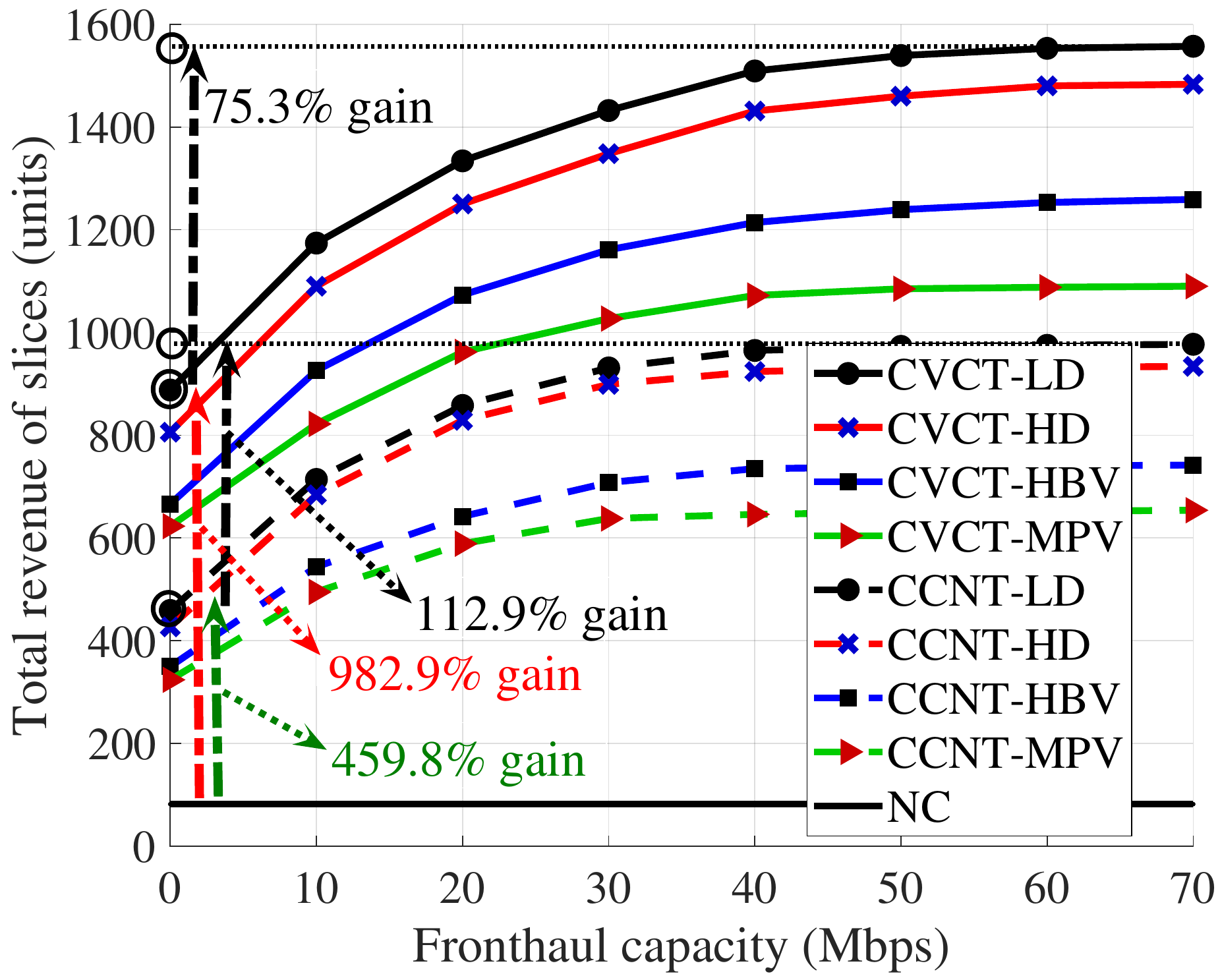}
   \label{Fig07Rev}
}
\caption
{Impact of fronthaul capacities between RRSs on performance of CPSs in different schemes.}
\label{Fig07}
\end{figure*}

Fig. \ref{Fig07} shows the impact of the fronthaul capacity limitation between RRSs on the performance of CPSs in the CVCT and CCNT schemes.
Generally, larger fronthaul capacities increase the cooperative communication capability. In this way, the systems performance will significantly be improved (see Fig. \ref{Fig07Rev}).

According to Fig. \ref{Fig07Rev}, for the LD strategy, the CVCT technology with $R^\text{max}_{b',b}=70$ Mbps outperforms the total revenue of slices by nearly $75.3\%$ compared to the NoCoop where $R^\text{max}_{b',b}=0$. In this regard, the systems performance is improved nearby $112.9\%$ in the CCNT scheme which is caused only by the cooperative caching technology. Besides, the caching capability without any cooperation causes nearby $459.8\%$ improvement in the total revenue of slices. In addition, the joint caching and transcoding capability at RRSs without any cooperation improves the system performance by nearly 10-fold compared to the NC scheme. From this result, it can be concluded that the self-transcoding capability in the system improves the total revenue of slices up to 5-fold alone.

\subsection{Effect of Zipf Parameter}
\begin{figure}
\centering
\subfigure[Total number of unique requests vs. Zipf parameter for different video categorizes.]{
   \includegraphics[scale=0.41]{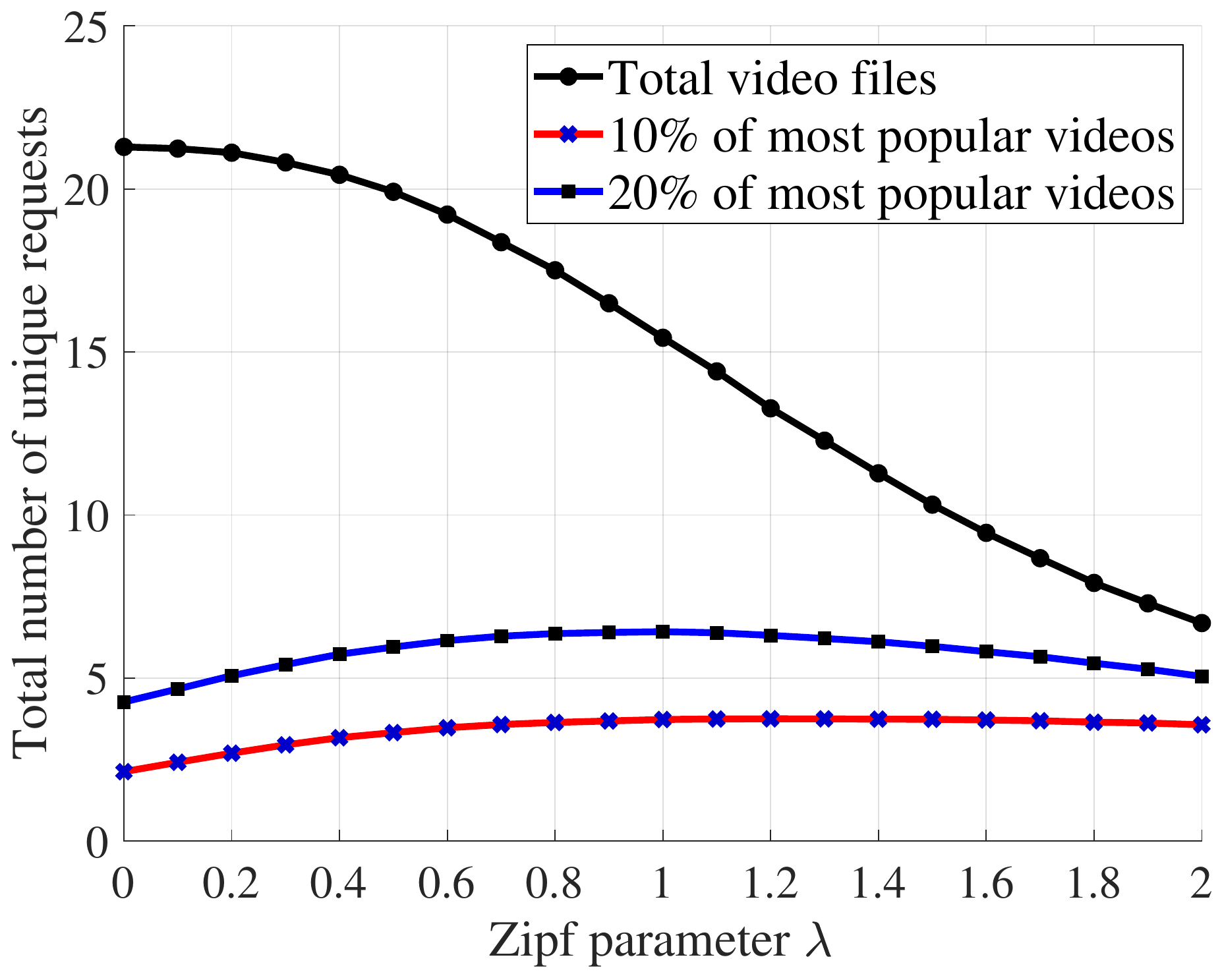}
   \label{Fig08rquestUnique}
}
\subfigure[Request percentage of most ranked videos vs. Zipf parameter for different video categorizes.]{
   \includegraphics[scale=0.41]{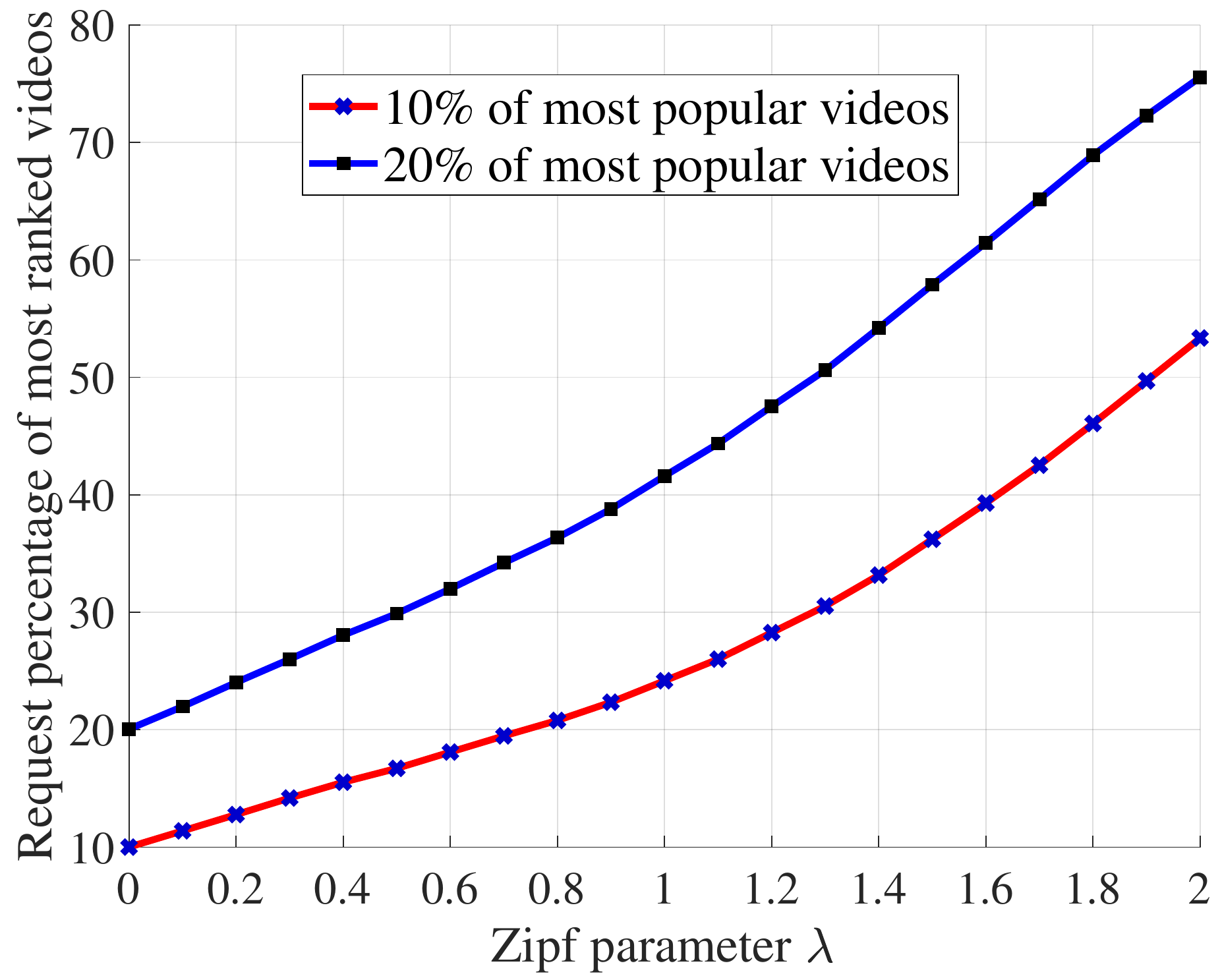}
   \label{Fig08requPercentage}
}
\subfigure[Total provisioning cost of slices vs. Zipf parameter.]{
   \includegraphics[scale=0.41]{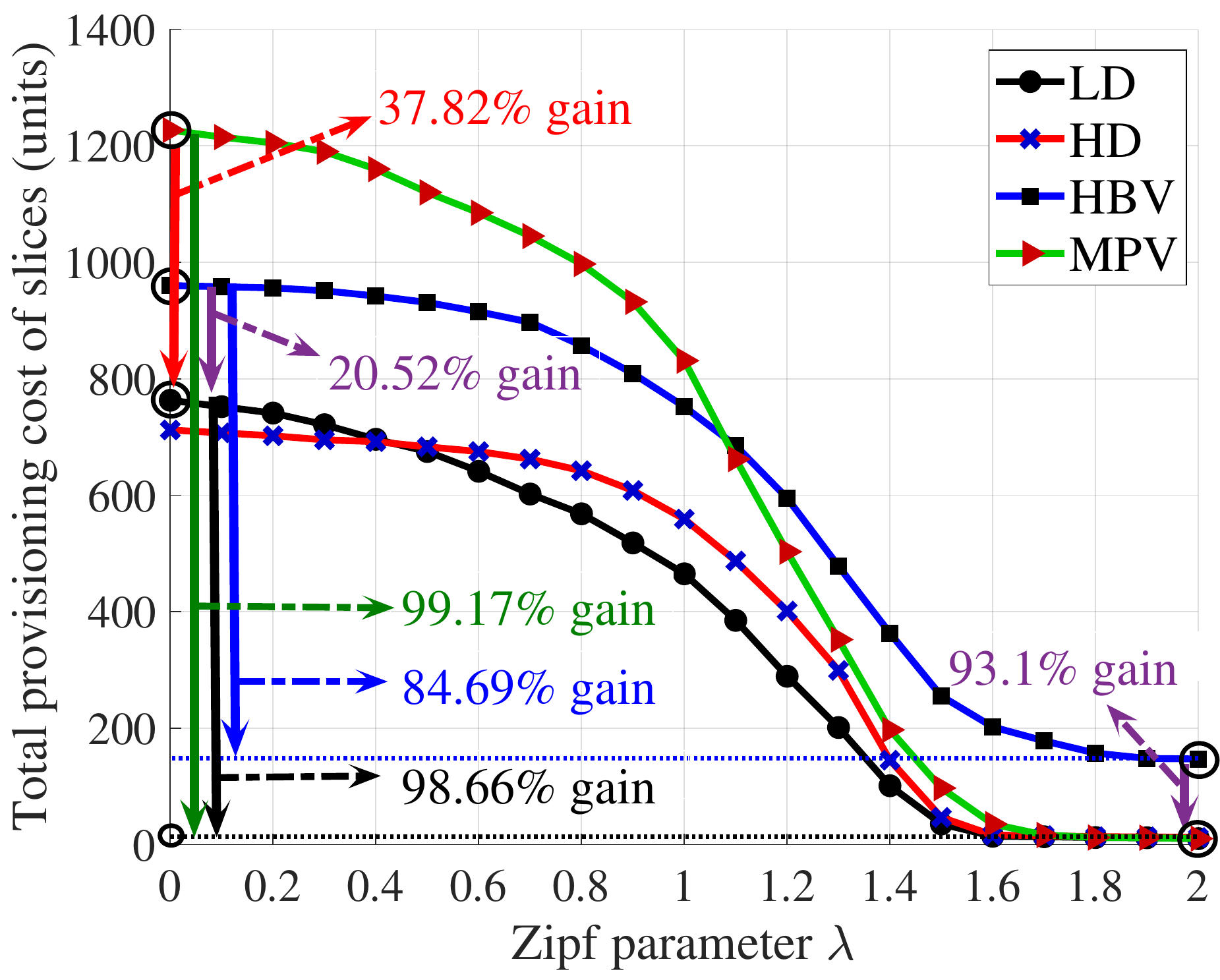}
   \label{Fig08BSCost}
}
\subfigure[Total revenue of slices vs. Zipf parameter.]{
   \includegraphics[scale=0.41]{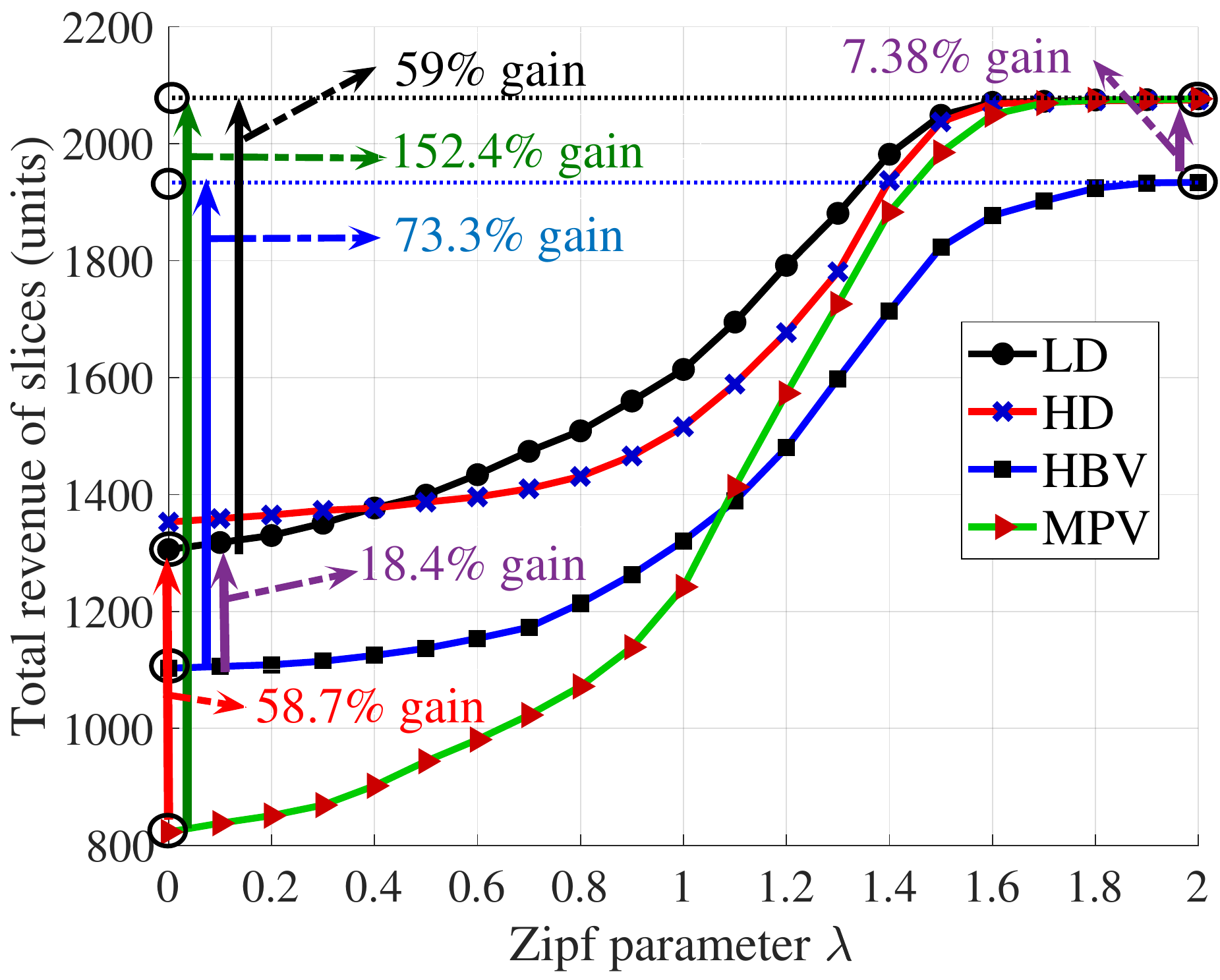}
   \label{Fig08Rev}
}
\caption
{Effect of the Zipf parameter on performance of different CPSs.}
\label{Fig08}
\end{figure}

Fig. \ref{Fig08} shows the effect of Zipf parameter $\lambda$ on the performance of different CPSs.
As shown in Fig. \ref{Fig08rquestUnique}, for $\lambda=1.5$, $U=30$, and $VL=40$, there are nearly $10$ unique requests. For the aforementioned setting, Fig. \ref{Fig08requPercentage} shows that nearly $57.9\%$ and $36.22\%$ of these requests are for $20\%$ and $10\%$ of most popular videos, respectively.
These results are averaged over $10000$ sets of requests. Accordingly, Figs. \ref{Fig08rquestUnique} and \ref{Fig08requPercentage} show that when $\lambda$ increases, the diversity of requests decreases in the system. In this line, the request percentage of the most ranked videos increases, exponentially in the system. In this regard, the backhaul, fronthaul, and processing resource usages are reduced which decrease the total provisioning cost of slices (shown in Fig. \ref{Fig08BSCost}). As a result, the total revenue of slices is improved that is shown in Fig. \ref{Fig08Rev}.

Based on Figs. \ref{Fig08BSCost} and \ref{Fig08Rev}, it can be derived that the HD strategy is more compatible than that of LD when $\lambda$ tends to zero. Besides, MPV is more affected by the Zipf parameter, since it is only a baseline popular strategy. Specifically, when $\lambda$ varies from 0 to 2, MPV outperforms by about $152.4\%$. For $\lambda=0$, i.e., when users uniformly request videos, HBV is more compatible than that of MPV. This is because, in this situation, the VPD factor is not dominant at all. In addition, the performance gaps in terms of the total revenue of slices between the LD and HBV strategies and the LD and MPV strategies are nearly $18.4\%$ and $57.8\%$, respectively. Interestingly, these results show that MPV is not compatible for high diversity situations, whereas HBV can be a good solution with its very low complexity structure.
On the other hand, when $\lambda$ is large enough, i.e., when only a few videos are frequently requested by users, the performance gaps between our proposed LD and HD strategies and MPV decrease significantly in the system. Actually, MPV is close to an almost optimal strategy when $\lambda$ is too large. It is noted that there still exists a performance gap (nearly 7.38\% in terms of total revenue of slices) between HBV and other strategies when $\lambda$ is too large, since HBV does not consider the VPD.

\subsection{Comparison Between MC-NOMA and OMA}
\begin{figure*}
\centering
\subfigure[Average data rate of each user vs. total number of subcarriers.]{
   \includegraphics[scale=0.34]{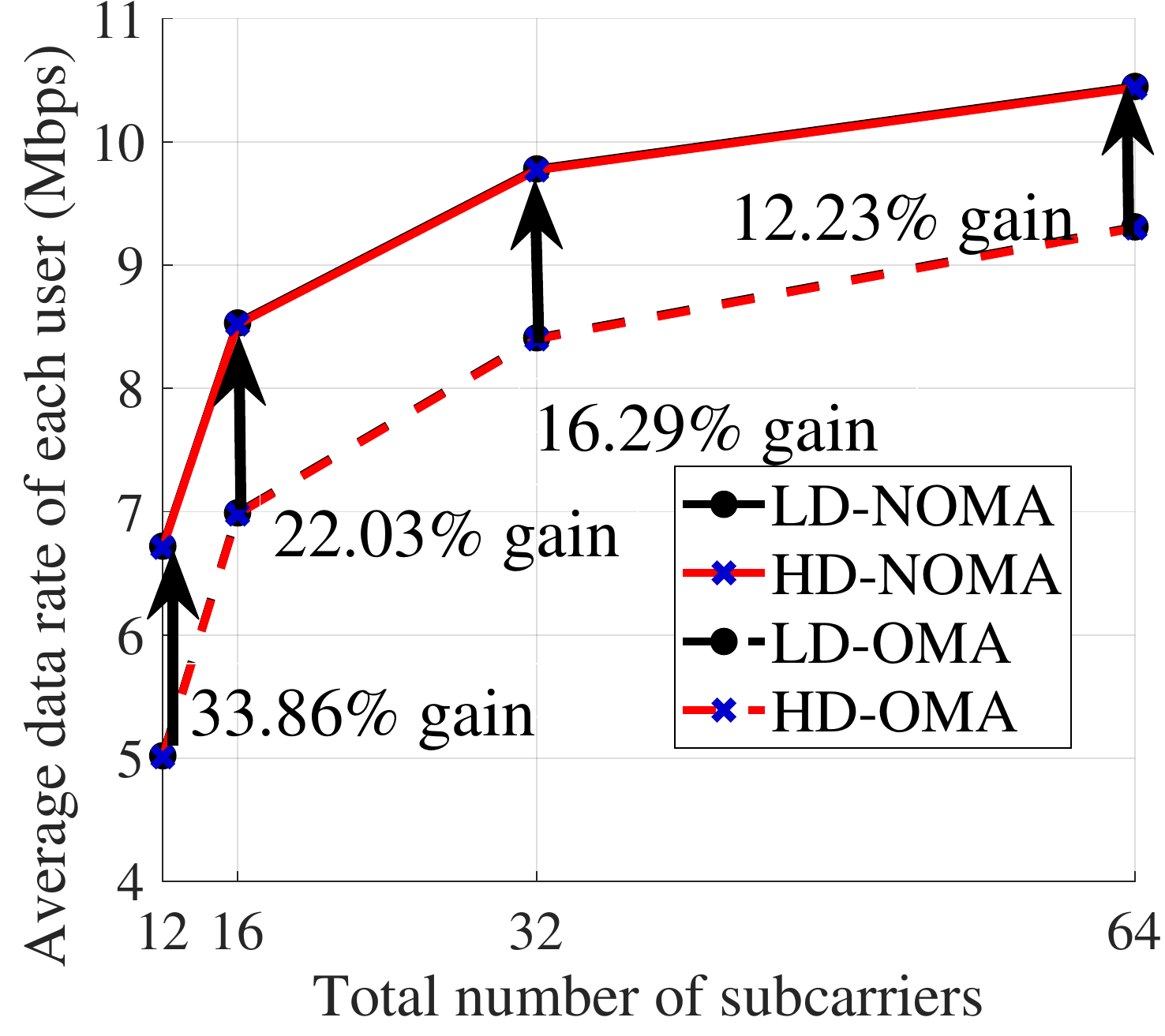}
   \label{Fig09rate}
}
\subfigure[Total reward of slices vs. total number of subcarriers.]{
   \includegraphics[scale=0.34]{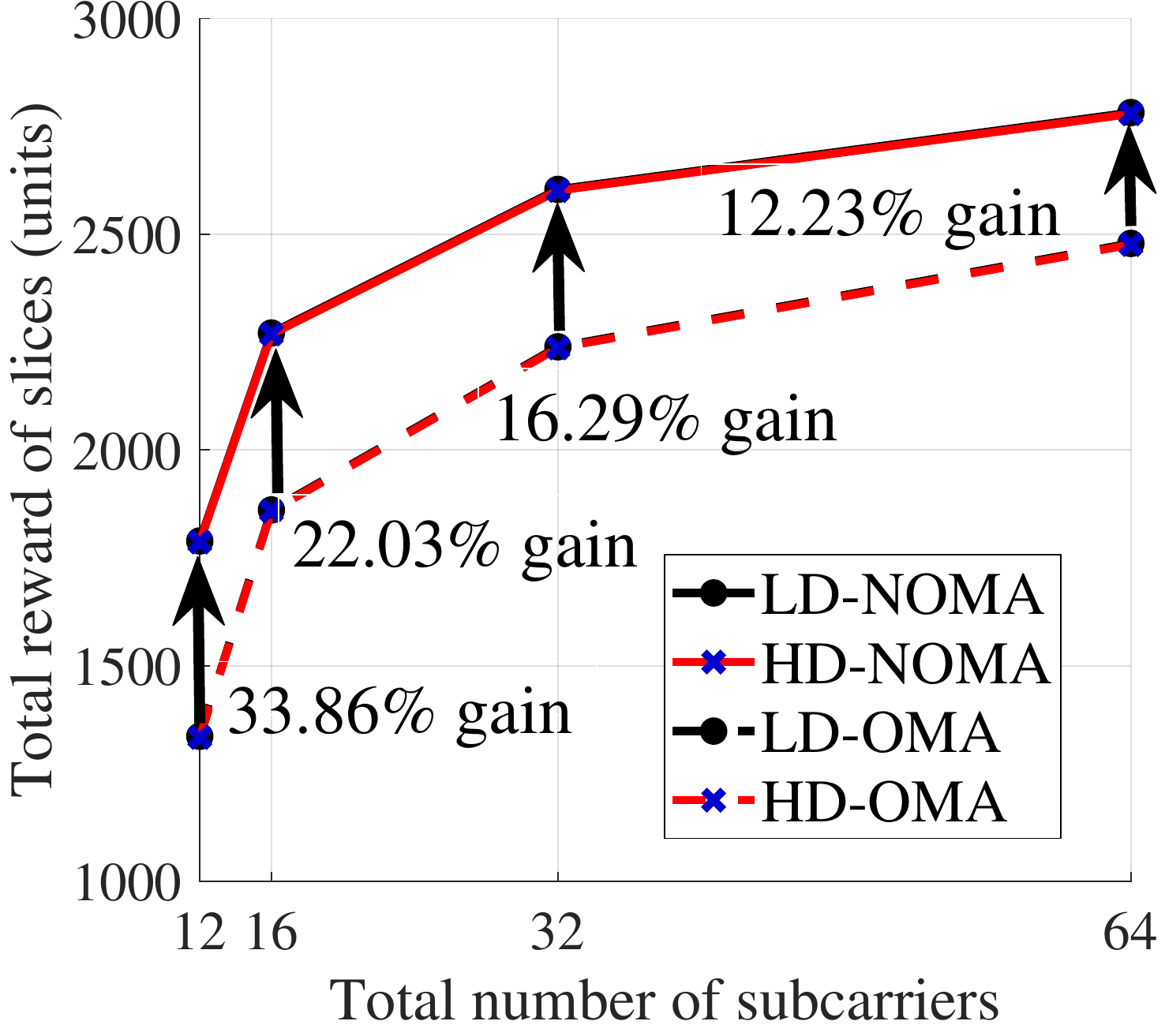}
   \label{Fig09reward}
}
\subfigure[Total provisioning cost of slices vs. total number of subcarriers.]{
   \includegraphics[scale=0.34]{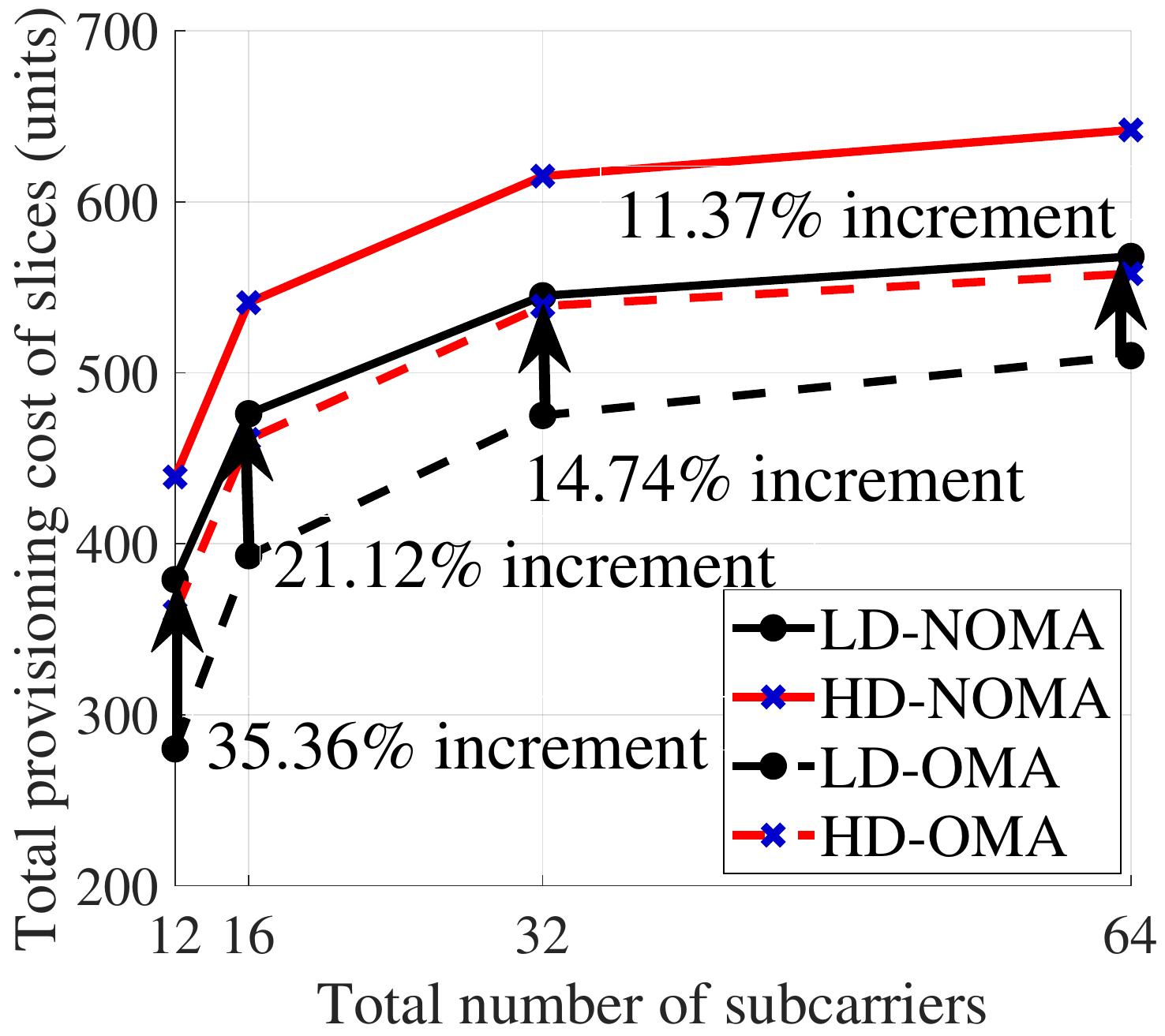}
   \label{Fig09BSCost}
}
\subfigure[Total bandwidth cost of slices vs. total number of subcarriers.]{
   \includegraphics[scale=0.34]{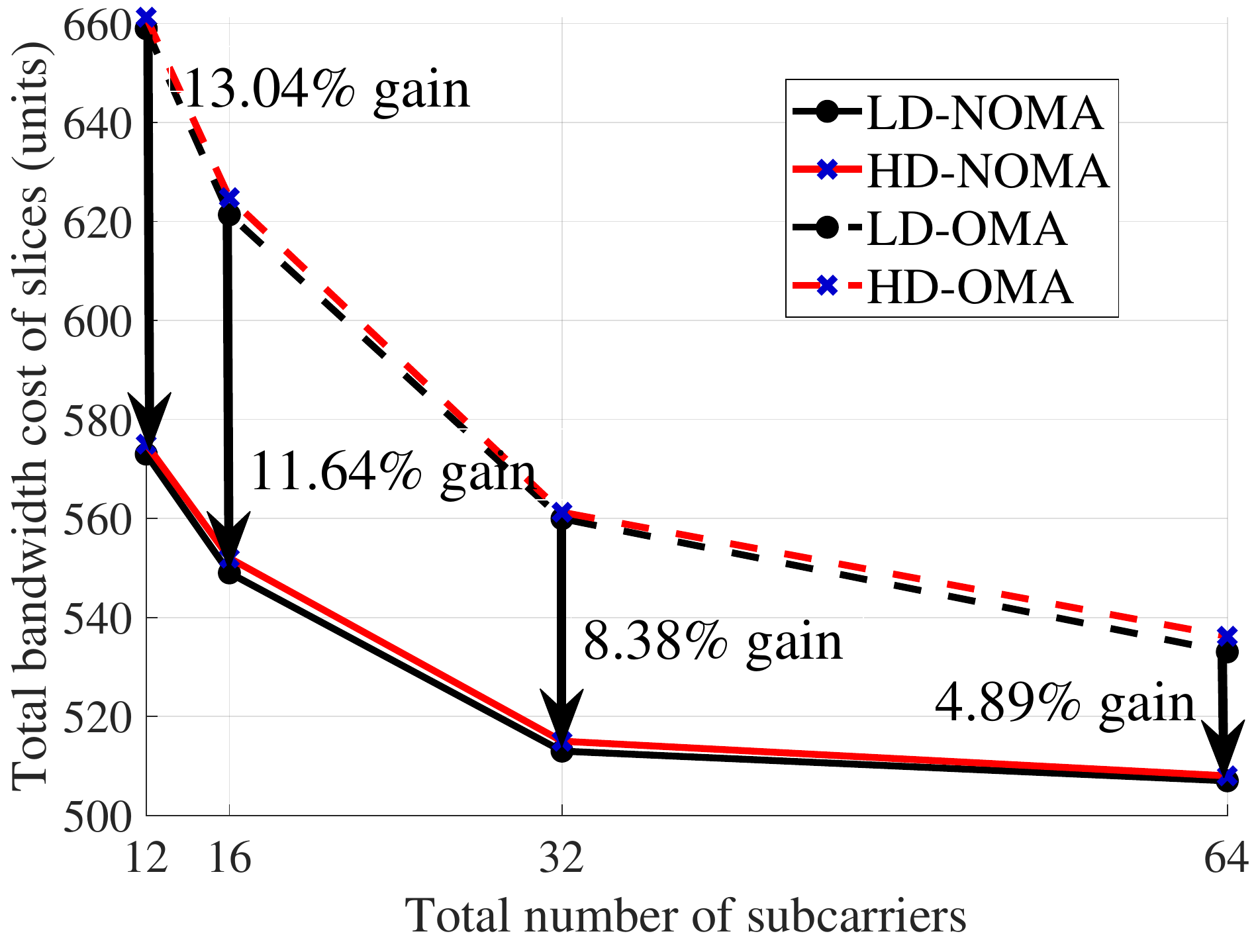}
   \label{Fig09subcarrierCost}
}
\subfigure[Total revenue of slices vs. total number of subcarriers.]{
   \includegraphics[scale=0.34]{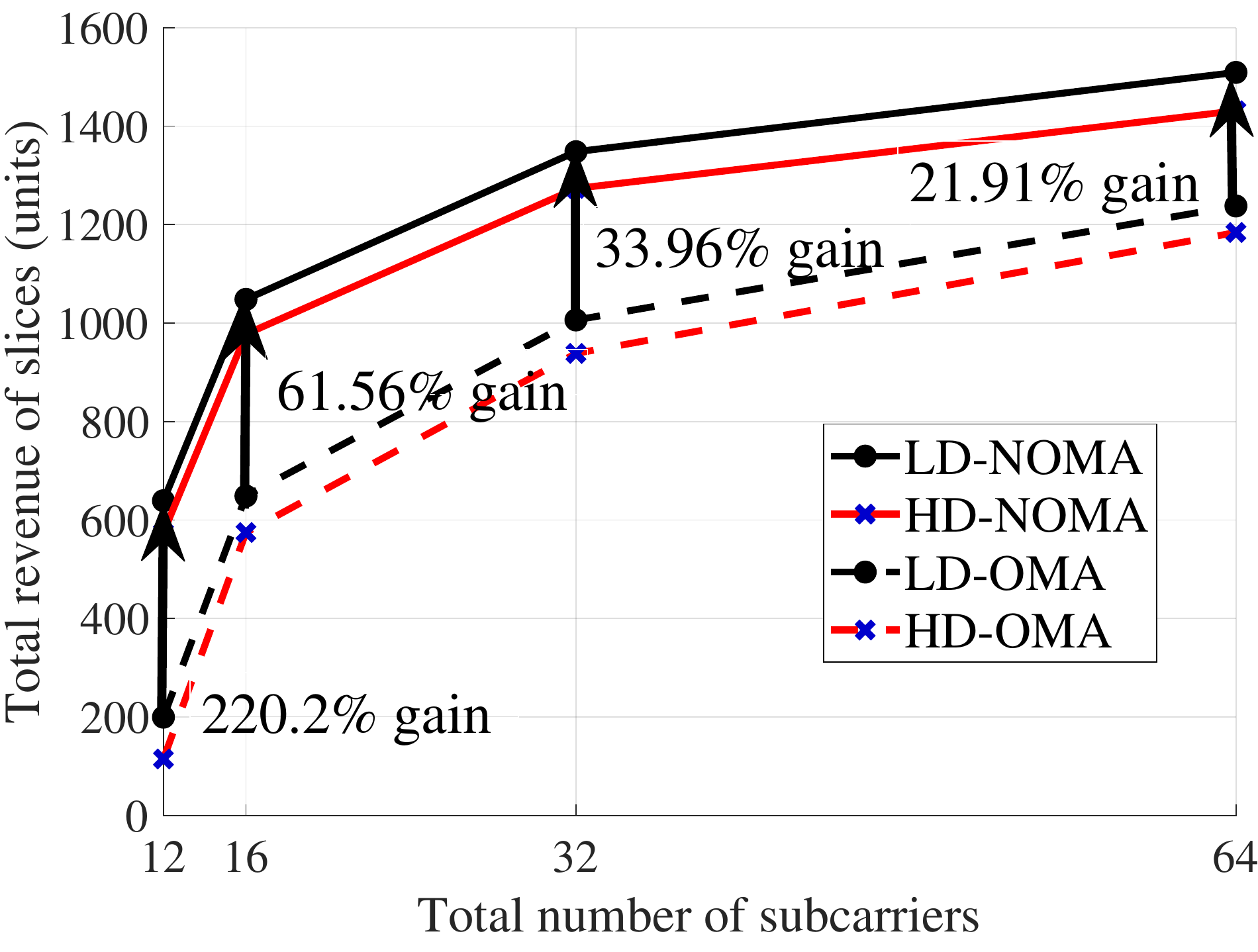}
   \label{Fig09revenue}
}
\caption
{Effect of the total number of subcarriers on performance of the LD and HD strategies in the MC-NOMA and OMA schemes.}
\label{Fig09}
\end{figure*}

Fig. \ref{Fig09} compares the performance of MC-NOMA and OMA in the LD and HD strategies for different values of total number of subcarriers $N$.
Generally, increasing $N$ in multicarrier wireless networks improves the spectral efficiency of users. In this regard, the average data rate of users increases, as shown in Fig. \ref{Fig09rate}. This improvement outperforms the total reward of slices (see Fig. \ref{Fig09reward}). Paradoxically, increasing the data rate of users needs more backhaul, fronthaul, and processing resources due to the parallel delay constraints \eqref{constraint event 2 access transc}, \eqref{constraint event 3 access front}, \eqref{constraint event 4 fronthaul access}, \eqref{constraint event 5 transcoding access}, and \eqref{constraint event 6 backhaul access}.
In this regard, improving the users access data rate leads to increasing the provisioning cost of slices which is shown in Fig. \ref{Fig09BSCost}. Interestingly, by increasing $N$, the expensive wireless bandwidth usage is significantly reduced, which corresponds to a decrease in the total bandwidth cost. (see Fig. \ref{Fig09subcarrierCost}). This is because increasing $N$ improves the flexibility of bandwidth allocation in the multicarrier systems. Hence, more opportunities are provided for slices to assign the available subcarriers to the users and satisfy the QoS requirements. The bandwidth cost reduction and slice reward increments combined have a greater effect on the revenue of slices than the increment of the provisioning costs. Hence, increasing $N$ improves the total revenue of slices as shown in Fig. \ref{Fig09revenue}.

MC-NOMA is expected to increase the spectral efficiency more than OMA by performing SIC at receivers shown in Fig. \ref{Fig09rate}. In this regard, the total reward of slices is improved in the system (see Fig. \ref{Fig09reward}). Although MC-NOMA causes more provisioning costs than OMA because of increasing access data rates in parallel systems (see Fig. \ref{Fig09BSCost}), it reduces the total bandwidth cost of slices much more than OMA. This is because each slice can use the same subcarriers for its own users in each cell, while the price of each subcarrier bandwidth is paid once in the cell by that slice (see \eqref{revenue MVNO}). This opportunity is only instantiated by MC-NOMA, since it provides the reuse of orthogonal subcarriers at each RRS. As shown in Fig. \ref{Fig09rate}, for $N=12$, the average data rate of each user in MC-NOMA is improved by nearly $33.86\%$ compared to OMA. This is because when OMA is applied to a system with critically low number of subcarriers, system performance is degraded owing to the flexibility of the bandwidth reuse at each RRS being completely eliminated. Finally, MC-NOMA outperforms OMA by nearly 220.2\% when $N = 12$; and when $N = 64$, this result is reduced to $21.91\%$.

\subsection{Comparison Between Our Proposed Resource Allocation Frameworks}
\begin{figure*}
\centering
\subfigure[Total reward/revenue of slices vs. path loss exponent $\varpi$.]{
   \includegraphics[scale=0.33]{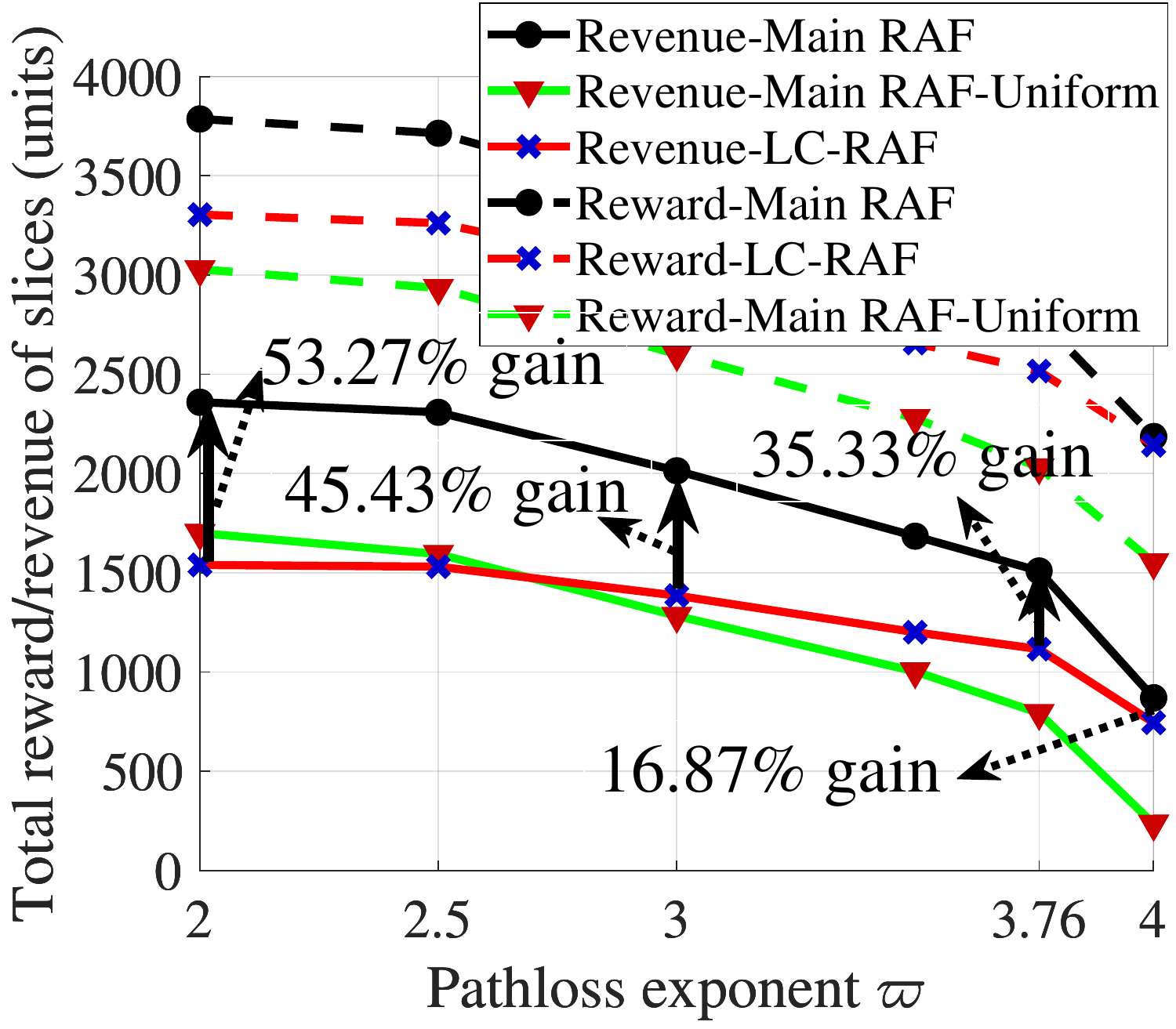}
   \label{Fig10Performance}
}
\subfigure[Delivery computational complexity order vs. total number of users. In this simulation, we set $t^0=1$, $\varrho=1$ and $\epsilon^0=\text{e}$. Moreover, we consider $3$ iterations for both Algorithm \ref{Alg iterative main} and the SCA approach.]{
   \includegraphics[scale=0.33]{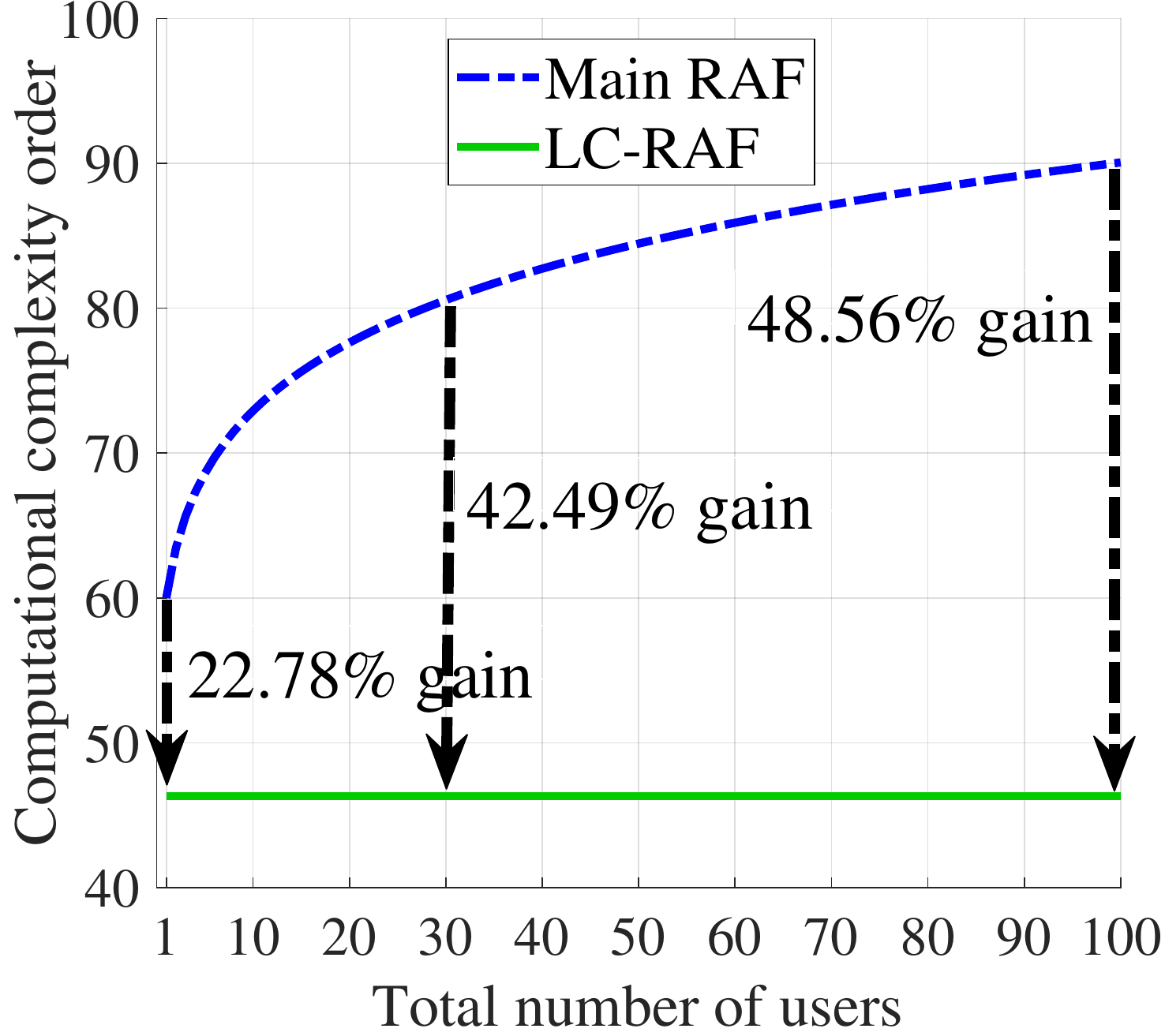}
   \label{Fig10Complexityuser}
}
\subfigure[Delivery computational complexity order vs. total number of unique videos. In this simulation, we set $t^0=1$, $\varrho=1$ and $\epsilon^0=\text{e}$. Moreover, we consider $3$ iterations for both Algorithm \ref{Alg iterative main} and the SCA approach.]{
   \includegraphics[scale=0.33]{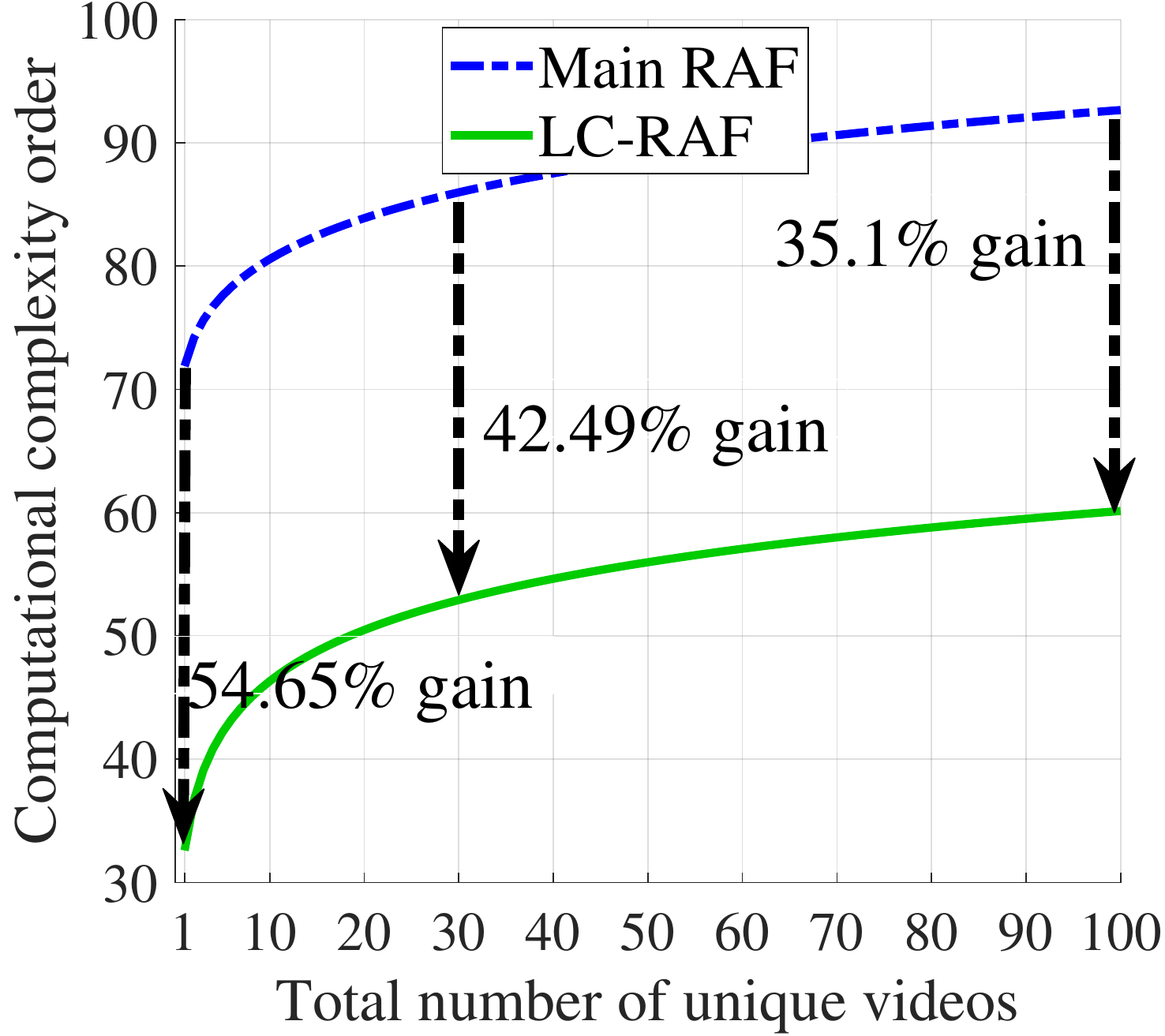}
   \label{Fig10Complexityvideo}
}

\caption
{Performance comparison of our proposed RAFs in terms of total revenue of slices and computational complexity of the delivery algorithm for the LD strategy in uniform and non-uniform deployment of users.}
\label{Fig10}
\end{figure*}

Fig. \ref{Fig10} compares the performance of our proposed LC-RAF to that of Fig. \ref{Fig00structure} in terms of total revenue of slices and computational complexity order of the delivery algorithm.
In Fig. \ref{Fig10Performance} we assume that the path loss is modeled as $128.1 + 10\varpi \log_{10} \left(d_{b,u}\right) + z_{b,u}$ in dB where the path loss exponent $\varpi$ varies from $2$ to $4$. Obviously, when $\varpi$ increases, the data rate of users decreases in the system. Accordingly, the total reward of slices degrades, which reduces the total revenue of slices. Interestingly, as seen in Fig. \ref{Fig10Performance}, the performance gap between the main and low-complexity frameworks is reduced from $53.27\%$ to $16.87\%$ when $\varpi$ varies from $2$ to $4$. This is because for larger values of $\varpi$, the flexibility of the user association process as well as the impact of the variations of small-scale fading decrease in the network. In this regard, users are more restricted; instead of choosing the RRS that has the requested video, each user is associated with the RRS that can provide the required data rate without consuming extra bandwidth and power resources. On the other hand, the CVCT system can significantly compensate for this restriction to prevent the extra backhaul resource usages. From Fig. \ref{Fig10Performance}, it can be observed that our proposed LC-RAF is a good choice for environments with larger values of $\varpi$.

As can be seen in Figs. \ref{Fig10Complexityuser} and \ref{Fig10Complexityvideo}, the computational complexity of the delivery algorithm in the LC-RAF is reduced by nearly $48.56\%$ and $35.1\%$, when the total number of users $U$ and unique videos $V$ are large enough, respectively. Interestingly, as shown in Fig. \ref{Fig10Complexityuser}, the complexity of the delivery algorithm in the LC-RAF does not depend on the number of users, which means this framework can be a good choice for dense environments.

We investigate the performance of our proposed LD strategy in the main RAF when all the $30$ users are uniformly distributed in the coverage area of HP-RRS in Fig. \ref{Fig10Performance}. In this scenario, the density of users in the coverage area of LP-RRSs decreases compared to that of considered in Fig. \ref{Fig03}. Therefore, more users are expected to be associated with the HP-RRS due to their long distances to LP-RRSs. Because of the limited transmit power of the HP-RRS and also increasing in the number of served users by the HP-RRS with poor wireless channel conditions, the average data rate of users will be decreased in the system. To this end, the slice incomes will be degraded when all the users are uniformly distributed in the coverage area of the network. On the other hand, the system is enforced to consume more radio resources to compensate the poor channel conditions of users due to the long distances to HP-RRS. Therefore, the slice radio resource costs also increase which degrade the slice revenues. This situation will be worst when the path loss exponent increases.

\subsection{Comparison Between Our Proposed Proactive DACPSs and DACRS}
\begin{figure}
\centering
\includegraphics[scale=0.41]{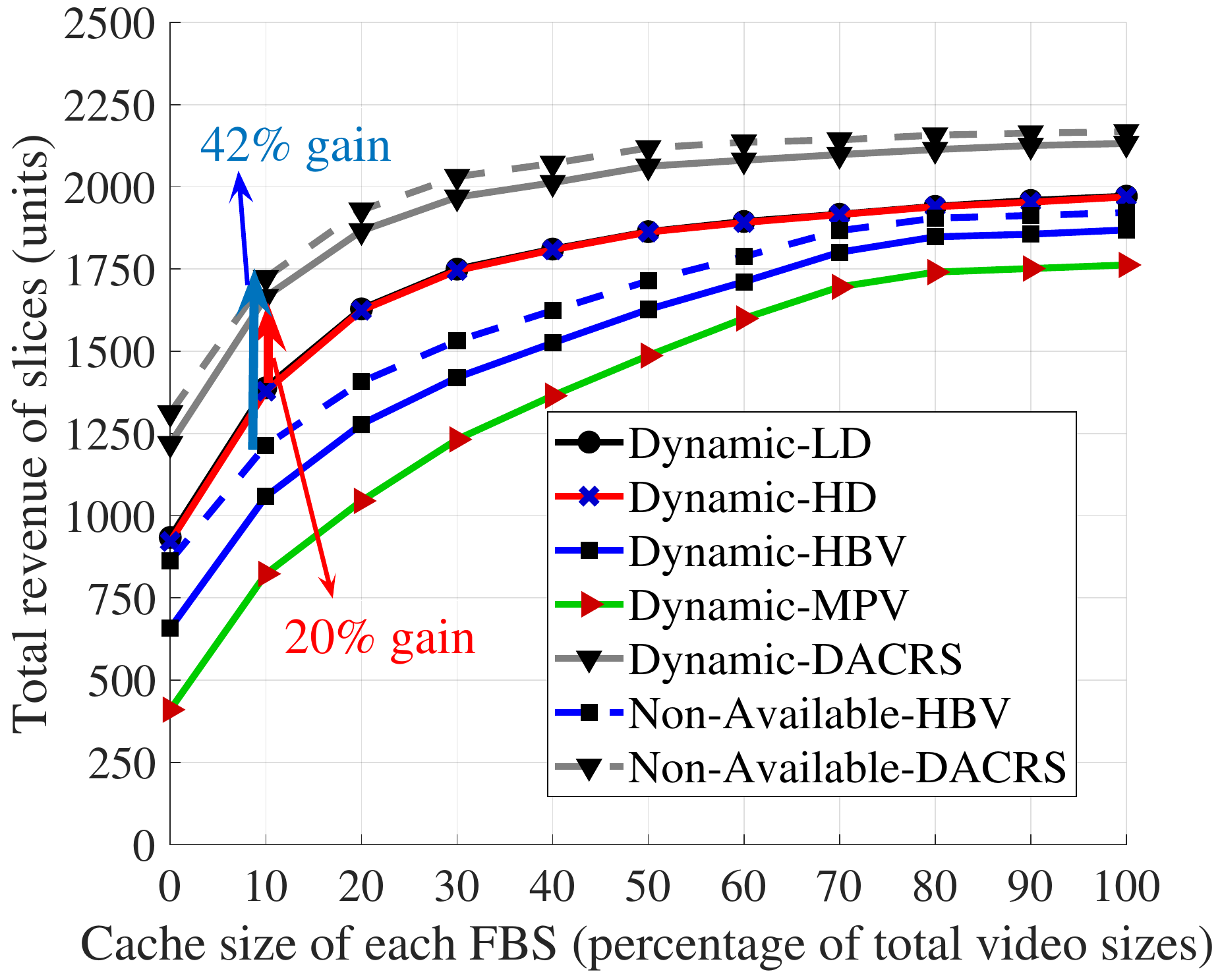}
\caption{Performance comparison of our proposed DACPSs and DACRS in terms of total revenue of slices in two scenarios as 1) Non-available CDI and VPD, where CDI and VPD are not available in Phase 1; 2) Dynamic CDI and VPD, where CDI and VPD are available in Phase 1 but dynamically change during Phase 2.}
\label{Fig11Rev}
\end{figure}

We investigate the performance of our proposed DACRS in terms of total revenue of slices in Fig. \ref{Fig11Rev}. In this regard, we consider two scenarios where in the first scenario, it is assumed that the fixed VPD and CDI are not available at the schedular in Phase 1, and for the second scenario, it is assumed that the CDI and VPD are available in Phase 1 but dynamically change during Phase 2. Since our proposed DACPSs and the existing MPV strategy are conducted based on the VPD, we cannot utilize these strategies for the first scenario.
For the first scenario (namely Non-available CDI and VPD), we compare the performance of our proposed DACRS with the heuristic HBV strategy. For our proposed DACRS strategy, we assume that when the VPD is not available in Phase 1, the HBV strategy is adopted in Phase 1, and the storages are updated during Phase 2 according to our DACRS. As shown, our proposed DACPS outperforms the system performance nearly $42\%$ compared to the proactive HBV strategy. In the second scenario (namely Dynamic CDI and VPD), we compared the proactive DACPSs, MPV, and HBV with our proposed DACRS, where for our proposed DACPS, we adopt the proactive LD strategy in Phase 1, and the storages are updated during Phase 2 according to our proposed DACRS. As shown in Fig. \ref{Fig11Rev}, DACRS improves the slice revenues up to $20\%$ compared to our proposed proactive DACPSs. This is because, the caching decision with our proactive strategies cannot be updated with the changes of VPD and CDI during Phase 2 while DACRS can update the storages according to the new request diversity and network channel conditions.

\section{Summary and Concluding Remarks}\label{Section Conclusion}
In this paper, we investigated the idea of developing a DACPS in a parallel CVCT system followed by a limited backhaul and fronthaul MC-NOMA-assisted HV-MEC. In this network, we first proposed a RAF based on the network operational time. In Phase 1, we designed two diversity-based schemes where in each scheme, we maximized the estimated average revenue of slices subject to the minimum required access data rate of each user owned by each slice and some fundamental system constraints. In order to find an efficient solution for these large-scale and NP-hard problems, we proposed a solution algorithm based on the alternating optimization method. To reduce the computational complexity of the delivery algorithm, we proposed a LC-RAF where the radio resource allocation policy obtained in Phase 1 is adopted for all time slots of Phase 2. Numerical assessments showed that our proposed DACPSs improve the average system delivery cost between $25\%$ and $40\%$ compared to the conventional baseline bitrate/popular video CPSs. It is also observed that the integration of cooperative caching and cooperative transcoding capabilities can improve the system revenue up to 17-fold, which is notable considering each technology alone outperforms the system revenue only 6 to 10-fold. Moreover, we showed that our proposed LC-RAF can be a good choice for dense environments with high levels of path loss where the distance of users to RRSs is more pivotal for the user association decision than the placement of videos in order to satisfy the QoS of users. Finally, to tackle the non-availability and/or dynamically changes of VPD and CDI during the delivery phase, we proposed a DACRS in which the caching decision is readopted at the end of each time slot according to the last requests of users, arrival CSI and availability of videos at storages. It is shown that our proposed DACRS improves the slice revenues up to $20\%$ compared to our proposed proactive DACPSs for a scenario where CDI and VPD are dynamically changed through the delivery phase.

\appendices

\section{Equivalent Transformation of \eqref{Main Problem jointpower}}\label{appendix transformation power problem}
The optimization problem \eqref{Main Problem jointpower} can be transformed into the following equivalent form as
\begin{subequations}\label{Main Problem jointpower transform}
\begin{align}\label{obf Main Problem jointpower transform}
&\max_{ \boldsymbol{\phi} , \boldsymbol{p} , \boldsymbol{r}^\text{BH}, \boldsymbol{r}^\text{FH} }\hspace{.0 cm} 	
\bar{\textbf{\$}}^\text{LD}_\text{tot}
\\
\textrm{s.t.}\hspace{.0cm}~~
& \text{\eqref{Constraint maximum transmit power BS}-\eqref{channel capacity constraint FH BS}, \eqref{constraint subcarrier binary}, \eqref{constraint minimum rate access erg},}  \nonumber
\\
& \label{constraint SIC user transform}
\mathbb{E}_{\boldsymbol{h}} \bigg{ \{ } h^{n}_{b,u} \left( I^{\text{Intra},n}_{b,u'} + I^{\text{Inter},n}_{b,u'}  +  \sigma^{n}_{b,u'} \right) \bigg{ \} } >
\mathbb{E}_{\boldsymbol{h}} \bigg{ \{ }   h^{n}_{b,u'} \bigg( \sum\limits_{v \in \mathcal{U}, v \neq u' \atop \scriptstyle h^{n}_{b,v} > h^{n}_{b,u'}} p^{n}_{b,v} h^{n}_{b,u} + I^{\text{Inter},n}_{b,u}  +  \sigma^{n}_{b,u} \bigg) \bigg{ \} }, \nonumber \\
&~~~~~~~~~~~~~~~~~~~~\forall  b \in \mathcal{B},u,u' \in \mathcal{U}, n \in \mathcal{N}, \tau^{n}_{b,u},\tau^{n}_{b,u'} > 0, \mathbb{E}_{\boldsymbol{h}} \big{ \{ }  h^{n}_{b,u} \big{ \} } > \mathbb{E}_{\boldsymbol{h}} \big{ \{ } h^{n}_{b,u'} \big{ \} },
\\
& \label{constraint event 4 transcoding fronthaul transform problem}
t^{v_h,v_l}_{b',b} \frac{r^{v_l}_{b',b}} {s_{v_l}} \leq t^{v_h,v_l}_{b',b} \frac{ \phi^{v_h,v_l}_{b'}  } { \eta^{v_h,v_l} N^{v_h,v_l}_{\text{Cycle}} },
\\
& \label{constraint event 5 fronthaul transcoding transform problem}
w^{v_h,v_l}_{b',b} \frac{ \phi^{v_h,v_l}_{b}  } { \eta^{v_h,v_l} N^{v_h,v_l}_{\text{Cycle}} } \leq w^{v_h,v_l}_{b',b} \frac{r^{v_h}_{b',b}} {s_{v_h}} ,
\\
& \label{constraint event 2 average access transc transform}
y^{v_h,v_l}_{b} \frac{ \bar{r}^\text{Ac}_{b,u}}{s_{v_l}} \leq \frac{ \phi^{v_h,v_l}_b  } { \eta^{v_h,v_l} N^{v_h,v_l}_{\text{Cycle}} }  y^{v_h,v_l}_{b} ,
\\
& \label{constraint event 3 average access front transform}
z^{v_l}_{b',b} \bar{r}^\text{Ac}_{b,u} \leq r^{v_l}_{b',b} z^{v_l}_{b',b},
\\
& \label{constraint event 4 fronthaul average access transform}
t^{v_h,v_l}_{b',b} \bar{r}^\text{Ac}_{b,u} \leq t^{v_h,v_l}_{b',b} r^{v_l}_{b',b},
\\
& \label{constraint event 5 transcoding average access transform}
w^{v_h,v_l}_{b',b} \frac{\bar{r}^\text{Ac}_{b,u}} {s_{v_l}} \leq w^{v_h,v_l}_{b',b} \frac{ \phi^{v_h,v_l}_{b}  } { \eta^{v_h,v_l} N^{v_h,v_l}_{\text{Cycle}} },
\\
& \label{constraint event 6 backhaul average access transform}
o^{v_l}_{b} \bar{r}^\text{Ac}_{b,u} \leq o^{v_l}_{b} r^{v_l}_{0,b},
\\
& \label{constraint relax processing allocation}
\phi^{v_h,v_l}_{b} \geq 0.
\end{align}
\end{subequations}
In this line, to tackle the fractional constraints \eqref{constraint event 4 transcoding fronthaul}, \eqref{constraint event 5 fronthaul transcoding}, and \eqref{constraint event 2 average access transc}, we transform them into equivalent forms as \eqref{constraint event 4 transcoding fronthaul transform problem}-\eqref{constraint event 6 backhaul average access transform}.
Note that \eqref{constraint event 2 average access transc transform}-\eqref{constraint event 6 backhaul average access transform} hold even $\bar{r}^\text{Ac}_{b,u}=0$, i.e., user $u$ is not connected to RRS $b$. We also note that in this transformation, \eqref{constraint event 4 transcoding fronthaul transform problem} and \eqref{constraint event 5 fronthaul transcoding transform problem} are turned into linear forms (with relaxed $\phi^{v_h,v_l}_{b}$). Constraints \eqref{SIC user ergodic}, \eqref{constraint minimum rate access erg}, and \eqref{constraint event 2 average access transc transform}-\eqref{constraint event 6 backhaul average access transform} are in non-convex forms due to the fractional form of SINR function $\gamma^{n}_{b,u}$, non-concavity and non-convexity of $\bar{r}^\text{Ac}_{b,u}$, respectively. However, \eqref{SIC user ergodic} can be transformed into an equivalent linear form with respect to $\boldsymbol{p}$ as \eqref{constraint SIC user transform} meaning that \eqref{SIC user ergodic} should be held when $\tau^{n}_{b,u} \tau^{n}_{b,u'} \neq 0$.

\section{The Proposed SCA Algorithm for Solving \eqref{Main Problem jointpower transform}}\label{appendix SCA power}
To tackle the non-convexity of both \eqref{obf Main Problem jointpower transform} and \eqref{constraint minimum rate access erg} and all constraints in \eqref{constraint event 2 average access transc transform}-\eqref{constraint event 6 backhaul average access transform}, the access data rate function $\bar{r}^\text{Ac}_{b,u}$ should be transformed into concave and convex forms, respectively.
To approximate the access data rate function $\bar{r}^\text{Ac}_{b,u}$ in \eqref{obf Main Problem jointpower transform} and \eqref{constraint minimum rate access erg} to a concave format at each iteration $\kappa_2$ of the SCA algorithm, we first define
\begin{align}\label{rate MENHA}
r^{n}_{b,u} = f^{n}_{b,u} - g^{n}_{b,u},
\end{align}
where $f^{n}_{b,u}$ and $g^{n}_{b,u}$ are concave functions with respect to $\boldsymbol{p}$. Moreover, $f^{n}_{b,u}$ and $g^{n}_{b,u}$ are formulated, respectively, by
\begin{align}\label{f function}
f^{n}_{b,u} = \tau^{n}_{b,u} W_s \log_2 \left( I^{\text{Intra},n}_{b,u} + I^{\text{Inter},n}_{b,u}  +  \sigma^{n}_{b,u} + p^{n}_{b,u} h^{n}_{b,u} \right),
\end{align}
\begin{align}\label{g function}
g^{n}_{b,u} = \tau^{n}_{b,u} W_s \log_2 \left( I^{\text{Intra},n}_{b,u} + I^{\text{Inter},n}_{b,u}  +  \sigma^{n}_{b,u} \right).
\end{align}
Then, we approximate $g^{n}_{b,u}(\boldsymbol{p}_{\kappa_2})$ at each iteration $\kappa_2$ by its first order Taylor series approximation around $\boldsymbol{p}_{\kappa_2-1}$ as follows \cite{Jointsubchannelassignment,LimitedFeedback}:
\begin{align}\label{g approximated}
g^{n}_{b,u}(\boldsymbol{p}_{\kappa_2}) \approx  g^{n}_{b,u}(\boldsymbol{p}_{\kappa_2-1}) +
\nabla g^{n}_{b,u}(\boldsymbol{p}_{\kappa_2-1})
(\boldsymbol{p}_{\kappa_2} - \boldsymbol{p}_{\kappa_2-1}),
\end{align}
where $\nabla g^{n}_{b,u}(\boldsymbol{p}_{\kappa_2-1})$ is a vector of length $UB$ and its entry is obtained by
\begin{equation}\label{g nabla}
\nabla g^{n}_{b,u} (\boldsymbol{p})  =
\left\{
  \begin{array}{ll}
    0, & \hbox{$\forall i=b, u' \in \mathcal{U}/\{u\}, h^{n}_{b,u'} \leq h^{n}_{b,u}$;} \\
    \frac{ \tau^{n}_{b,u} W_\text{s} h^{n}_{b,u} }{ (\ln2) \left( I^{\text{Intra},n}_{b,u} + I^{\text{Inter},n}_{b,u}  +  \sigma^{n}_{b,u} \right) }, & \hbox{$\forall i=b, u' \in \mathcal{U}/\{u\}, h^{n}_{b,u'} > h^{n}_{b,u}$;} \\
    \frac{ \tau^{n}_{b,u} W_\text{s} h^{n}_{i,u} }{ (\ln2) \left( I^{\text{Intra},n}_{b,u} + I^{\text{Inter},n}_{b,u}  +  \sigma^{n}_{b,u} \right) }, & \hbox{$\forall i \neq b, u' \in \mathcal{U}/\{u\}$.}
  \end{array}
\right.
\end{equation}
Therefore, the concave approximated function of $\bar{r}^{n}_{b,u}$ at each iteration $\kappa_2$ is expressed by
\begin{align}\label{rate approximate}
\hat{\bar{r}}^{n}_{b,u} \left(\boldsymbol{p}_{\kappa_2}\right) \approx \mathbb{E}_{\boldsymbol{h}} \left\{ f^{n}_{b,u} \left(\boldsymbol{p}_{\kappa_2}\right) - g^{n}_{b,u}\left(\boldsymbol{p}_{\kappa_2-1}\right) -
\nabla g^{n}_{b,u}\left(\boldsymbol{p}_{\kappa_2-1}\right)
\left(\boldsymbol{p}_{\kappa_2} - \boldsymbol{p}_{\kappa_2-1}\right) \right\}.
\end{align}
Besides, in order to approximate $\bar{r}^\text{Ac}_{b,u}$ in \eqref{constraint event 2 average access transc transform}-\eqref{constraint event 6 backhaul average access transform} to a convex form at each iteration $\kappa_2$ of the SCA algorithm, we first define $\bar{r}^\text{Ac}_{b,u}$ as a D.C. function in \eqref{rate MENHA}.
Then, we approximate $f^{n}_{b,u}(\boldsymbol{p}_{\kappa_2})$ at each iteration $\kappa_2$ by its first order Taylor series approximation around $\boldsymbol{p}_{\kappa_2-1}$ as
\begin{align}\label{f approximated}
f^{n}_{b,u}(\boldsymbol{p}_{\kappa_2}) \approx  f^{n}_{b,u}(\boldsymbol{p}_{\kappa_2-1}) +
\nabla f^{n}_{b,u}(\boldsymbol{p}_{\kappa_2-1})
(\boldsymbol{p}_{\kappa_2} - \boldsymbol{p}_{\kappa_2-1}),
\end{align}
where $\nabla f^{n}_{b,u}(\boldsymbol{p}_{\kappa_2-1})$ is a vector of length $UB$ and its entry is expressed by
\begin{equation}\label{f nabla}
\nabla f^{n}_{b,u} (\boldsymbol{p})  =
\left\{
  \begin{array}{ll}
    \frac{ \tau^{n}_{b,u} W_\text{s} h^{n}_{b,u} }{ (\ln2) \left( I^{\text{Intra},n}_{b,u} + I^{\text{Inter},n}_{b,u}  +  \sigma^{n}_{b,u} + p^{n}_{b,u} h^{n}_{b,u} \right) }, & \hbox{$\forall i=b, u' =u$;} \\
    0, & \hbox{$\forall i=b, u' \in \mathcal{U}/\{u\}, h^{n}_{b,u'} \leq h^{n}_{b,u}$;} \\
    \frac{ \tau^{n}_{b,u} W_\text{s} h^{n}_{b,u} }{ (\ln2) \left( I^{\text{Intra},n}_{b,u} + I^{\text{Inter},n}_{b,u}  +  \sigma^{n}_{b,u} + p^{n}_{b,u} h^{n}_{b,u} \right) }, & \hbox{$\forall i=b, u' \in \mathcal{U}/\{u\}, h^{n}_{b,u'} > h^{n}_{b,u}$;} \\
    \frac{ \tau^{n}_{b,u} W_\text{s} h^{n}_{i,u} }{ (\ln2) \left( I^{\text{Intra},n}_{b,u} + I^{\text{Inter},n}_{b,u}  +  \sigma^{n}_{b,u} + p^{n}_{b,u} h^{n}_{b,u} \right) }, & \hbox{$\forall i \neq b, u' \in \mathcal{U}/\{u\}$.}
  \end{array}
\right.
\end{equation}
Hence, the convex approximated form of $\bar{r}^{n}_{b,u}$ at each iteration $\kappa_2$ is given by
\begin{align}\label{rate approximate f}
\tilde{\bar{r}}^{n}_{b,u} \left(\boldsymbol{p}_{\kappa_2}\right) \approx \mathbb{E}_{\boldsymbol{h}} \left\{ f^{n}_{b,u}\left(\boldsymbol{p}_{\kappa_2-1}\right) +
\nabla f^{n}_{b,u}\left(\boldsymbol{p}_{\kappa_2-1}\right)
\left(\boldsymbol{p}_{\kappa_2} - \boldsymbol{p}_{\kappa_2-1}\right) - g^{n}_{b,u} \left(\boldsymbol{p}_{\kappa_2}\right) \right\}.
\end{align}
By substituting $\hat{\bar{r}}^{n}_{b,u}$ in \eqref{obf Main Problem jointpower transform} and \eqref{constraint minimum rate access erg} and $\tilde{\bar{r}}^{n}_{b,u}$ in \eqref{constraint event 2 average access transc transform}-\eqref{constraint event 6 backhaul average access transform}, the optimization problem \eqref{Main Problem jointpower transform} is transformed into an approximated DCP form which can be easily solved by utilizing efficient convex programming solutions, such as the Lagrange dual method or the CVX software \cite{CVXmatlab,Lecturesconvexoptimization,7762913,7812683,7488289}.

\section{Equivalent Transformation of \eqref{Main Problem caching subcarrier}}\label{appendix Proposition 1}
The resulting IDCP form of \eqref{Main Problem caching subcarrier} is formulated as
\begin{subequations}\label{Main Problem caching subcarrier transf epi2}
\begin{align}\label{obf Main Problem caching subcarrier transf epi2}
&\min_{ \boldsymbol{\Upsilon}, \boldsymbol{\rho}, \boldsymbol{\theta}, \boldsymbol{\tau}, \boldsymbol{\vartheta}, \tilde{\boldsymbol{\vartheta}} , \boldsymbol{\nu} , \hat{\boldsymbol{\Upsilon}}, \breve{\boldsymbol{\Upsilon}}
 }\hspace{.0 cm}
\sum\limits_{m \in \mathcal{M}} \bar{\textbf{\$}}^\text{slice,Epi,2}_m
\\
\textrm{s.t.}\hspace{.0cm}~~
& \text{
\eqref{user association one BS}, \eqref{constraint cases storage}, \eqref{user association subcarrier}, \eqref{processing capacity constraint eRRH}-\eqref{channel capacity constraint FH BS}, \eqref{constraint untranscodable}-\eqref{constraint subcarrier binary}, \eqref{constraint cache size BS}, \eqref{constraint minimum rate access erg}, \eqref{constraint rho binary}, \eqref{constraint event 4 transcoding fronthaul transform problem}, \eqref{constraint event 5 fronthaul transcoding transform problem},}    \nonumber \\
\label{constraint event 2 average access transc transform2 epi}
& \frac{ \mathbb{E}_{\boldsymbol{h}} \bigg{ \{ }  \sum_{n=1}^{N} \breve{y}^{v_h,v_l,n}_{b,u} W_\text{s} \log_2 ( 1 + \gamma^{n}_{b,u} ) \bigg{ \} } }{s_{v_l}} \leq \frac{ \phi^{v_h,v_l}_b  } { \eta^{v_h,v_l} N^{v_h,v_l}_{\text{Cycle}} }  y^{v_h,v_l}_{b} , \forall u \in \mathcal{U}_m,
\\
\label{constraint event 3 average access transc transform2 epi}
&  \mathbb{E}_{\boldsymbol{h}} \bigg{ \{ }  \sum_{n=1}^{N} \breve{z}^{v_l,n}_{b',b,u} W_\text{s} \log_2 ( 1 + \gamma^{n}_{b,u} ) \bigg{ \} } \leq r^{v_l}_{b',b} z^{v_l}_{b',b}, \forall u \in \mathcal{U}_m,
\\
\label{constraint event 4 average access transc transform2 epi}
& \frac{  \mathbb{E}_{\boldsymbol{h}} \bigg{ \{ }   \sum_{n=1}^{N} \breve{t}^{v_h,v_l,n}_{b',b,u}  W_\text{s} \log_2 ( 1 + \gamma^{n}_{b,u} )  \bigg{ \} } }{s_{v_l}} \leq t^{v_h,v_l}_{b',b} r^{v_l}_{b',b}, \forall u \in \mathcal{U}_m,
\\
\label{constraint event 5 average access transc transform2 epi}
& \frac{  \mathbb{E}_{\boldsymbol{h}} \bigg{ \{ }  \sum_{n=1}^{N} \breve{w}^{v_h,v_l,n}_{b',b,u}  W_\text{s} \log_2 ( 1 + \gamma^{n}_{b,u} )   \bigg{ \} } }{s_{v_l}} \leq w^{v_h,v_l}_{b',b} \frac{ \phi^{v_h,v_l}_{b}  } { \eta^{v_h,v_l} N^{v_h,v_l}_{\text{Cycle}} }, \forall u \in \mathcal{U}_m,
\\
\label{constraint event 6 average access transc transform2 epi}
&   \mathbb{E}_{\boldsymbol{h}} \bigg{ \{ }  \sum_{n=1}^{N} \breve{o}^{v_l,n}_{b,u}  W_\text{s} \log_2 ( 1 + \gamma^{n}_{b,u} )   \bigg{ \} } \leq o^{v_l}_{b} r^{v_l}_{0,b}, \forall u \in \mathcal{U}_m,
\\
\label{SIC user transform for subcarriers}
& \mathbb{E}_{\boldsymbol{h}} \bigg{ \{ } \tau^{n}_{b,u} \frac { p^{n}_{b,u'} h^{n}_{b,u} }  { \sum\limits_{v \in \mathcal{U}, v \neq u' \atop \scriptstyle h^{n}_{b,v} > h^{n}_{b,u'}} p^{n}_{b,v} h^{n}_{b,u} + I^{\text{Inter},n}_{b,u}  +  \sigma^{n}_{b,u} }  \bigg{ \} }
>
\mathbb{E}_{\boldsymbol{h}} \big{ \{ }  \tau^{n}_{b,u'} \gamma^{n}_{b,u'}  \big{ \} }, \forall  b \in \mathcal{B} ,
u,u' \in \mathcal{U}, n \in \mathcal{N}, \nonumber \\
&~~~~~~~~~~~~~~~~~~~~~~~~~~~~~~~~~~~~~~~~~~~~~~~~~~~~~p^{n}_{b,u},p^{n}_{b,u'} > 0, \mathbb{E}_{\boldsymbol{h}} \big{ \{ }  h^{n}_{b,u} \big{ \} } > \mathbb{E}_{\boldsymbol{h}} \big{ \{ } h^{n}_{b,u'} \big{ \} },
\\
\label{all cases constraint average transformed1}
& \vartheta_b \geq \theta_{b,u}, \forall b \in \mathcal{B},u \in \mathcal{U},~~~~~~~\vartheta_b \in \{0,1\},
\\
\label{all cases constraint average transformed2}
& x^{v_l}_{b} +
\sum\limits_{\scriptstyle v_h \in \mathcal{V} \atop \scriptstyle h > l} y^{v_h,v_l}_{b} +
\sum\limits_{\scriptstyle b' \in \mathcal{B} \atop \scriptstyle b' \neq b} z^{v_l}_{b',b} +
\sum\limits_{\scriptstyle b' \in \mathcal{B} \atop \scriptstyle b' \neq b} \sum\limits_{\scriptstyle v_h \in \mathcal{V} \atop \scriptstyle h > l} \left( t^{v_h,v_l}_{b',b} + w^{v_h,v_l}_{b',b} \right) + o^{v_l}_{b} = \vartheta_b, \forall b \in \mathcal{B}, v_l \in \mathcal{V},
\\
\label{constraint epi objective 1}
& \tilde{\vartheta}_{b,m} \geq \theta_{b,u}, \forall b \in \mathcal{B}, u \in \mathcal{U}_m,~~~~~~~\tilde{\vartheta}_{b,m} \in \{0,1\},
\\
\label{constraint epi objective 2}
& \nu^{n}_{b,m} \geq \tau^{n}_{b,u}, \forall b \in \mathcal{B}, n \in \mathcal{N}, u \in \mathcal{U}_m,~~~~~~~\nu^{n}_{b,m} \in \{0,1\},
\\
\label{constraint revenue adding x theta 1}
& \hat{x}^{v_l}_{b,m} \leq x^{v_l}_{b},~~~\hat{x}^{v_l}_{b,m} \leq \tilde{\vartheta}_{b,m},~~~\hat{x}^{v_l}_{b,m} \geq x^{v_l}_{b} + \tilde{\vartheta}_{b,m} -1,~~~\hat{x}^{v_l}_{b,m} \in \{0,1\},
\\
\label{constraint revenue adding y theta 1}
& \hat{y}^{v_h,v_l}_{b,m} \leq y^{v_h,v_l}_{b},~~~\hat{y}^{v_h,v_l}_{b,m} \leq \tilde{\vartheta}_{b,m},~~~\hat{y}^{v_h,v_l}_{b,m} \geq y^{v_h,v_l}_{b} + \tilde{\vartheta}_{b,m} - 1,~~~\hat{y}^{v_h,v_l}_{b} \in \{0,1\},
\\
\label{constraint revenue adding z theta 1}
& \hat{z}^{v_l}_{b',b,m} \leq z^{v_l}_{b',b},~~~\hat{z}^{v_l}_{b',b,m} \leq \tilde{\vartheta}_{b,m},~~~\hat{z}^{v_l}_{b',b,m} \geq z^{v_l}_{b',b} + \tilde{\vartheta}_{b,m} -1,~~~\hat{z}^{v_l}_{b',b,m} \in \{0,1\},
\\
\label{constraint revenue adding t theta 1}
& \hat{t}^{v_h,v_l}_{b',b,m} \leq t^{v_h,v_l}_{b',b},~~~\hat{t}^{v_h,v_l}_{b',b,m} \leq \tilde{\vartheta}_{b,m},~~~\hat{t}^{v_h,v_l}_{b',b,m} \geq t^{v_h,v_l}_{b',b} + \tilde{\vartheta}_{b,m} -1 ,~~~\hat{t}^{v_h,v_l}_{b',b,m} \in \{0,1\},
\\
\label{constraint revenue adding w theta 1}
& \hat{w}^{v_h,v_l}_{b',b,m} \leq w^{v_h,v_l}_{b',b},~~~\hat{w}^{v_h,v_l}_{b',b,m} \leq \tilde{\vartheta}_{b,m},~~~\hat{w}^{v_h,v_l}_{b',b,m} \geq w^{v_h,v_l}_{b',b} + \tilde{\vartheta}_{b,m} -1 ,~~~\hat{w}^{v_h,v_l}_{b',b,m} \in \{0,1\},
\\
\label{constraint revenue adding o theta 1}
& \hat{o}^{v_l}_{b,m} \leq o^{v_l}_{b},~~~\hat{o}^{v_l}_{b,m} \leq \tilde{\vartheta}_{b,m},~~~\hat{o}^{v_l}_{b,m} \geq o^{v_l}_{b} + \tilde{\vartheta}_{b,m} -1,~~~\hat{o}^{v_l}_{b,m} \in \{0,1\},
\\
\label{constraint delay adding y theta 1}
& \breve{y}^{v_h,v_l,n}_{b,u} \leq y^{v_h,v_l}_{b},~~~\breve{y}^{v_h,v_l,n}_{b,u} \leq \tau^{n}_{b,u},~~~\breve{y}^{v_h,v_l,n}_{b,u} \geq y^{v_h,v_l}_{b} + \tau^{n}_{b,u} -1,~~~\breve{y}^{v_h,v_l,n}_{b,u} \in \{0,1\},
\\
\label{constraint delay adding z theta 1}
& \breve{z}^{v_l,n}_{b',b,u} \leq z^{v_l}_{b',b},~~~\breve{z}^{v_l,n}_{b',b,u} \leq \tau^{n}_{b,u},~~~\breve{z}^{v_l,n}_{b',b,u} \geq z^{v_l}_{b',b} + \tau^{n}_{b,u} -1 ,~~~\breve{z}^{v_l,n}_{b',b,u} \in \{0,1\},
\\
\label{constraint delay adding t theta 1}
& \breve{t}^{v_h,v_l,n}_{b',b,u} \leq t^{v_h,v_l}_{b',b},~~~\breve{t}^{v_h,v_l,n}_{b',b,u} \leq \tau^{n}_{b,u},~~~\breve{t}^{v_h,v_l,n}_{b',b,u} \geq t^{v_h,v_l}_{b',b} + \tau^{n}_{b,u} -1 ,~~~\breve{t}^{v_h,v_l,n}_{b',b,u} \in \{0,1\},
\\
\label{constraint delay adding w theta 1}
& \breve{w}^{v_h,v_l,n}_{b',b,u} \leq w^{v_h,v_l}_{b',b},~~\breve{w}^{v_h,v_l,n}_{b',b,u} \leq \tau^{n}_{b,u},~~\breve{w}^{v_h,v_l,n}_{b',b,u} \geq w^{v_h,v_l}_{b',b} + \tau^{n}_{b,u} -1 ,~~\breve{w}^{v_h,v_l,n}_{b',b,u} \in \{0,1\},
\\
\label{constraint delay adding o theta 1}
& \breve{o}^{v_l,n}_{b,u} \leq o^{v_l}_{b},~~~\breve{o}^{v_l,n}_{b,u} \leq \tau^{n}_{b,u},~~~\breve{o}^{v_l,n}_{b,u} \geq o^{v_l}_{b} + \tau^{n}_{b,u} -1 ,~~~\breve{o}^{v_l,n}_{b,u} \in \{0,1\},
\end{align}
\end{subequations}
where
$
\bar{\textbf{\$}}^\text{slice,Epi,2}_m = \sum_{ u \in \mathcal{U}_m } \sum\limits_{b \in \mathcal{B}} \bar{r}^\text{Ac}_{b,u} \psi_m
- \sum_{ u \in \mathcal{U}_m } \sum\limits_{b \in \mathcal{B}} \sum_{ n \in \mathcal{N} } \left( p^{n}_{b,u} \alpha^\text{Pow}_b \right)
- \sum\limits_{b \in \mathcal{B}} \sum_{ n \in \mathcal{N} } \left( \nu^{n}_{b,m} W_\text{s} \alpha^\text{Sub}_b \right)     \\
- \sum_{b \in \mathcal{B}} \sum\limits_{v_l \in \mathcal{V}} \Delta_{v_l}
\bigg(
\hat{x}^{v_l}_{b,m} \left( s_{v_l} \mu^\text{Cache}_b \right) +
\sum\limits_{\scriptstyle v_h \in \mathcal{V} \atop \scriptstyle h > l} \hat{y}^{v_h,v_l}_{b,m} \left( s_{v_h} \mu^\text{Cache}_b + \phi^{v_h,v_l}_b \mu^\text{Proc}_b \right) +
\sum\limits_{\scriptstyle b' \in \mathcal{B} \atop \scriptstyle b' \neq b} \hat{z}^{v_l}_{b',b,m} \left( s_{v_l} \mu^\text{Cache}_{b'} + r^{v_l}_{b',b} \alpha^\text{FH} \right) +  \\
\sum\limits_{\scriptstyle b' \in \mathcal{B} \atop \scriptstyle b' \neq b} \sum\limits_{\scriptstyle v_h \in \mathcal{V} \atop \scriptstyle h > l} \hat{t}^{v_h,v_l}_{b',b,m} \left( s_{v_h} \mu^\text{Cache}_{b'} + \phi^{v_h,v_l}_{b'} \mu^\text{Proc}_{b'} + r^{v_l}_{b',b} \alpha^\text{FH} \right) +
\sum\limits_{\scriptstyle b' \in \mathcal{B} \atop \scriptstyle b' \neq b} \sum\limits_{\scriptstyle v_h \in \mathcal{V} \atop \scriptstyle h > l} \hat{w}^{v_h,v_l}_{b',b,m} \big( s_{v_h} \mu^\text{Cache}_{b'} + r^{v_h}_{b',b} \alpha^\text{FH} +     \\
\phi^{v_h,v_l}_{b} \mu^\text{Proc}_{b} \big) +
\hat{o}^{v_l}_{b,m} \left( r^{v_l}_{0,b} \alpha^\text{BH} \right)
\bigg)
$,
$\boldsymbol{\vartheta}=[\vartheta_b]$, $\tilde{\boldsymbol{\vartheta}}=[\tilde{\vartheta}_{b,m}]$, $\boldsymbol{\nu}=[\nu^{n}_{b,m}]$,
$\hat{\boldsymbol{\Upsilon}}=[\hat{\boldsymbol{x}},\hat{\boldsymbol{y}},\hat{\boldsymbol{z}},\hat{\boldsymbol{t}},\hat{\boldsymbol{w}},\hat{\boldsymbol{o}}]$,
$\hat{\boldsymbol{x}}=[\hat{x}^{v_l}_{b,m}]$, $\hat{\boldsymbol{y}}=[\hat{y}^{v_h,v_l}_{b,m}]$, $\hat{\boldsymbol{z}}=[\hat{z}^{v_l}_{b',b,m}]$, $\hat{\boldsymbol{t}}=[\hat{t}^{v_h,v_l}_{b',b,m}]$, $\hat{\boldsymbol{w}}=[\hat{w}^{v_h,v_l}_{b',b,m}]$, $\hat{\boldsymbol{o}}=[\hat{o}^{v_l}_{b,m}]$,
$\breve{\boldsymbol{\Upsilon}}=[\breve{\boldsymbol{y}},\breve{\boldsymbol{z}},\breve{\boldsymbol{t}},\breve{\boldsymbol{w}},\breve{\boldsymbol{o}}]$,
$\breve{\boldsymbol{y}}=[\breve{y}^{v_h,v_l,n}_{b,u}]$, $\breve{\boldsymbol{z}}=[\breve{z}^{v_l,n}_{b',b,u}]$, $\breve{\boldsymbol{t}}=[\breve{t}^{v_h,v_l,n}_{b',b,u}]$, $\breve{\boldsymbol{w}}=[\breve{w}^{v_h,v_l,n}_{b',b,u}]$, and $\breve{\boldsymbol{o}}=[\breve{o}^{v_l,n}_{b,u}]$.
In order to overcome the nonlinearity challenges in \eqref{Main Problem caching subcarrier}, we first transform \eqref{SIC user} into an equivalent integer linear form in \eqref{SIC user transform for subcarriers} which should be held when $p^{n}_{b,u},p^{n}_{b,u'} > 0$ and $h^{n}_{b,u} > h^{n}_{b,u'}$.
In addition, by introducing a new binary variable $\vartheta_b$, where
\begin{align}
\vartheta_b \geq \theta_{b,u}, \forall b \in \mathcal{B},~~~~~~~\vartheta_b \in \{0,1\},
\end{align}
constraint \eqref{all cases constraint average} can be transformed into an equivalent form as \eqref{all cases constraint average transformed2}
which is in an integer linear form with respect to the new variable $\vartheta_b$ and all variables in $\boldsymbol{\Upsilon}$ \cite{Boydconvex}.
In the same line, by defining new variables $\tilde{\vartheta}_{b,m}$ and $\nu^{n}_{b,m}$ such that \eqref{constraint epi objective 1} and \eqref{constraint epi objective 2} are held, respectively, the term $\bar{\textbf{\$}}^\text{slice,UB}_m$ in \eqref{obf Main Problem caching subcarrier} is transformed into the following equivalent form as
\begin{multline}\label{revenue MVNO average epi}
\bar{\textbf{\$}}^\text{slice,Epi,1}_m = \sum_{ u \in \mathcal{U}_m } \sum\limits_{b \in \mathcal{B}} \bar{r}^\text{Ac}_{b,u} \psi_m
- \sum_{ u \in \mathcal{U}_m } \sum\limits_{b \in \mathcal{B}} \sum_{ n \in \mathcal{N} } \left( p^{n}_{b,u} \alpha^\text{Pow}_b \right)
- \sum\limits_{b \in \mathcal{B}} \sum_{ n \in \mathcal{N} } \left( \nu^{n}_{b,m} W_\text{s} \alpha^\text{Sub}_b \right) -     \\
\sum_{b \in \mathcal{B}} \tilde{\vartheta}_{b,m} \bar{\$}^{\text{Cost,RRS}}_{b},
\end{multline}
which is still in the integer nonlinear form, because of binary bilinear products generated by multiplications of $\tilde{\vartheta}_{b,m}$ in all scheduling variables in $\boldsymbol{\Upsilon}$ in the term $\tilde{\vartheta}_{b,m} \bar{\$}^{\text{Cost,RRS}}_{b}$. To address this challenge, we utilize the following lemma to linearize each binary bilinear product.
\\
\textbf{Lemma 1.} \emph{Assume that $x \in \{0,1\}$ and $y \in \{0,1\}$ are binary variables. The binary variable $z \in \{0,1\}$ can be substituted with the binary bilinear product $xy$ if $z \leq x$, $z \leq y$, and $z \geq x+y-1$.}
\begin{proof}
For each two binary variables $x$ and $y$, the following equality is always satisfied: $xy=\min\{x,y\}$. By utilizing the epigraph technique and introducing a new binary variable $z \in \{0,1\}$ such that $z \leq x$, $z \leq y$, and $z \geq x+y-1$, the integer linear term $z$ can be replaced with the integer nonlinear term $xy$.
\end{proof}
Based on \textbf{Lemma 1}, by adding constraints \eqref{constraint revenue adding x theta 1}-\eqref{constraint revenue adding o theta 1},
the binary variables $\hat{y}^{v_h,v_l}_{b,m}$, $\hat{z}^{v_l}_{b',b,m}$, $\hat{t}^{v_h,v_l}_{b',b,m}$, $\hat{w}^{v_h,v_l}_{b',b,m}$, and $\hat{o}^{v_l}_{b,m}$ are replaced with the binary bilinear products $y^{v_h,v_l}_{b} \tilde{\vartheta}_{b,m}$, $z^{v_l}_{b',b} \tilde{\vartheta}_{b,m}$, $t^{v_h,v_l}_{b',b} \tilde{\vartheta}_{b,m}$, $w^{v_h,v_l}_{b',b} \tilde{\vartheta}_{b,m}$, and $o^{v_l}_{b} \tilde{\vartheta}_{b,m}$ in \eqref{revenue MVNO average epi}, respectively, which turn \eqref{revenue MVNO average epi} into a linear integer form.

To cope with the nonlinearity of delay functions in \eqref{constraint event 4 transcoding fronthaul} and \eqref{constraint event 5 fronthaul transcoding}, and average access delay functions in \eqref{constraint event 2 average access transc}, we utilize the transformation method presented in Appendix \ref{appendix transformation power problem}. To this end, we first transform
\eqref{constraint event 4 transcoding fronthaul} and \eqref{constraint event 5 fronthaul transcoding} into \eqref{constraint event 4 transcoding fronthaul transform problem} and \eqref{constraint event 5 fronthaul transcoding transform problem}, respectively, where both of them are in integer linear forms. Besides, we transform constraints in
\eqref{constraint event 2 average access transc} into \eqref{constraint event 2 average access transc transform}-\eqref{constraint event 6 backhaul average access transform}
which are still in integer nonlinear forms with respect to $\boldsymbol{\Upsilon}$ and $\boldsymbol{\theta}$. According to \textbf{Lemma 1}, by adding constraints \eqref{constraint delay adding y theta 1}-\eqref{constraint delay adding o theta 1},
the binary variables $\breve{y}^{v_h,v_l,n}_{b,u}$, $\breve{z}^{v_l,n}_{b',b,u}$, $\breve{t}^{v_h,v_l,n}_{b',b,u}$, $\breve{w}^{v_h,v_l,n}_{b',b,u}$, and $\breve{o}^{v_l,n}_{b,u}$ are replaced with the binary bilinear products $y^{v_h,v_l}_{b} \tau^{n}_{b,u}$, $z^{v_l}_{b',b} \tau^{n}_{b,u}$, $t^{v_h,v_l}_{b',b} \tau^{n}_{b,u}$, $w^{v_h,v_l}_{b',b} \tau^{n}_{b,u}$, and $o^{v_l}_{b} \tau^{n}_{b,u}$ in \eqref{constraint event 2 average access transc transform}-\eqref{constraint event 6 backhaul average access transform}, respectively where the result constraints \eqref{constraint event 2 average access transc transform2 epi}-\eqref{constraint event 6 average access transc transform2 epi} are in integer linear forms.

\section{Equivalent Transformation of \eqref{Main Problem caching HD}}\label{appendix transformation HD cache placement problem}
The first step of Algorithm \ref{Alg iterative main}, i.e., finding joint $\boldsymbol{p}$, $\boldsymbol{\phi}$, $\boldsymbol{r}^\text{FH}$ and $\boldsymbol{r}^\text{BH}$, is similar to the first step of solving \eqref{Main Problem caching} except that in the second step, i.e., finding joint $\boldsymbol{\Upsilon}$, $\boldsymbol{\rho}$, $\boldsymbol{\tau}$ and $\boldsymbol{\theta}$, $\theta_{b,u}$ is directly multiplied to all scheduling variables in $\boldsymbol{\Upsilon}$. Therefore, by using \textbf{Lemma 1}, \eqref{Main Problem caching HD} is transformed into an equivalent IDCP problem formulated as
\begin{subequations}\label{Main Problem caching subcarrier transf epi HD}
\begin{align}\label{obf Main Problem caching subcarrier transf epi  HD}
&\min_{ \boldsymbol{\Upsilon}, \boldsymbol{\rho}, \boldsymbol{\theta}, \boldsymbol{\tau}, \boldsymbol{\vartheta}, \boldsymbol{\nu} , \breve{\boldsymbol{\Upsilon}}, \tilde{\boldsymbol{\Upsilon}}}
\hspace{.0 cm}
\sum\limits_{m \in \mathcal{M}} \bar{\textbf{\$}}^\text{slice,Epi,LB}_m
\\
\textrm{s.t.}\hspace{.0cm}~~
& \text{
\eqref{user association one BS},\eqref{constraint cases storage},\eqref{user association subcarrier},\eqref{processing capacity constraint eRRH}-\eqref{channel capacity constraint FH BS},\eqref{constraint untranscodable}-\eqref{constraint subcarrier binary},\eqref{constraint cache size BS},\eqref{constraint minimum rate access erg},\eqref{constraint rho binary},\eqref{constraint event 4 transcoding fronthaul transform problem},\eqref{constraint event 5 fronthaul transcoding transform problem},\eqref{constraint event 2 average access transc transform2 epi}-\eqref{all cases constraint average transformed2},\eqref{constraint epi objective 2},\eqref{constraint delay adding y theta 1}-\eqref{constraint delay adding o theta 1},} \nonumber \\
\label{constraint MVNO average revenue nonnegative epi HD}
& \tilde{x}^{v_l}_{b,u} \leq x^{v_l}_{b},~~~\tilde{x}^{v_l}_{b,u} \leq \theta_{b,u},~~~\tilde{x}^{v_l}_{b,u} \geq x^{v_l}_{b} + \theta_{b,u} -1,~~~\tilde{x}^{v_l}_{b,u} \in \{0,1\},
\\
\label{constraint delay adding2 y theta 1}
& \tilde{y}^{v_h,v_l}_{b,u} \leq y^{v_h,v_l}_{b},~~~\tilde{y}^{v_h,v_l}_{b,u} \leq \theta_{b,u},~~~\tilde{y}^{v_h,v_l}_{b,u} \geq y^{v_h,v_l}_{b} + \theta_{b,u} -1,~~~\tilde{y}^{v_h,v_l}_{b,u} \in \{0,1\},
\\
\label{constraint delay adding2 z theta 1}
& \tilde{z}^{v_l}_{b',b,u} \leq z^{v_l}_{b',b},~~~\tilde{z}^{v_l}_{b',b,u} \leq \theta_{b,u},~~~\tilde{z}^{v_l}_{b',b,u} \geq z^{v_l}_{b',b} + \theta_{b,u} -1 ,~~~\tilde{z}^{v_l}_{b',b,u} \in \{0,1\},
\\
\label{constraint delay adding2 t theta 1}
& \tilde{t}^{v_h,v_l}_{b',b,u} \leq t^{v_h,v_l}_{b',b},~~~\tilde{t}^{v_h,v_l}_{b',b,u} \leq \theta_{b,u},~~~\tilde{t}^{v_h,v_l}_{b',b,u} \geq t^{v_h,v_l}_{b',b} + \theta_{b,u} -1 ,~~~\tilde{t}^{v_h,v_l}_{b',b,u} \in \{0,1\},
\\
\label{constraint delay adding2 w theta 1}
& \tilde{w}^{v_h,v_l}_{b',b,u} \leq w^{v_h,v_l}_{b',b},~~~\tilde{w}^{v_h,v_l}_{b',b,u} \leq \theta_{b,u},~~~\tilde{w}^{v_h,v_l}_{b',b,u} \geq w^{v_h,v_l}_{b',b} + \theta_{b,u} -1 ,~~~\tilde{w}^{v_h,v_l}_{b',b,u} \in \{0,1\},
\\
\label{constraint delay adding2 o theta 1}
& \tilde{o}^{v_l}_{b,u} \leq o^{v_l}_{b},~~~\tilde{o}^{v_l}_{b,u} \leq \theta_{b,u},~~~\tilde{o}^{v_l}_{b,u} \geq o^{v_l}_{b} + \theta_{b,u} -1 ,~~~\tilde{o}^{v_l}_{b,u} \in \{0,1\},
\end{align}
\end{subequations}
where $\tilde{\boldsymbol{\Upsilon}}=[\tilde{\boldsymbol{x}},\tilde{\boldsymbol{y}},\tilde{\boldsymbol{z}},\tilde{\boldsymbol{t}},\tilde{\boldsymbol{w}},\tilde{\boldsymbol{o}}]$,
$\tilde{\boldsymbol{x}}=[\tilde{x}^{v_l}_{b,u}]$, $\tilde{\boldsymbol{y}}=[\tilde{y}^{v_h,v_l}_{b,u}]$, $\tilde{\boldsymbol{z}}=[\tilde{z}^{v_l}_{b',b,u}]$, $\tilde{\boldsymbol{t}}=[\tilde{t}^{v_h,v_l}_{b',b,u}]$, $\tilde{\boldsymbol{w}}=[\tilde{w}^{v_h,v_l}_{b',b,u}]$, $\tilde{\boldsymbol{o}}=[\tilde{o}^{v_l}_{b,u}]$ and
$
\bar{\textbf{\$}}^\text{slice,Epi,LB}_m = \sum_{ u \in \mathcal{U}_m } \sum\limits_{b \in \mathcal{B}} \bar{r}^\text{Ac}_{b,u} \psi_m
- \sum_{ u \in \mathcal{U}_m } \sum\limits_{b \in \mathcal{B}} \sum_{ n \in \mathcal{N} } \left( p^{n}_{b,u} \alpha^\text{Pow}_b \right)
-\sum\limits_{b \in \mathcal{B}} \sum_{ n \in \mathcal{N} } \left( \nu^{n}_{b,m} W_\text{s} \alpha^\text{Sub}_b \right) -
\\
\sum_{b \in \mathcal{B}} \sum_{ u \in \mathcal{U}_m } \sum\limits_{v_l \in \mathcal{V}} \Delta_{v_l}
\bigg(
\tilde{x}^{v_l}_{b,u} \left( s_{v_l} \mu^\text{Cache}_b \right) +
\sum\limits_{\scriptstyle v_h \in \mathcal{V} \atop \scriptstyle h > l} \tilde{y}^{v_h,v_l}_{b,u} \left( s_{v_h} \mu^\text{Cache}_b + \phi^{v_h,v_l}_b \mu^\text{Proc}_b \right) +
\\
\sum\limits_{\scriptstyle b' \in \mathcal{B} \atop \scriptstyle b' \neq b} \tilde{z}^{v_l}_{b',b,u} ( s_{v_l} \mu^\text{Cache}_{b'} + r^{v_l}_{b',b} \alpha^\text{FH} ) +
\sum\limits_{\scriptstyle b' \in \mathcal{B} \atop \scriptstyle b' \neq b} \sum\limits_{\scriptstyle v_h \in \mathcal{V} \atop \scriptstyle h > l} \tilde{t}^{v_h,v_l}_{b',b,u} \left( s_{v_h} \mu^\text{Cache}_{b'} + \phi^{v_h,v_l}_{b'} \mu^\text{Proc}_{b'} + r^{v_l}_{b',b} \alpha^\text{FH} \right) +
\\
\sum\limits_{\scriptstyle b' \in \mathcal{B} \atop \scriptstyle b' \neq b} \sum\limits_{\scriptstyle v_h \in \mathcal{V} \atop \scriptstyle h > l} \tilde{w}^{v_h,v_l}_{b',b,u} \big( s_{v_h} \mu^\text{Cache}_{b'} + r^{v_h}_{b',b} \alpha^\text{FH} +
\phi^{v_h,v_l}_{b} \mu^\text{Proc}_{b} \big) +
\tilde{o}^{v_l}_{b,u} \left( r^{v_l}_{0,b} \alpha^\text{BH} \right)
\bigg).
$

\section{PROOF OF PROPOSITION 2}\label{appendix proposition 2}
Each iteration $\kappa_1$ in the proposed Algorithm \ref{Alg iterative main} for solving \eqref{Main Problem caching} consists of two main steps. In the first step, we find $\left( \boldsymbol{p}_{\kappa_1}, \boldsymbol{\phi}_{\kappa_1}, \boldsymbol{r}^\text{FH}_{\kappa_1}, \boldsymbol{r}^\text{BH}_{\kappa_1} \right)$ by solving \eqref{Main Problem caching} for a fixed $\left( \boldsymbol{\Upsilon}_{\kappa_1-1} , \boldsymbol{\rho}_{\kappa_1-1}, \boldsymbol{\tau}_{\kappa_1-1}, \boldsymbol{\theta}_{\kappa_1-1} \right)$ using an equivalent transformation method and subsequently applying the iterative SCA Algorithm \ref{Alg SCA power caching} which provides a local optimum solution for \eqref{Main Problem caching} that is proved in \emph{\textbf{Proposition 3}}. In this line, we have the following relation
\begin{multline}\label{Iter1}
\bar{\textbf{\$}}_\text{tot} \left( \boldsymbol{p}_{\kappa_1-1}, \boldsymbol{\phi}_{\kappa_1-1}, \boldsymbol{r}^\text{FH}_{\kappa_1-1}, \boldsymbol{r}^\text{BH}_{\kappa_1-1}, \boldsymbol{\Upsilon}_{\kappa_1-1} , \boldsymbol{\rho}_{\kappa_1-1}, \boldsymbol{\tau}_{\kappa_1-1}, \boldsymbol{\theta}_{\kappa_1-1} \right)
\leq    \\
\bar{\textbf{\$}}_\text{tot} \left( \boldsymbol{p}_{\kappa_1}, \boldsymbol{\phi}_{\kappa_1}, \boldsymbol{r}^\text{FH}_{\kappa_1}, \boldsymbol{r}^\text{BH}_{\kappa_1}, \boldsymbol{\Upsilon}_{\kappa_1-1} , \boldsymbol{\rho}_{\kappa_1-1}, \boldsymbol{\tau}_{\kappa_1-1}, \boldsymbol{\theta}_{\kappa_1-1} \right).
\end{multline}
In the second step of iteration $\kappa_1$, we find $\left( \boldsymbol{\Upsilon}_{\kappa_1} , \boldsymbol{\rho}_{\kappa_1}, \boldsymbol{\tau}_{\kappa_1}, \boldsymbol{\theta}_{\kappa_1} \right)$ by solving \eqref{Main Problem caching} for the given $\left( \boldsymbol{p}_{\kappa_1}, \boldsymbol{\phi}_{\kappa_1}, \boldsymbol{r}^\text{FH}_{\kappa_1}, \boldsymbol{r}^\text{BH}_{\kappa_1} \right)$ from the previous step. In doing so, the INLP problem \eqref{Main Problem caching} is transformed into a IDCP from in \eqref{Main Problem caching subcarrier transf epi2} which is efficiently solved by using CVX with the internal solver MOSEK. Using the fact that CVX with the MOSEK solver improves the objective function \eqref{obf Main Problem caching subcarrier transf epi2} or the objective function of its equivalent problem \eqref{Main Problem caching} for the given $\left( \boldsymbol{p}_{\kappa_1}, \boldsymbol{\phi}_{\kappa_1}, \boldsymbol{r}^\text{FH}_{\kappa_1}, \boldsymbol{r}^\text{BH}_{\kappa_1} \right)$, the following relation holds
\begin{multline}\label{Iter2}
\bar{\textbf{\$}}_\text{tot} \left( \boldsymbol{p}_{\kappa_1}, \boldsymbol{\phi}_{\kappa_1}, \boldsymbol{r}^\text{FH}_{\kappa_1}, \boldsymbol{r}^\text{BH}_{\kappa_1}, \boldsymbol{\Upsilon}_{\kappa_1-1} , \boldsymbol{\rho}_{\kappa_1-1}, \boldsymbol{\tau}_{\kappa_1-1}, \boldsymbol{\theta}_{\kappa_1-1} \right)
\leq    \\
\bar{\textbf{\$}}_\text{tot} \left( \boldsymbol{p}_{\kappa_1}, \boldsymbol{\phi}_{\kappa_1}, \boldsymbol{r}^\text{FH}_{\kappa_1}, \boldsymbol{r}^\text{BH}_{\kappa_1}, \boldsymbol{\Upsilon}_{\kappa_1} , \boldsymbol{\rho}_{\kappa_1}, \boldsymbol{\tau}_{\kappa_1}, \boldsymbol{\theta}_{\kappa_1} \right).
\end{multline}
According to \eqref{Iter1} and \eqref{Iter2}, it can be easily observed that after each iteration $\kappa_1$ of Algorithm \ref{Alg iterative main}, the objective function \eqref{obf Main Problem caching} is either improved or remains constant. Essentially, when the iterations in Algorithm \ref{Alg iterative main} continue, it will converge to a locally optimal solution. Note that since the solution approach in the first step of Algorithm \ref{Alg iterative main}, i,e, finding $\left( \boldsymbol{p}_{\kappa_1}, \boldsymbol{\phi}_{\kappa_1}, \boldsymbol{r}^\text{FH}_{\kappa_1}, \boldsymbol{r}^\text{BH}_{\kappa_1} \right)$ basically follows the SCA approach which converges to a locally optimal solution, the globally optimal solution $\left( \boldsymbol{p}_{\kappa_1}, \boldsymbol{\phi}_{\kappa_1}, \boldsymbol{r}^\text{FH}_{\kappa_1}, \boldsymbol{r}^\text{BH}_{\kappa_1} \right)$ can not be guaranteed and the performance gap from this locally optimal solution and the global solution is also unknown. To this end, their corresponding solutions $\left( \boldsymbol{\Upsilon}_{\kappa_1} , \boldsymbol{\rho}_{\kappa_1}, \boldsymbol{\tau}_{\kappa_1}, \boldsymbol{\theta}_{\kappa_1} \right)$ maybe different. Accordingly, alternate Algorithm \ref{Alg iterative main} cannot guarantee the global optimality of the solution for \eqref{Main Problem caching} and what can be proved is only the local optimality of the solution. Nevertheless, in \cite{Jointsubchannelassignment,FastglobaloptimalpowerDC} is noted that the SCA method often empirically converges to a globally optimal solution.

\section{PROOF OF PROPOSITION 3}\label{appendix proposition 3}
In order to prove \emph{\textbf{Proposition 3}}, we first prove that the proposed SCA-based Algorithm \ref{Alg SCA power caching} generates a sequence of improved feasible solutions in equivalent problem of \eqref{Main Problem jointpower} which is formulated in \eqref{Main Problem jointpower transform}. As discussed, in each iteration $\kappa_2$ of the SCA approach, $g^{n}_{b,u}(\boldsymbol{p}_{\kappa_2})$ is approximated with its first order Taylor series in \eqref{constraint minimum rate access} and $f^{n}_{b,u}(\boldsymbol{p}_{\kappa_2})$ is approximated with its first order Taylor series in \eqref{constraint event 2 average access transc transform}-\eqref{constraint event 6 backhaul average access transform}. Indeed, $\nabla g^{n}_{b,u}(\boldsymbol{p}_{\kappa_2-1})$ and $\nabla f^{n}_{b,u}(\boldsymbol{p}_{\kappa_2-1})$ are supergradient functions of $g^{n}_{b,u}(\boldsymbol{p}_{\kappa_2-1})$ and $f^{n}_{b,u}(\boldsymbol{p}_{\kappa_2-1})$ for iteration $\kappa_2-1$, respectively. Since $g^{n}_{b,u}(\boldsymbol{p}_{\kappa_2})$ and $f^{n}_{b,u}(\boldsymbol{p}_{\kappa_2})$ are concave, it must be hold
\begin{align}\label{gradian inequ1}
g^{n}_{b,u}(\boldsymbol{p}_{\kappa_2}) \leq  g^{n}_{b,u}(\boldsymbol{p}_{\kappa_2-1}) + \nabla g^{n}_{b,u}(\boldsymbol{p}_{\kappa_2-1}) (\boldsymbol{p}_{\kappa_2} - \boldsymbol{p}_{\kappa_2-1}),
\end{align}
\begin{align}\label{gradian inequ1 2}
f^{n}_{b,u}(\boldsymbol{p}_{\kappa_2}) \leq  f^{n}_{b,u}(\boldsymbol{p}_{\kappa_2-1}) + \nabla f^{n}_{b,u}(\boldsymbol{p}_{\kappa_2-1}) (\boldsymbol{p}_{\kappa_2} - \boldsymbol{p}_{\kappa_2-1}),
\end{align}
at each iteration $\kappa_2$, respectively. Using \eqref{rate approximate} and \eqref{rate approximate f}, the ergodic access data rate function $\bar{r}^{n}_{b,u} \left(\boldsymbol{p}_{\kappa_2}\right)$ in \eqref{rate MENHA} is approximated to a concave form $\hat{\bar{r}}^{n}_{b,u} \left(\boldsymbol{p}_{\kappa_2}\right)$ in \eqref{constraint minimum rate access} and also to a convex form $\tilde{\bar{r}}^{n}_{b,u} \left(\boldsymbol{p}_{\kappa_2}\right)$ in \eqref{constraint event 2 average access transc transform}-\eqref{constraint event 6 backhaul average access transform} at each iteration $\kappa_2$. According to \eqref{rate MENHA} and \eqref{gradian inequ1} it can be easily shown that the following relations hold
\begin{multline}\label{gradian inequ2}
\bar{r}^{n}_{b,u} \left(\boldsymbol{p}_{\kappa_2}\right) = \mathbb{E}_{\boldsymbol{h}} \left\{ f^{n}_{b,u}(\boldsymbol{p}_{\kappa_2}) - g^{n}_{b,u}(\boldsymbol{p}_{\kappa_2}) \right\} \geq  \\ \hat{\bar{r}}^{n}_{b,u} \left(\boldsymbol{p}_{\kappa_2}\right) = \mathbb{E}_{\boldsymbol{h}} \left\{ f^{n}_{b,u} \left(\boldsymbol{p}_{\kappa_2}\right) - g^{n}_{b,u}\left(\boldsymbol{p}_{\kappa_2-1}\right) -
\nabla g^{n}_{b,u}\left(\boldsymbol{p}_{\kappa_2-1}\right)
\left(\boldsymbol{p}_{\kappa_2} - \boldsymbol{p}_{\kappa_2-1}\right) \right\}.
\end{multline}
Moreover, based on \eqref{rate MENHA} and \eqref{gradian inequ1 2}, we have
\begin{multline}\label{gradian inequ2 2}
\bar{r}^{n}_{b,u} \left(\boldsymbol{p}_{\kappa_2}\right) = \mathbb{E}_{\boldsymbol{h}} \left\{ f^{n}_{b,u}(\boldsymbol{p}_{\kappa_2}) - g^{n}_{b,u}(\boldsymbol{p}_{\kappa_2}) \right\} \leq  \\ \tilde{\bar{r}}^{n}_{b,u} \left(\boldsymbol{p}_{\kappa_2}\right) = \mathbb{E}_{\boldsymbol{h}} \left\{ f^{n}_{b,u}\left(\boldsymbol{p}_{\kappa_2-1}\right) +
\nabla f^{n}_{b,u}\left(\boldsymbol{p}_{\kappa_2-1}\right)
\left(\boldsymbol{p}_{\kappa_2} - \boldsymbol{p}_{\kappa_2-1}\right) - g^{n}_{b,u} \left(\boldsymbol{p}_{\kappa_2}\right) \right\}.
\end{multline}
Besides, after solving the convex approximated problem of \eqref{Main Problem jointpower transform} in each iteration $\kappa_2$, the following inequalities hold
\begin{align}\label{constraint minimum rate access approximated}
\sum_{b \in \mathcal{B}} \sum_{n=1}^{N} \hat{\bar{r}}^{n}_{b,u} \left(\boldsymbol{p}_{\kappa_2}\right) \geq R^\text{min}_{m}, \forall m \in \mathcal{M}, u \in \mathcal{U}_m,
\end{align}
\begin{align}\label{constraint minimum rate access approximated1}
y^{v_h,v_l}_{b} \frac{ \sum_{n=1}^{N} \tilde{\bar{r}}^{n}_{b,u} \left(\boldsymbol{p}_{\kappa_2}\right) }{s_{v_l}} \leq \frac{ \phi^{v_h,v_l}_b  } { \eta^{v_h,v_l} N^{v_h,v_l}_{\text{Cycle}} }  y^{v_h,v_l}_{b} ,
\end{align}
\begin{align}\label{constraint minimum rate access approximated2}
z^{v_l}_{b',b} \sum_{n=1}^{N} \tilde{\bar{r}}^{n}_{b,u} \left(\boldsymbol{p}_{\kappa_2}\right) \leq r^{v_l}_{b',b} z^{v_l}_{b',b},
\end{align}
\begin{align}\label{constraint minimum rate access approximated3}
t^{v_h,v_l}_{b',b} \sum_{n=1}^{N} \tilde{\bar{r}}^{n}_{b,u} \left(\boldsymbol{p}_{\kappa_2}\right) \leq t^{v_h,v_l}_{b',b} r^{v_l}_{b',b},
\end{align}
\begin{align}\label{constraint minimum rate access approximated4}
w^{v_h,v_l}_{b',b} \frac{\sum_{n=1}^{N} \tilde{\bar{r}}^{n}_{b,u} \left(\boldsymbol{p}_{\kappa_2}\right)} {s_{v_l}} \leq w^{v_h,v_l}_{b',b} \frac{ \phi^{v_h,v_l}_{b}  } { \eta^{v_h,v_l} N^{v_h,v_l}_{\text{Cycle}} },
\end{align}
and
\begin{align}\label{constraint minimum rate access approximated5}
o^{v_l}_{b} \sum_{n=1}^{N} \tilde{\bar{r}}^{n}_{b,u} \left(\boldsymbol{p}_{\kappa_2}\right) \leq o^{v_l}_{b} r^{v_l}_{0,b}.
\end{align}
From \eqref{gradian inequ2}-\eqref{constraint minimum rate access approximated5}, it can be concluded that
\begin{align}\label{constraint minimum rate main}
\sum_{b \in \mathcal{B}} \bar{r}^{n}_{b,u} \left(\boldsymbol{p}_{\kappa_2}\right) \geq R^\text{min}_{m}, \forall m \in \mathcal{M}, u \in \mathcal{U}_m,
\end{align}
\begin{align}\label{constraint minimum rate main1}
y^{v_h,v_l}_{b} \frac{ \bar{r}^{n}_{b,u} \left(\boldsymbol{p}_{\kappa_2}\right) }{s_{v_l}} \leq \frac{ \phi^{v_h,v_l}_b  } { \eta^{v_h,v_l} N^{v_h,v_l}_{\text{Cycle}} }  y^{v_h,v_l}_{b} ,
\end{align}
\begin{align}\label{constraint minimum rate main2}
z^{v_l}_{b',b} \bar{r}^{n}_{b,u} \left(\boldsymbol{p}_{\kappa_2}\right) \leq r^{v_l}_{b',b} z^{v_l}_{b',b},
\end{align}
\begin{align}\label{constraint minimum rate main3}
t^{v_h,v_l}_{b',b} \bar{r}^{n}_{b,u} \left(\boldsymbol{p}_{\kappa_2}\right) \leq t^{v_h,v_l}_{b',b} r^{v_l}_{b',b},
\end{align}
\begin{align}\label{constraint minimum rate main4}
w^{v_h,v_l}_{b',b} \frac{ \bar{r}^{n}_{b,u} \left(\boldsymbol{p}_{\kappa_2}\right) } {s_{v_l}} \leq w^{v_h,v_l}_{b',b} \frac{ \phi^{v_h,v_l}_{b}  } { \eta^{v_h,v_l} N^{v_h,v_l}_{\text{Cycle}} },\end{align}
\begin{align}\label{constraint minimum rate main5}
o^{v_l}_{b} \bar{r}^{n}_{b,u} \left(\boldsymbol{p}_{\kappa_2}\right) \leq o^{v_l}_{b} r^{v_l}_{0,b},
\end{align}
which means the optimal solution generated for the convex approximated problem of \eqref{Main Problem jointpower transform} at each iteration $\kappa_2$ is feasible with respect to the nonconvex problem \eqref{Main Problem jointpower transform}.
Moreover, based on \eqref{gradian inequ2}, we have
\begin{multline}\label{inequality func 1}
\sum_{m \in \mathcal{M}} \bigg( \sum_{ u \in \mathcal{U}_m } \sum\limits_{b \in \mathcal{B}} \bar{r}^\text{Ac}_{b,u} \left(\boldsymbol{p}^{*}_{\kappa_2}\right) \psi_m
- \sum_{ u \in \mathcal{U}_m } \sum\limits_{b \in \mathcal{B}} \sum_{ n \in \mathcal{N} } \left( p^{n,(\kappa_2)}_{b,u} \alpha^\text{Pow}_b \right)
- \sum\limits_{b \in \mathcal{B}} \sum_{ n \in \mathcal{N} } \left( \max\limits_{ u \in \mathcal{U}_m } \{ \tau^{n}_{b,u} \} W_\text{s} \alpha^\text{Sub}_b
\right)     \\
- \sum_{b \in \mathcal{B}} \min \left\{ \sum_{u \in \mathcal{U}_m} \theta_{b,u} , 1 \right\} \bar{\$}^{\text{Cost,RRS}}_{b} \bigg)
\geq
\sum_{m \in \mathcal{M}} \bigg( \sum_{ u \in \mathcal{U}_m } \sum\limits_{b \in \mathcal{B}} \sum\limits_{n \in \mathcal{N}} \left( \hat{\bar{r}}^{n}_{b,u} \left(\boldsymbol{p}^{*}_{\kappa_2}\right) \right) \psi_m
- \sum_{ u \in \mathcal{U}_m } \sum\limits_{b \in \mathcal{B}} \sum_{ n \in \mathcal{N} } \left( p^{n}_{b,u} \alpha^\text{Pow}_b \right) \\
- \sum\limits_{b \in \mathcal{B}} \sum_{ n \in \mathcal{N} } \left( \max\limits_{ u \in \mathcal{U}_m } \{ \tau^{n}_{b,u} \} W_\text{s} \alpha^\text{Sub}_b
\right) - \sum_{b \in \mathcal{B}} \min \left\{ \sum_{u \in \mathcal{U}_m} \theta_{b,u} , 1 \right\} \bar{\$}^{\text{Cost,RRS}}_{b} \bigg).
\end{multline}
By using the fact that the DCP approximated problem of \eqref{Main Problem jointpower transform} can be efficiently solved at each iteration $\kappa_2$, where the globally optimal solution is in the feasible region of \eqref{Main Problem jointpower transform}, it can be derived
\begin{multline}\label{inequality func 2}
\sum_{m \in \mathcal{M}} \bigg( \sum_{ u \in \mathcal{U}_m } \sum\limits_{b \in \mathcal{B}} \sum\limits_{n \in \mathcal{N}} \left( \hat{\bar{r}}^{n}_{b,u} \left(\boldsymbol{p}^{*}_{\kappa_2}\right) \right) \psi_m
- \sum_{ u \in \mathcal{U}_m } \sum\limits_{b \in \mathcal{B}} \sum_{ n \in \mathcal{N} } \left( p^{n,(\kappa_2)^*}_{b,u} \alpha^\text{Pow}_b \right) -
\sum\limits_{b \in \mathcal{B}} \sum_{ n \in \mathcal{N} } \left( \max\limits_{ u \in \mathcal{U}_m } \{ \tau^{n}_{b,u} \} W_\text{s} \alpha^\text{Sub}_b
\right)
\\
- \sum_{b \in \mathcal{B}} \min \left\{ \sum_{u \in \mathcal{U}_m} \theta_{b,u} , 1 \right\} \bar{\$}^{\text{Cost,RRS}}_{b} \bigg)
=
\max_{ \boldsymbol{\phi}_{\kappa_2} , \boldsymbol{p}_{\kappa_2} , \boldsymbol{r}^\text{BH}_{\kappa_2}, \boldsymbol{r}^\text{FH}_{\kappa_2} }
\sum_{m \in \mathcal{M}} \bigg( \sum_{ u \in \mathcal{U}_m } \sum\limits_{b \in \mathcal{B}} \sum\limits_{n \in \mathcal{N}} \left( \hat{\bar{r}}^{n}_{b,u} \left(\boldsymbol{p}_{\kappa_2}\right) \right) \psi_m
-
\\
\sum_{ u \in \mathcal{U}_m } \sum\limits_{b \in \mathcal{B}} \sum_{ n \in \mathcal{N} } \left( p^{n,(\kappa_2)}_{b,u} \alpha^\text{Pow}_b \right) -
\sum\limits_{b \in \mathcal{B}} \sum_{ n \in \mathcal{N} } \left( \max\limits_{ u \in \mathcal{U}_m } \{ \tau^{n}_{b,u} \} W_\text{s} \alpha^\text{Sub}_b
\right) - \sum_{b \in \mathcal{B}} \min \left\{ \sum_{u \in \mathcal{U}_m} \theta_{b,u} , 1 \right\} \bar{\$}^{\text{Cost,RRS}}_{b} \bigg)
\geq
\\
\sum_{m \in \mathcal{M}} \bigg( \sum_{ u \in \mathcal{U}_m } \sum\limits_{b \in \mathcal{B}} \sum\limits_{n \in \mathcal{N}} \left(
\mathbb{E}_{\boldsymbol{h}} \left\{ f^{n}_{b,u} \left(\boldsymbol{p}_{\kappa_2-1}\right) - g^{n}_{b,u}\left(\boldsymbol{p}_{\kappa_2-1}\right) -
\nabla g^{n}_{b,u}\left(\boldsymbol{p}_{\kappa_2-1}\right)
\left(\boldsymbol{p}_{\kappa_2-1} - \boldsymbol{p}_{\kappa_2-1}\right) \right\}
\right) \psi_m -
\\
\sum_{ u \in \mathcal{U}_m } \sum\limits_{b \in \mathcal{B}} \sum_{ n \in \mathcal{N} } \left( p^{n,(\kappa_2-1)}_{b,u} \alpha^\text{Pow}_b \right)
- \sum\limits_{b \in \mathcal{B}} \sum_{ n \in \mathcal{N} } \left( \max\limits_{ u \in \mathcal{U}_m } \{ \tau^{n}_{b,u} \} W_\text{s} \alpha^\text{Sub}_b
\right) -
\sum_{b \in \mathcal{B}} \min \left\{ \sum_{u \in \mathcal{U}_m} \theta_{b,u} , 1 \right\} \bar{\$}^{\text{Cost,RRS}}_{b} \bigg)
=
\\
\sum_{m \in \mathcal{M}} \bigg( \sum_{ u \in \mathcal{U}_m } \sum\limits_{b \in \mathcal{B}} \sum\limits_{n \in \mathcal{N}}
\left(
\mathbb{E}_{\boldsymbol{h}} \left\{ f^{n}_{b,u} \left(\boldsymbol{p}_{\kappa_2-1}\right) - g^{n}_{b,u}\left(\boldsymbol{p}_{\kappa_2-1}\right) \right\}
\right)
\psi_m - \sum_{ u \in \mathcal{U}_m } \sum\limits_{b \in \mathcal{B}} \sum_{ n \in \mathcal{N} } \left( p^{n,(\kappa_2-1)}_{b,u} \alpha^\text{Pow}_b \right) -
\\
\sum\limits_{b \in \mathcal{B}} \sum_{ n \in \mathcal{N} } \left( \max\limits_{ u \in \mathcal{U}_m } \{ \tau^{n}_{b,u} \} W_\text{s} \alpha^\text{Sub}_b
\right) - \sum_{b \in \mathcal{B}} \min \left\{ \sum_{u \in \mathcal{U}_m} \theta_{b,u} , 1 \right\} \bar{\$}^{\text{Cost,RRS}}_{b} \bigg)
=
\\
\sum_{m \in \mathcal{M}} \bigg( \sum_{ u \in \mathcal{U}_m } \sum\limits_{b \in \mathcal{B}} \sum\limits_{n \in \mathcal{N}} \left( \bar{r}^{n}_{b,u} \left(\boldsymbol{p}_{\kappa_2-1}\right) \right) \psi_m
- \sum_{ u \in \mathcal{U}_m } \sum\limits_{b \in \mathcal{B}} \sum_{ n \in \mathcal{N} } \left( p^{n,(\kappa_2-1)}_{b,u} \alpha^\text{Pow}_b \right) -
\\
\sum\limits_{b \in \mathcal{B}} \sum_{ n \in \mathcal{N} } \left( \max\limits_{ u \in \mathcal{U}_m } \{ \tau^{n}_{b,u} \} W_\text{s} \alpha^\text{Sub}_b
\right) - \sum_{b \in \mathcal{B}} \min \left\{ \sum_{u \in \mathcal{U}_m} \theta_{b,u} , 1 \right\} \bar{\$}^{\text{Cost,RRS}}_{b} \bigg).
\end{multline}
According to \eqref{inequality func 1} and \eqref{inequality func 2}, it is obvious that
\begin{multline}\label{inequality func 3}
\sum_{m \in \mathcal{M}} \bigg( \sum_{ u \in \mathcal{U}_m } \sum\limits_{b \in \mathcal{B}} \bar{r}^\text{Ac}_{b,u} \left(\boldsymbol{p}^{*}_{\kappa_2}\right) \psi_m
- \sum_{ u \in \mathcal{U}_m } \sum\limits_{b \in \mathcal{B}} \sum_{ n \in \mathcal{N} } \left( p^{n,(\kappa_2)}_{b,u} \alpha^\text{Pow}_b \right)
- \sum\limits_{b \in \mathcal{B}} \sum_{ n \in \mathcal{N} } \left( \max\limits_{ u \in \mathcal{U}_m } \{ \tau^{n}_{b,u} \} W_\text{s} \alpha^\text{Sub}_b
\right)     \\
- \sum_{b \in \mathcal{B}} \min \left\{ \sum_{u \in \mathcal{U}_m} \theta_{b,u} , 1 \right\} \bar{\$}^{\text{Cost,RRS}}_{b} \bigg)
\geq
\sum_{m \in \mathcal{M}} \bigg( \sum_{ u \in \mathcal{U}_m } \sum\limits_{b \in \mathcal{B}} \sum\limits_{n \in \mathcal{N}} \left( \bar{r}^{n}_{b,u} \left(\boldsymbol{p}_{\kappa_2-1}\right) \right) \psi_m -
\\
\sum_{ u \in \mathcal{U}_m } \sum\limits_{b \in \mathcal{B}} \sum_{ n \in \mathcal{N} } \left( p^{n,(\kappa_2-1)}_{b,u} \alpha^\text{Pow}_b \right) -
\sum\limits_{b \in \mathcal{B}} \sum_{ n \in \mathcal{N} } \left( \max\limits_{ u \in \mathcal{U}_m } \{ \tau^{n}_{b,u} \} W_\text{s} \alpha^\text{Sub}_b
\right)
- \sum_{b \in \mathcal{B}} \min \left\{ \sum_{u \in \mathcal{U}_m} \theta_{b,u} , 1 \right\} \bar{\$}^{\text{Cost,RRS}}_{b} \bigg),
\end{multline}
which means after each iteration $\kappa_2$ of the proposed SCA approach for solving \eqref{Main Problem jointpower transform}, the objective function \eqref{obf Main Problem jointpower transform} which is exactly the same as \eqref{obf Main Problem jointpower} is improved (increased) or remains constant. Accordingly, the proposed algorithm for solving \eqref{Main Problem jointpower} will converge to a local optimum solution.



\hyphenation{op-tical net-works semi-conduc-tor}
\bibliographystyle{IEEEtran}
\bibliography{IEEEabrv,Bibliography}

\end{document}